%% file: main.tex
\documentclass[11pt,a4paper]{article}

\usepackage{fullpage}

\usepackage{amssymb,amsmath} 
\usepackage{MnSymbol} 
\usepackage{amsthm}

\usepackage{diagbox}
\usepackage{array}

\usepackage[colorlinks=true, allcolors=blue]{hyperref}
\usepackage{cleveref}

\usepackage[final]{graphicx}

\usepackage{float}
\usepackage[vlined,linesnumbered,ruled]{algorithm2e}

\usepackage{soul}
\usepackage{xspace}
\usepackage[dvipsnames]{xcolor}
\definecolor{revisioncyan}{RGB}{190,235,245}

\usepackage{paralist}
\usepackage{subcaption}

\usepackage{pifont}
\usepackage{tikz}
\usetikzlibrary{positioning}
\usetikzlibrary{shapes}
\usetikzlibrary{shapes.symbols,patterns}
\usetikzlibrary{calc,through,backgrounds}
\usetikzlibrary{decorations.pathreplacing}
\usetikzlibrary{arrows}
\usetikzlibrary{matrix}
\usetikzlibrary{shadows}
\usetikzlibrary{fit}

\usepackage[most]{tcolorbox}

\newtheorem{theorem}{Theorem}
\newtheorem{lemma}{Lemma}
\newtheorem{observation}[lemma]{Observation}
\newtheorem{corollary}[theorem]{Corollary}

\crefname{observation}{observation}{observations}
\Crefname{observation}{Observation}{Observations}

\theoremstyle{definition}

\crefname{algocf}{algorithm}{algorithms}
\Crefname{algocf}{Algorithm}{Algorithms}

\newcommand{\paren}[1]{\left( #1 \right)}

\newcommand{\set}[1]{\left\{ #1 \right\}}

\newcommand{\ceil}[1]{\left\lceil #1 \right\rceil}
\newcommand{\floor}[1]{\left\lfloor #1 \right\rfloor}
\newcommand{\abs}[1]{\left| #1 \right|}

\newcommand{\calB}{\mathcal{B}}
\newcommand{\calF}{\mathcal{F}}
\newcommand{\calG}{\mathcal{G}}
\newcommand{\calP}{\mathcal{P}}
\newcommand{\calT}{\mathcal{T}}

\newcommand{\nBTwo}{\omega_{\geq 2}}

\newcommand{\opt}{\text{opt}}

\tikzset{
  thesisDVertex/.style={circle,draw=black,fill=black,text=white,
    font=\scriptsize,minimum size=12pt,inner sep=0pt,line width=0.4pt},
  thesisNVertex/.style={circle,draw=black,fill=white,text=black,
    font=\tiny,minimum size=9.5pt,inner sep=0pt,line width=0.35pt}
}

\newcommand{\StarTree}[7]{
  \begin{scope}[shift={(#1,#2)}]
    \node(B#3) [thesisNVertex] at (-4,3) {#5};
    \node(C#3) [thesisNVertex] at (-1,-4) {#6};
    \node(D#3) [thesisNVertex] at (2,3) {#7};
    \node(A#3) [thesisDVertex] at (-1,0) {#4};
    \begin{scope}[every edge/.style={draw=black,ultra thick}]
      \path (A#3) edge (B#3);
      \path (A#3) edge (C#3);
      \path (A#3) edge (D#3);
    \end{scope}
  \end{scope}
}

\newcommand{\StarTreeSharedLeft}[7]{
  \begin{scope}[shift={(#1,#2)}]
    \node(C#3) [thesisNVertex] at (-1,-4) {#6};
    \node(D#3) [thesisNVertex] at (2,3) {#7};
    \node(A#3) [thesisDVertex] at (-1,0) {#5};
    \begin{scope}[every edge/.style={draw=black,ultra thick}]
      \path (A#3) edge (#4);
      \path (A#3) edge (C#3);
      \path (A#3) edge (D#3);
    \end{scope}
  \end{scope}
}

\newcommand{\degrlong}{\textsc{Degree Realization}\xspace}
\newcommand{\degr}{\textsc{DegR}\xspace}

\newcommand{\clique}{\textsc{Max Clique}\xspace}
\newcommand{\indset}{\textsc{Max Independent Set}\xspace}

\newcommand{\vc}{\textsc{Min Vertex Cover}\xspace}
\newcommand{\totvc}{\textsc{Min Total Vertex Cover}\xspace}
\newcommand{\pairvc}{\textsc{Min Paired Vertex Cover}\xspace}

\newcommand{\domset}{\textsc{Min Dominating Set}\xspace}
\newcommand{\totdomset}{\textsc{Min Total Dominating Set}\xspace}
\newcommand{\pairdomset}{\textsc{Min Paired Dominating Set}\xspace}
\newcommand{\edgedomset}{\textsc{Min Edge Dominating Set}\xspace}

\newcommand{\edgemerge}{\textsf{EdgeMerge}\xspace}
\newcommand{\leafmerge}{\textsf{LeafMerge}\xspace}
\newcommand{\edgeaddition}{\textsf{EdgeAddition}\xspace}

\newcommand{\vertexcover}{\textsf{Realization-VC}\xspace}
\newcommand{\dominatingset}{\textsf{Realization-DS}\xspace}

\newcommand{\vertexcoverghr}{\textsf{Realization-VC-GHR}\xspace}

\newcommand{\taut}{\tau_t}
\newcommand{\taup}{\tau_p}

\newcommand{\gammat}{\gamma_t}
\newcommand{\gammap}{\gamma_p}
\newcommand{\gammae}{\gamma_e}

\newcommand{\gammakk}{\bar{\gamma}}

\newcommand{\addnl}{\ell_{NL}}

\soulregister\vc0
\soulregister\totvc0
\soulregister\pairvc0
\soulregister\edgedomset0
\soulregister\degrlong0
\soulregister\totdomset0
\soulregister\pairdomset0
\soulregister\dominatingset0
\soulregister\vertexcover0
\soulregister\vertexcoverfromforest0
\soulregister\dominatingsetfromforest0
\soulregister\edgemerge0
\soulregister\leafmerge0
\soulregister\vertexcoverghr0
\soulregister\log0
\soulregister\cite7
\soulregister\Cref7

\newcommand{\dror}[1]{{\color{blue}\underline{\bf{Dror:}} #1}}

\def\omri#1{{\color{ForestGreen}\underline{\bf{Omri:}} #1}}

\begin{document}

\title{\textbf{Computing Forest Degree Realizations \\ with Minimum Cardinality Domination}%
\thanks{Supported in part by a US-Israel BSF grant (2022205).}
}

\author{%
Amotz Bar-Noy%
\thanks{Department of Computer Science, 
City University of New York, 
 USA;
{\tt amotz@sci.brooklyn.cuny.edu}}
\and
Omri Kosary%
\thanks{Department of Computer Science, Technion, Haifa, Israel;
{\tt omri.kosary@campus.technion.ac.il}}
\and
David Peleg%
\thanks{Department of Computer Science and Applied Mathematics, 
Weizmann Institute of Science, Rehovot, Israel;
{\tt david.peleg@weizmann.ac.il}}
\and
Dror Rawitz%
\thanks{Faculty of Engineering, Bar Ilan University, Ramat-Gan, Israel;
{\tt dror.rawitz@biu.ac.il}}
}

\maketitle

\begin{abstract}
In the \degrlong problem with respect to a family $\calP$ of graphs
the input is a non-increasing sequence $d = (d_1,\ldots,d_n)$ 
of positive integers, and the goal is to decide whether there exists a simple undirected graph 
$G \in \calP$, whose degrees correspond to $d$, i.e., such that $\deg(G) = d$. 
In this paper we consider the version of \degrlong in which the realization is required to be a forest 
(i.e., $\calP$ is the family of forests).

We consider \emph{optimized} \degrlong in which the goal is to obtain a realization that minimizes 
an objective function $f$. That is, the goal is to find a realization $G$ that minimizes $f(G)$ 
among the realizations of the given input sequence.
More specifically, we focus on the following functions: the size of an optimal \emph{vertex cover} and 
the size of an optimal \emph{dominating set}. 
We also consider the \emph{total} and \emph{paired} versions of both \vc and \domset.
We provide characterizations and linear time realization algorithms for all the above-mentioned problems.
Since \pairvc is closely related to \edgedomset, our results for \degrlong 
with respect to the former extend to the latter.
\end{abstract}


\section{Introduction}



In the \degrlong (\degr) problem the input is a non-increasing sequence $d = (d_1,\ldots,d_n)$ 
of $n$ positive integers, and the goal is to decide whether there exists a simple undirected graph 
$G$, whose degrees correspond to $d$, i.e., such that $\deg(G) = d$. 
If there exists such a graph $G$, then $d$ is called \emph{graphic}.
Erd\H{o}s and Gallai~{\cite{EG60}} gave necessary and sufficient conditions for the realizability of a given 
sequence, which imply a decision algorithm whose running time is $O(n)$.
Havel~\cite{havel55} and Hakimi~\cite{Hakimi62} presented an algorithm for finding a realization, 
whose running time depends linearly on the sum of all degrees, i.e., has running time $O(\sum_i d_i)$.



Given a sequence $d$, it is sometimes required that the realizing graph $G$ of $d$ has a certain property.
For example, it may be the case that $G$ has to be bipartite. 
We formally define a property as the collection of graphs that satisfy the property.
The problem of \degr with respect to a graph family $\calP$ is similar to standard \degr
but with one more requirement, which is that the realization $G$ belongs to $\calP$.
Clearly, when $\calP = \calG$, where $\calG$ is the collection of all graphs, we revert to regular \degr.
In this work, we are mainly interested in the family $\calF$ of forests.
%
Gupta et al.~\cite{GJT07} proved that a sequence $d$ is realizable by a tree if and only if $\sum_i d_i = 2n-2$.  
This result readily extends to forests, namely a sequence can be realized using a forest if and only if 
$\sum_i d_i$ is even and $\sum_i d_i \leq 2n-2$. Such a sequence is called \emph{forest-graphic} or \emph{forestic}.
The above result is constructive and implies a realization algorithm for forests.
In fact, it can be shown that a forestic sequence always has a realization consisting of a caterpillar and a matching
(see, e.g., \cite{BBPRR25-2pbe}).



A sequence $d$ may have several realizations, some of which are more desirable than others.
Let $f: \calG \to \mathbb{N}$ be a function that assigns \emph{scores} to graphs.
We may prefer a realization, which minimizes or maximizes the score.
Formally, given a sequence $d$ the goal is to find a realization $G$ of $d$, which 
minimizes or maximizes $f(G)$. That is, to compute a realization optimizing the following function:
\[
f_{\opt}(d) = \opt \set{f(G) : G \text{ is a realization of } d}
~,
\]
where $\opt \in \set{\min,\max}$ and to provide a graph $G$ such that $f(G) = f_{\opt}(d)$.
Such problems are called \emph{optimized} realization problems.
For example, consider the size of a minimum cardinality \emph{dominating set} in $G$,
which is denoted by $\gamma(G)$. That is,
\[ 
\gamma(G) = \min \set{\abs{D} : D \subseteq V \text{ is a dominating set in } G}
~.
\]
The optimized \degr problem with respect to $\gamma_{\min}$ is to find a realization $G$ of $d$ 
with minimum size optimal dominating set among the realizations of $d$.
In this work we only consider minimization problems, thus we use $f(d)$ instead of $f_{\min}(d)$
(e.g., $\gamma(d)$ instead of $\gamma_{\min}(d)$).
One may also add a property, constraining the realization to a graph family $\calP$.
For example, the optimized realization problem with respect to the family $\calF$ 
and the function $\gamma$, or with respect to $\gamma^{\calF}$, 
is to find a forest realization of $d$ with a minimum size optimal dominating set 
among the forest realizations of $d$.

Given a graph $G = (V,E)$, a dominating set (vertex cover) $D \subseteq V$ is called \emph{total} if
the induced subgraph $G[D]$ does not contain isolated vertices.
A dominating set (vertex cover) is called \emph{paired} if $G[D]$ contains a perfect matching.
In this paper we consider \degr by forests with respect to the following score functions:
\begin{compactitem}
\item $\tau(G)$ and $\gamma(G)$ -- the sizes of an optimal vertex cover and dominating set.
\item $\taut(G)$ and $\gammat(G)$ -- the sizes of an optimal total vertex cover and total dominating set.
\item $\taup(G)$ and $\gammap(G)$ -- the sizes of an optimal paired vertex cover and paired dominating set.
\end{compactitem}

The most relevant paper is by Gentner et al.~\cite{GHR18},
who worked on optimized degree realization with respect to $\tau^\calF$ and $\gamma^\calF$.
They proved that a forestic sequence $d$ is realizable by a graph with 
a vertex cover of size $r$ if and only if $\sum_{i=1}^r d_i \geq \sum_{i=r+1}^n d_i$. 
Hence, 
\[
\tau^{\calF}(d) = \min \set{ k : \sum_{i=1}^k d_i \geq \sum_{i=k+1}^n d_i }
\]
Their proof is constructive in the sense that it can lead to a realization algorithm.
They also provided the following condition for \degr with respect to $\gamma^{\calF}$:
If $d$ is a forestic sequence 
then 
\[
\gamma^\calF(d) = \min\set{\tau^\calF(d) ,\gammakk(d)}
\]
where  
\begin{align*}
\gammakk(d)
& = \min \set{ k : \sum_{i=1}^k d_i \geq n-k \text{ and } 0 <  \sum_{i=k+1}^n d_i - \sum_{i=1}^k d_i \leq 2(\nBTwo-k) -2 }
\end{align*}
and $\omega_1$ is the multiplicity of $1$ in $d$ and $\nBTwo = n - \omega_1$.
(If $\gammakk(d)$ is undefined, then $\gamma^\calF(d) = \tau^\calF(d)$.)
It is important to note that the above characterizations were obtained due to a \emph{prefix lemma}
that states that, given a sequence $d$, if there is a forest realization of $d$ with a
dominating set (vertex cover) of size $t$, 
then there is a realization with a dominating set (vertex cover) that consists of the $t$ highest degree vertices.


\subsection{Our Results}

In this work we study several optimized \degr by forests with respect to several 
domination problems. More specifically, we consider the optimized \degr 
in forests for \vc and \domset, their total versions, and their paired versions.
For each of the problems we present both a characterization and a linear time realization algorithm.

Our main tool for designing realization algorithms is a framework that works as follows.
It starts with a star forest where the star centers constitute a dominating set (or a vertex cover). 
Then, it uses three simple operations that combine two trees into one:
\begin{inparaenum}[(i)]
\item merging two vertices,
\item merging two edges, and 
\item adding an edge.
\end{inparaenum}
We note that our results for the variants of \vc are used to obtain our results for the variants of \domset.

The above framework relies on the fact that the dominating set (vertex cover) is composed of vertices that 
correspond to a prefix of the given sequence.
Our results for \degr with respect to $\tau^{\calF}$ and $\gamma^{\calF}$
rely on the prefix lemmas that were proven in~\cite{GHR18}.
We provide prefix lemmas for \degr with respect to $\taut^{\calF}$ and $\gammat^{\calF}$.

We show that, given a forestic sequence $d$, if $\taut^{\calF}(d)$ is even, then $\taup^{\calF}(d) = \taut^{\calF}(d)$, 
and that if $\gammat^{\calF}(d)$ is even, then $\gammap^{\calF}(d) = \gammat^{\calF}(d)$.
In fact, in this case our realization algorithms for the total versions compute a realization with a paired solution.
In addition, we prove that if $\taut^{\calF}(d)$ is odd, then $\taup^{\calF}(d) = \taut^{\calF}(d) +1$.
Similarly, we prove that if $\gammat^{\calF}(d)$ is odd, then $\gammap^{\calF}(d) = \gammat^{\calF}(d) +1$.
In this case, we show that one may add one vertex to a total solution in order to obtain a paired solution
for the realizations that are computed by our algorithms.
The added vertex is not arbitrary and should be chosen carefully.
It is known that the minimum size of an \emph{edge dominating set} equals 
the minimum size of a \emph{maximal matching}.
Moreover, $\taup(G) = 2\gammae(G)$, for every graph $G$, 
where $\gammae$ is the edge domination number (see \Cref{sec:preliminaries} for more details).
Hence, our results extend to \degr with respect to $\gammae^{\calF}$.

Finally, in Appendix~\ref{sec:vc-previous} we give a realization algorithm that is implied  
by the characterization of \degr with respect to $\tau^{\calF}$ from~{\cite{GHR18}} 
and show that it can be implemented to run in linear time.


\subsection{Related Work}


Rao~\cite{Rao79} studied \degr by a graph containing a clique of size $r$ (i.e., $K_r$), for any $r$. 
He gave a characterization of sequences that are realizable with respect to this property. 
His result was based on a \emph{prefix lemma} that showed that if there exists a realization 
with a $K_r$, then there exists a realization such that the top $r$ degrees induce a $K_r$.
Corresponding realization algorithms were given in~\cite{Rao79b,Yin12} (see also \cite{GHR16}).
These realization algorithms can be used for finding a realization with the largest clique, 
namely for solving the optimized realization problem with respect to $\omega_{\max}$,
where $\omega(G)$ is the \emph{clique number} of $G$.
As noted in~\cite{GHR16}, the result for \clique implies a similar result for \indset, since
\(
\alpha_{\max}(d_1,\ldots,d_n) = \omega_{\max}(n-1-d_n,\ldots,n-1-d_1)
\),
where $\alpha(G)$ is the \emph{independence number} of $G$.
In addition, recall that $\alpha(G) + \tau(G) = n$, for every graph $G$,
where $\tau(G)$ is the minimum vertex cover size of $G$.
It follows that $\tau_{\min}(d) = n - \alpha_{\max}(d)$, for any sequence $d$,
which means that the same result also applies to realization with respect to \vc.
A general prefix lemma was given in~\cite{GJL99} where it was shown 
that if a sequence $d$ has a realization $G$ with a subgraph $H$ with $t$ vertices, 
then there exists a realization $G'$ such that $H$ is a subgraph of $G'$ whose vertices 
correspond to the top $t$ degrees.
A characterization for \degr with respect to $\nu_{\max}$ was given in~\cite{EKMT24}, 
where $\nu(G)$ is the matching number of $G$, 
while a corresponding realization algorithm was provided in~\cite{BKPR26-matching}.
The latter result was based on a stronger version of the prefix lemma from~\cite{GJL99}.
Recently, a characterization and a realization algorithm for \degr with respect to $\gamma_{\min}$
were given in~\cite{BKPR26-domset}. These results were also based on a prefix lemma that was shown in~\cite{GHR18}.


Optimized \degr was studied in several graph families.
As mentioned above, \degr with respect to $\gamma_{\min}^\calF$ was considered in~\cite{GHR18}.
Characterizations with respect to $\gamma_{\max}^\calF$ and $\alpha_{\min}^\calF$ were given in~\cite{GHR16}.
A characterization for \degr with respect to $\nu^{\calT}_{\max}$, where $\calT$ is the family of trees, 
was given in~\cite{BockRautenbach19}. The latter paper also contains a characterization and a realization algorithm
for \degr with respect to $\nu^{\calB}_{\max}$, where $\calB$ is the family of bipartite graphs, assuming 
that the input is a bi-sequence. The latter results were obtained using a strong and bipartite version of a prefix lemma.
Also, these results imply a characterization and a realization algorithm with respect to $\tau^{\calB}_{\min}$ and
$\alpha^{\calB}_{\max}$.

\vc 
cannot be approximated within a factor of $\sqrt{2} - o(1)$, 
unless $\text{P} = \text{NP}$~{\cite{KMS17}}, and if the Unique Games Conjecture is true, 
then \vc cannot be approximated within any constant factor less than 2~{\cite{KhotRegev08}}.
On the other hand, it is known to be approximable within a factor of $2$~\cite{NemhauserTrotter1975,Bar-YehudaEven1981,Clark1983,Hochbaum1982}, and 
$2 - o(1)$~\cite{Bar-YehudaEven1985,MonienSpeckenmeyer1985,Halperin2002,Karakostas05}.
It was shown that \totvc is APX-hard and that it has a 2-approximation algorithm in~\cite{FernauManlove09}.
In~\cite{PoonWang17} it was shown that \totvc is APX-hard even in connected sub-cubic graphs.
\pairvc is closely related to edge domination:
Duarte et al.~\cite{DuarteOliveiraSouza21} showed that $\taup(G) = 2\gammae(G)$, for any graph $G$.
Hence, solving \pairvc is equivalent to solving \edgedomset.
As for \edgedomset, it is NP-hard~\cite{YannakakisGavril80}, and it cannot be approximated within $1.18$,
unless $\text{P} = \text{NP}$~\cite{EMPX15}. Also, if the Unique Games Conjecture is true, 
it has no polynomial-time approximation with any constant ratio smaller than 2~\cite{DudyczLewandowskiMarcinkowski19}.
A $(2-o(1))$-approximation algorithm was given in~\cite{GotthilfLewensteinRainshmidt09}.

\domset  is not approximable within $(1-o(1)) \ln n$, unless $\text{P}=\text{NP}$~\cite{DinurSteurer14}.
Since \domset and \totdomset are special cases of \textsc{Min Set Cover}, 
they can be approximated within $O(\log n)$, where $n = \abs{V}$~\cite{Joh74,Lov75,Chv79,Slavik96}.
\totdomset is NP-hard to approximate within $(1-o(1))\ln n$~\cite{WangKitamuraIzumiMasuzawa25}. 
For more details on \totdomset see~\cite{Henning09,HenningYeo13}.
\pairdomset admits a $(1+\ln(2\Delta))$-approximation, where $\Delta$ is the maximum degree~\cite{ChenLuZeng09}. 
The problem is NP-hard to approximate within $(1-o(1))\ln n$~\cite{WangKitamuraIzumiMasuzawa25}. 
For further background, see~\cite{McCoy13}.




\section{Preliminaries}
\label{sec:preliminaries}

\subsection{Definitions and Notation}

Let $G = (V,E)$ be a simple graph, where $V = \set{v_1, \ldots, v_n}$.
Let $N(v_i) = \set{v_j : (v_i,v_j) \in E}$ be the neighborhood of vertex $v_i \in V$,
and let $\deg(v_i) = |N(v_i)|$ be the degree of $v_i$, for every $v_i \in V$.
The degree sequence of $G$ is defined as $\deg(G) = (\deg_G(v_1), \ldots, \deg_G(v_n))$.
In this paper we assume that a graph $G$ has no isolated vertices.
Given subsets $X, Y \subseteq V$, let $E(X,Y) = \set{(x,y) \in E : x \in X, y \in Y}$.

Let $d = (d_1, \ldots,d_n)$ be a non-increasing sequence of positive integers.
If there exists a graph $G$ such that $\deg(G) = d$, then it is said that $G$ \emph{realizes} $d$.
A sequence $d$ that has a realization $G$ is called $\emph{graphic}$.
We refer to $\sum d = \sum_i d_i$ as the \emph{volume} of $d$.
Observe that if $d$ is graphic, then a realization $G$ of $d$ has $m = \frac{1}{2} \sum d$ edges.
Clearly, a degree of a vertex is at most $n-1$ and the sum of the degrees in a graph is even. 
Also, zero degree means an isolated vertex. Hence, it is enough to consider \emph{normal} sequences, 
where a sequence $d$ is called \emph{normal} if $d_i \in \set{1,\ldots,n-1}$, for every $i$, 
and the \emph{volume} $\sum d$ is even.
%
Given a normal sequence $d$ let $\omega_k$ denote the multiplicity of $k$ in $d$, namely 
$\omega_k \triangleq |\set{i : d_i = k}|$. Also, define $\omega_{\geq k} \triangleq \sum_{j \geq k} \omega_j$.
For the sake of brevity, we use $a^k$ as a shorthand for a subsequence of $k$ consecutive $a$'s 
(e.g., $2^3$ represents $2,2,2$).

A sequence $d$  that can be realized by a forest is called \emph{forestic}.
Given a sequence $d$ and a forest $F$ realizing $d$, the number of trees in $F$ is 
$\kappa \triangleq n - \frac{1}{2} \sum d$. 



\subsection{Simple Observations}

This section contains several simple observations that will be useful in the sequel.

\begin{observation}
\label{obs:vcIsds}
A vertex cover of a graph $G$ (with no isolated vertices) is also a dominating set of $G$.
\end{observation}

\begin{observation}
\label{obs:tvcIstds}
A total (paired) vertex cover of a graph $G$ (with no isolated vertices) is also a total (paired) dominating set of $G$.
\end{observation}

\begin{observation}
\label{obs:pairedlstotal}
A paired vertex cover (dominating set) $D$ of graph $G$ is also a total vertex cover (dominating set) of graph $G$.
\end{observation}


The next observation has to do with the size of a dominating set.

\begin{observation}
\label{obs:ds-size}
Let $d$ be a graphic sequence with a realization $G$ such that $\deg(v_i) = d_i$.
A dominating set $D$ in $G$ satisfies $\sum_{v_i \in D} d_i \geq n - |D|$.
\end{observation}

A trivial condition on the size of a total dominating set is that it is at least $2$.
Another condition is given in the next observation.

\begin{observation}
\label{obs:tds-odd}
Let $d$ be a graphic sequence with a realization $G$ such that $\deg(v_i) = d_i$.
A total dominating set $D$ in $G$ satisfies 
\(
\sum_{v_i \in D} d_i \geq n + 2\ceil{|D|/2} - |D|
\).
\end{observation}
\begin{proof}
A total dominating set $D$ must dominate all vertices not in $D$, hence
$|E(D,V \setminus D)| \geq n - |D|$.
Also, $G[D]$ contains no isolated vertices, therefore $\sum_{v_i \in D} \deg_{G[D]}(v_i) \geq |D|$.
Since the sum of degrees is an even number we have that $
2|E(D,D)| = \sum_{v_i \in D} \deg_{G[D]}(v_i) \geq 2\ceil{|D|/2}$.
Overall we get
\[
\sum_{v_i \in D} d_i = |E(D,V\setminus D)|  + 2\cdot |E(D,D)| \geq n - |D| + 2\ceil{|D|/2} 
~.
\qedhere
\]
\end{proof}

Notice that $2\ceil{|D|/2} - |D| \in \set{0,1}$.
Observe that if $|D| = 1$, then  $\sum_{i \in D} d_i \leq d_1 \leq n-1$, while $n + 2\ceil{|D|/2} - |D| = n+1$.
Hence, the above bound implies that $|D| \geq 2$.

The following is implied by \Cref{obs:tvcIstds,obs:tds-odd}.

\begin{observation}
\label{obs:tvc-odd}
Let $d$ be a graphic sequence with a realization $G$ such that $\deg(v_i) = d_i$.
A total vertex cover $D$ in $G$ satisfies 
\(
\sum_{v_i \in D} d_i \geq n + 2\ceil{|D|/2} - |D|
\).
\end{observation}

Finally, we consider the connection between \pairvc and \edgedomset.
Clearly, any maximal matching is also an edge dominating set.
Yannakakis and Gavril~\cite{YannakakisGavril80} showed how to transform a minimum edge-dominating set 
into a maximal matching of the same cardinality (see also \cite{harary69}).
Duarte et al.~\cite{DuarteOliveiraSouza21} observed that the endpoints of a maximal matching induce
a paired vertex cover and that the matching induced by a paired vertex cover is maximal.

\begin{observation}[\cite{DuarteOliveiraSouza21}]
$\taup(G) = 2\gammae(G)$, for every simple graph $G$.
\end{observation}


\section{Domination Framework}
\label{sec:framework}

In this section, we describe a framework that, given a sequence $d$, 
constructs a realization that attains an optimal solution to several domination problems.

The framework consists of three operations on graphs and a definition of a graph 
which serves as the starting point.


\subsection{The initial Forest: Star Forest}

Given a subset $D \subseteq V$, we define an \emph{initial star forest} $S$,
which is composed of $|D|$ stars whose centers are the vertices of $D$.
More formally, given $D$, define $S = (V,E)$ as follows:
\begin{align*} 
V & = \bigcup_{v_i \in D} \paren{\set{v_i} \cup \set{v_i^j : j \in \set{1,\ldots,d_i} } }
&
E & = \bigcup_{v_i \in D} \bigcup_{j=1}^{d_i} \set{ (v_i,v_i^j) }
~.
\end{align*}

\begin{figure}[t]
\centering
\begin{tikzpicture}[scale=0.25] 
\begin{scope}[
every node/.style={thesisNVertex}
]
\node(B) at (-1,3) {1};
\node(C) at (2,0) {1};
\node(D) at (-1,-3) {1};
\node(E) at (-4,0) {1};

\node(G) at (5,3) {1};
\node(H) at (3,-2.5) {1}; %
\node(K) at (7,-2.5) {1}; %

\node(M) at (9,3) {1}; %

\end{scope}
\begin{scope}[
every node/.style={thesisDVertex}
]

\node(A) at (-1,0) {4};
\node(F) at (5,0) {3};
\node(L) at (9,0) {1}; %

\end{scope}
\begin{scope}[
every edge/.style={draw=black,ultra thick}
]
\path (A) [-] edge (B);
\path (A) [-] edge (C);
\path (A) [-] edge (D);
\path (A) [-] edge (E);
\path (F) [-] edge (G);
\path (F) [-] edge (H);
\path (F) [-] edge (K);
\path (L) [-] edge (M);

\end{scope}
\node[font=\scriptsize,above=1pt of C] {$x$};
\node[font=\scriptsize,left=1pt of H] {$y$};
\end{tikzpicture}
\caption{The initial star forest of $d = (4,3,1^9)$ with $3$ stars.}
\label{Figure:InitialStarForest}
\end{figure}
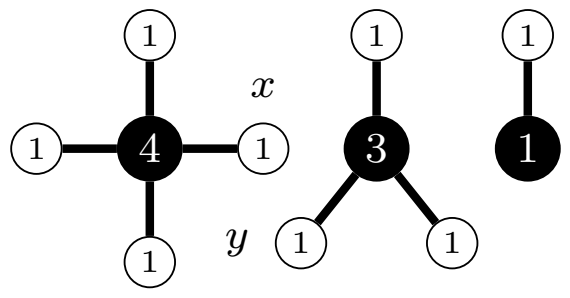

Let $n^S$, $d^S$, and $\kappa^S$ be the number of vertices, the degree sequence, and the number of trees of $S$, 
respectively.  An example is given in~\Cref{Figure:InitialStarForest}.


\begin{observation}
\label{obs:starforest}
\begin{inparaenum}[(i)]
\item $n^S = |D| + \sum_{v_i \in D} d_i$,  
\item $\sum d^S = 2 \cdot \sum_{v_i \in D} d_i$, and 
\item $\kappa^S = |D|$.
\end{inparaenum}
\end{observation}


\subsection{Edge Merge}

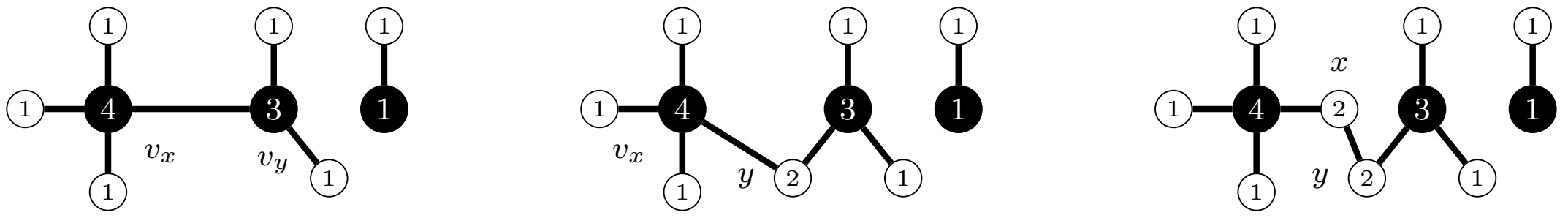
\begin{figure}[t]
\centering
\begin{tabular}{ccc}
\begin{subfigure}{0.3\textwidth}
\centering
\begin{tikzpicture}[scale=0.25] 
\begin{scope}[
every node/.style={thesisNVertex}
]

\node(B) at (-1,3) {1};
\node(D) at (-1,-3) {1};
\node(E) at (-4,0) {1};

\node(G) at (5,3) {1};
\node(K) at (7,-2.5) {1}; %

\node(M) at (9,3) {1}; %

\end{scope}
\begin{scope}[
every node/.style={thesisDVertex}
]

\node(A) at (-1,0) {4};
\node(F) at (5,0) {3};
\node(L) at (9,0) {1}; %

\end{scope}
\node[font=\scriptsize,below right=1pt and 1pt of A] {$v_x$};
\node[font=\scriptsize,below=1pt of F] {$v_y$};
\begin{scope}[
every edge/.style={draw=black,ultra thick}
]
\path (A) [-] edge (B);
\path (A) [-] edge (D);
\path (A) [-] edge (E);
\path (F) [-] edge (G);
\path (F) [-] edge (A);
\path (F) [-] edge (K);
\path (L) [-] edge (M);


\end{scope}
\end{tikzpicture}
\caption{Edge merge operation.}
\label{fig:edge_merge}
\end{subfigure}
&
\begin{subfigure}{0.3\textwidth}
\centering
\begin{tikzpicture}[scale=0.25] 
\begin{scope}[
every node/.style={thesisNVertex}
]

\node(B) at (-1,3) {1};
\node(D) at (-1,-3) {1};
\node(E) at (-4,0) {1};

\node(H) at (3,-2.5) {2};
\node(G) at (5,3) {1};
\node(K) at (7,-2.5) {1}; %

\node(M) at (9,3) {1}; %

\end{scope}
\begin{scope}[
every node/.style={thesisDVertex}
]

\node(A) at (-1,0) {4};
\node(F) at (5,0) {3};
\node(L) at (9,0) {1}; %

\end{scope}
\node[font=\scriptsize,left=1pt of H] {$y$};
\node[font=\scriptsize,below left=1pt and 1pt of A] {$v_x$};
\begin{scope}[
every edge/.style={draw=black,ultra thick}
]
\path (A) [-] edge (B);
\path (A) [-] edge (H);
\path (A) [-] edge (D);
\path (A) [-] edge (E);
\path (F) [-] edge (G);
\path (F) [-] edge (H);
\path (F) [-] edge (K);
\path (L) [-] edge (M);


\end{scope}
\end{tikzpicture}
\caption{Leaf merge operation.}
\label{fig:leaf_merge}
\end{subfigure}
&
\begin{subfigure}{0.3\textwidth}
\centering
\begin{tikzpicture}[scale=0.25] 
\begin{scope}[
every node/.style={thesisNVertex}
]

\node(B) at (-1,3) {1};
\node(C) at (2,0) {2};
\node(D) at (-1,-3) {1};
\node(E) at (-4,0) {1};

\node(G) at (5,3) {1};
\node(H) at (3,-2.5) {2}; %
\node(K) at (7,-2.5) {1}; %

\node(M) at (9,3) {1}; %

\end{scope}
\begin{scope}[
every node/.style={thesisDVertex}
]

\node(A) at (-1,0) {4};
\node(F) at (5,0) {3};
\node(L) at (9,0) {1}; %

\end{scope}

\node[font=\scriptsize,above=1pt of C] {$x$};
\node[font=\scriptsize,left=1pt of H] {$y$};
\begin{scope}[
every edge/.style={draw=black,ultra thick}
]
\path (A) [-] edge (B);
\path (A) [-] edge (C);
\path (A) [-] edge (D);
\path (A) [-] edge (E);
\path (F) [-] edge (G);
\path (F) [-] edge (H);
\path (F) [-] edge (K);
\path (L) [-] edge (M);
\path (C) [-] edge (H);

\end{scope}
\end{tikzpicture}
\caption{Edge addition operation.}
\label{fig:edge_addition}
\end{subfigure}
\end{tabular}
\caption{Examples of the three operations, each performed on the forest given in
\Cref{Figure:InitialStarForest}.}
\end{figure}

Let $F = (V,E)$ be a forest containing at least two trees, and let $D \subseteq V$.
Also, let $x$ and $y$ be two leaves that belong to two different trees.
Let $v_x, v_y \in D$ be the respective unique neighbors of $x$ and $y$.
The edge merge operation merges $x$ with $v_y$ and $y$ with $v_x$.
An example is given in~\Cref{fig:edge_merge}.

\smallskip

\begin{algorithm}[H]
\caption{\small \edgemerge}
\label{alg:EdgeMerge}
\DontPrintSemicolon
\SetKwInOut{Input}{input}
\SetKwInOut{Output}{output}
\Input{$(F,x,y)$}
\Output{$F'$}
$V' \gets V \setminus \set{x,y}$ \\
$E' \gets E \setminus \set{(x,v_x), (y,v_y)} \cup \set{(v_x,v_y)}$ \\
\Return $F' = (V',E')$
\end{algorithm}


\begin{observation}
\label{obs:edgemerge}
Let $F$ be a forest, and let $x, y \not\in D$ be leaves that belong to two different trees.
Let $F'$ be the graph obtained by an \edgemerge operation. Then,
\begin{inparaenum}[(i)]
\item $F'$ is a forest with $\kappa' = \kappa - 1$ trees,
\item $n' = n - 2$, 
\item $\sum d' = \sum d - 2$, and 
\item $\omega'_1 = \omega_1 - 2$, and $\omega'_i = \omega_i$, for every $i > 1$.
\end{inparaenum}
\end{observation}


\subsection{Leaf Merge}

Let $F = (V,E)$ be a forest containing at least two trees, and let $D \subseteq V$.
Let $x$ be a leaf and let $y \not\in D$, such that $x$ and $y$ belong to two different trees.
Let $v_x \in D$ be the only neighbor of $x$.
The leaf merge operation merges $x$ into $y$ .
An example is given in~\Cref{fig:leaf_merge}.

\smallskip

\begin{algorithm}[H]
\caption{\small \leafmerge}
\label{alg:LeafMerge}
\DontPrintSemicolon
\SetKwInOut{Input}{input}
\SetKwInOut{Output}{output}
\Input{$(F,x,y)$}
\Output{$F'$}
$V' \gets V \setminus \set{x} $ \\
$E' \gets E \setminus \set{(x,v_x)} \cup \set{(v_x,y)}$ \\
\Return $F' = (V',E')$
\end{algorithm}

\smallskip

\begin{observation}
\label{obs:leafMerge}
Let $F$ be a forest, and 
let $x, y \not\in D$ be two vertices that belong to two different trees, such that $x$ is a leaf.
Let $F'$ be the graph obtained by a \leafmerge operation. Then,
\begin{inparaenum}[(i)]
\item $F'$ is a forest with $\kappa' = \kappa - 1$ trees,
\item $n' = n - 1$, and
\item $\sum d' = \sum d$.
\end{inparaenum}
\end{observation}


\subsection{Edge Addition}

Let $F = (V,E)$ be a forest containing at least two trees, and let $D \subseteq V$.
Let $x \not\in D$ be a leaf and $y \not\in D$ be either a leaf or a vertex such that 
$x$ and $y$ belong to two different trees.
The edge addition operation connects $x$ with $y$.
An example is given in \Cref{fig:edge_addition}.

\smallskip

\begin{algorithm}[H]
\caption{\small \edgeaddition}
\label{alg:EdgeAddition}
\DontPrintSemicolon
\SetKwInOut{Input}{input}
\SetKwInOut{Output}{output}
\Input{$(F,x,y)$}
\Output{$F'$}

$V' \gets V $\\
$E' \gets E \cup \set{(x,y)}$ \\
\Return $F' = (V',E')$
\end{algorithm}

\smallskip

\begin{observation}
\label{obs:edgeaddition}
Let $F$ be a forest, and 
let $x, y \not\in D$ be two vertices that belong to two different trees, such that $x$ is a leaf.
Let $F'$ be the graph obtained by an edge addition operation. Then,
\begin{inparaenum}[(i)]
\item $F'$ is a forest with $\kappa' = \kappa - 1$ trees,
\item $n' = n$, and
\item $\sum d' = \sum d + 2$.
\end{inparaenum}
\end{observation}


\section{Optimized Vertex Cover in Forests}
\label{sec:vc}

In this section, we use our realization framework to design an efficient algorithm
that, given a forestic sequence $d$, computes a forest realization with a minimum vertex cover.
More specifically, the algorithm is composed of two steps. 
The first step computes $\tau = \tau^{\calF}(d)$ using the characterization given in~{\cite{GHR18}}.
Then, a realization is computed using our realization framework.

We provide an alternative proof of the characterization from~{\cite{GHR18}}.
In the next section we present a lower bound on the size of the minimum vertex cover, 
and the construction of our realization algorithm induces an equal upper bound.


\subsection{Properties of Vertex Cover}

Recall that, given a forestic sequence $d$, 
$\tau = \tau^{\calF}(d)$ is the minimum vertex cover size over all forest realizations of $d$. 
Gentner et al.~\cite{GHR18} proved the following theorem.

\begin{theorem}[\cite{GHR18}]
\label{thm:vc}
If $d$ is a forestic sequence then,
\(
\tau^{\calF}(d) = \min \set{t : \sum_{i=1}^t d_i \geq \sum_{i=t+1}^n d_i}
\).
\end{theorem}

The following is a necessary condition for the existence of a vertex cover of size $t$. 
This lemma was previously proven in~\cite{GHR18}, and we provide a proof for the sake of completeness.

\begin{lemma}[\cite{GHR18}]
\label{lemma:VCnecessary}
Let $d$ be a sequence of length $n$ and let $F = (V,E)$ be a forest realization of $d$ with a vertex cover of size $t$.
Then,
\(
\sum^t_{i=1} d_i \geq \sum_{i=t+1}^n d_i
\).
\end{lemma}
\begin{proof}
Let $D$ be a vertex cover of size $t$. It follows that
\[
\sum_{i=1}^t d_i 
\geq \sum_{v_i \in D} d_i 
\geq |E| 
=      \frac{1}{2} \sum_{v_i \in V} d_i 
=      \frac{1}{2} \paren{ \sum_{i=1}^t d_i + \sum_{i=t+1}^n d_i }
~. \qedhere
\]
\end{proof}

The next lemma shows that one may assume without loss of generality
that a vertex cover is composed of vertices with the largest degrees, 
namely vertices that correspond to a \emph{prefix} of the given sequence.
This lemma was previously proven in~\cite{GHR18} in \indset terms,
and we provide a proof in \vc terms for the sake of completeness.

\begin{lemma}[VC Prefix Lemma \cite{GHR18}]
\label{swappingVertexCoverLemma}
Let $d$ be a forestic sequence, and let $F$ be a forest realization of $d$, where $\deg(v_i) = d_i$.
Also, let $D$ be a vertex cover in $F$, such that $t = |D|$.
Then, there exists a realization $F'$ of $d$, such that $D' = \set{v_1,\ldots,v_t}$ is a vertex cover.
\end{lemma}
\begin{proof}
Let $F$ and $D$ be a forest realization of $d$  and a vertex cover that maximizes $\sum_{v_i \in D} d_i$.
If $\sum_{v_i \in D} d_i = \sum_{i=1}^t d_i$, then we are done.
Otherwise, there exist two vertices $v_j$ and $v_\ell$, such that $v_j \not\in D$, $v_\ell \in D$, and $d_j > d_\ell$.
We show that $F$ can be modified into a forest $F'$ such that the two vertices exchange their degrees and their labels, 
and $D \setminus \set{v_\ell} \cup \set{v_j}$ is a vertex cover.
This leads to a contradiction.

We obtain $F'$ by transferring $d_j - d_\ell$ edges from $v_j$ to $v_\ell$ without creating cycles.
Then we obtain our goal by switching labels between $v_j$ and $v_\ell$.
More formally, if $v_j$ and $v_\ell$ are not in the same tree of $F$, 
then $N(v_j) \cap N(v_\ell) = \emptyset$.
Otherwise, there is a path $v_j = v_{i_0}, v_{i_1}, \ldots, v_{i_k} = v_\ell$.
If $k > 2$ or $k = 1$, observe that $N(v_j) \cap N(v_\ell) = \emptyset$.
If $k = 2$, then $N(v_j) \cap N(v_\ell) = \set{v_{i_1}}$.
(See example with $k = 1$ in \Cref{fig:vc-prefix-a}.)
If the vertices are in the same tree, let $U \subseteq N(v_j) \setminus \set{v_{i_1}}$ such that $|U| = d_j - d_\ell$.
Otherwise, let $U \subseteq N(v_j)$ such that $|U| = d_j - d_\ell$.

Define:
\(
E' = E \setminus (\set{v_j} \times U) \cup (\set{v_\ell} \times U)
\).
Observe that no cycles were created, since the path from $v_j$ to $v_\ell$ remains as it was.
Also, deleting $(u,v_j)$ splits a tree into two trees, and thus adding $(u,v_{\ell})$ does not create a cycle.
The realization $F'$ is obtained by switching labels between $v_j$ and $v_\ell$.
(See example in \Cref{fig:vc-prefix-exchange}.)
Clearly, $F'$ is a realization of $d$. Moreover, since $D$ is a vertex cover in $F$, 
$D' = D \setminus \set{v_\ell} \cup \set{v_j}$ is a vertex cover in $F'$.
Since $\sum_{v_i \in D} d_i < \sum_{v_i \in D'} d_i$, we have a contradiction.
\end{proof}

\input{figures/VCPrefixLemmaExample.tex}

Given a forestic sequence $d$, \Cref{lemma:VCnecessary} implies that
\(
\tau^{\calF}(d) \geq \min \set{t : \sum_{i=1}^t d_i \geq \sum_{i=t+1}^n d_i}
\).
In the next section we rely on \Cref{swappingVertexCoverLemma} in order to 
present an algorithm which computes a realization with a vertex cover whose size is equal to the above lower bound. 
Hence, this can be viewed as an alternative (and constructive) proof of an upper bound on $\tau^{\calF}(d)$.


\subsection{Computing a Realization}

Now we present our algorithm for finding a realization $F$ of $d$, such that $\deg(v_i) = d_i$,
which contains the vertex cover $D = \set{v_1,\ldots, v_\tau}$, 
where $\tau = \tau^{\calF}(d)$.


The input to Algorithm~\vertexcover (\Cref{alg:VertexCover}) consists of a sequence $d$, 
a prefix size $\tau$, and an initial forest $F_0$.
More specifically, $\tau = \tau^{\calF}(d)$ and $F_0$ is the initial star forest with respect to 
$d$ and the vertex cover $D = \set{v_1,\ldots,v_\tau}$ that corresponds to the prefix $(d_1,\ldots, d_\tau)$.
The goal of the first while loop is to decrease the degree sum to $\sum d$ and the number of leaves.
This is done using \edgemerge operations.
After the first while loop, it remains to transform $\sum d - \sum_{i=1}^\tau d_i$ leaves into 
$n - \tau$ vertices with degrees $d_{\tau+1},\ldots,d_n$.
This is done using \leafmerge operations.
Running examples of the algorithm are given in \Cref{fig:running-vc} and \Cref{fig:running-vc2}.

\smallskip

\begin{algorithm}[ht]
\caption{\small\sf \vertexcover}
\label{alg:VertexCover}
\DontPrintSemicolon
\SetKwInOut{Input}{input}
\SetKwInOut{Output}{output}
\Input{$(d,\tau,F_0)$}
\Output{$F$}
$F \gets F_0$ \\
\While {$\sum \deg^F > \sum d$}{
    Let $x,y\in V(F)\setminus D$ be leaves in different trees of $F$ \\
    $F \gets \edgemerge(F, x,y)$
}
$k \gets \tau + 1$\\
\While{$n(F) > n$}{ \label{alg:vertexcoverStep2}
    Find $u \in V \setminus D$ such that $\deg^F(u) = 1$. \\
    \While {$\deg^F(u) < d_{k}$}{
        Find $v\in V \setminus D$ such that $\deg^F(v) = 1$ and $v$ is not in the same tree as $u$\\
        $F \gets \leafmerge(F,v, u)$
    }
    Assign $u$ to index $k$ and relabel $u$ as $v_k$ \\
    $k \gets k+1$ 
    \label{alg:vertexcoverStep2pp}
}
Relabel the remaining unassigned leaves as $v_k,\ldots,v_n$ \\
\Return $F$
\end{algorithm}

\smallskip

\input{figures/VertexCoverExample.tex}
\input{figures/VertexCoverExample2.tex}

In what follows we show that the algorithm computes a forest $F$ that realizes $d$
and that $D = \set{v_1,\ldots,v_\tau}$ is a vertex cover in $F$.

Let $F^0$ be the initial star forest.
For the analysis, let $F^j$ be $F$ after the $j$th operation (edge merges or leaf merges).
Define $n^j$, $d^j$, and $\kappa^j$ to be the number of vertices in $F^j$, 
the degree sequence of $F^j$, and the number of components in $F^j$, respectively.

\begin{observation}
\label{obs:VC-dtau-one}
Let $\tau = \tau^{\calF}(d)$.
If $d_\tau=1$, then the initial star forest is already a realization of $d$ with vertex cover $D$.
\end{observation}
\begin{proof}
We know that $\sum_{i=1}^{\tau} d_i \geq \sum_{i=\tau+1}^n d_i$.
Since $d_\tau = 1$ and due to minimality of $\tau$ and the parity of $\sum d$, 
it follows that $\sum_{i=1}^{\tau} d_i = \sum_{i=\tau+1}^n d_i$.
\Cref{obs:starforest} implies that the initial star forest is a realization of $d$.
\end{proof}

As a first step, we show that the surplus in the degree sum is bounded 
from above by the surplus in the number of vertices.
This also shows that we have enough trees in $F^0$ in order to perform the first while loop.

\begin{lemma}
\label{lemma:delta0}
$0 \leq \sum d^0 - \sum d \leq n^0 - n \leq 2(\tau - \kappa)$.
\end{lemma}
\begin{proof}
\Cref{obs:starforest} implies that
\[
\sum d^0 - \sum d
= \sum_{i=1}^{n^0} d^0_i - \sum_{i=1}^n d_i
=    2\sum_{i=1}^\tau d_i - \sum_{i=1}^n d_i
=    \sum_{i=1}^\tau d_i - \sum_{i=\tau+1}^n d_i
=    (n^0 - \tau) - \sum_{i=\tau+1}^n d_i
\leq n^0 - n
~.
\]
Also, since $\tau$ is the number of trees in $F^0$, we have that
\[
2(\tau - \kappa)
=    (2n^0 - \sum d^0) - (2n - \sum d)
=    2(n^0 - n) - (\sum d^0 -\sum d)
\geq n^0 - n
~. \qedhere
\]
\end{proof}

Next, we show that $F$ remains a forest during and after the first while loop.

\begin{lemma}
\label{lemma:VCalgFirstLoop}
$F^j$ is a forest throughout the execution of the first while loop
such that
\[
2(\kappa^j - \kappa) \geq n^j - n \geq \sum d^j - \sum d.
\]
\end{lemma}
\begin{proof}
We prove that $F^j$ is a forest such that $2(\kappa^j - \kappa) \geq n^j - n \geq \sum d^j - \sum d$,
by induction on the number of iterations of the first while loop.
In the base case, $F^0$ is the initial star forest, and thus 
$2(\kappa^0 - \kappa) \geq n^0 - n \geq \sum d^0 - \sum d$ due to \Cref{lemma:delta0}.

For the induction step, by the induction hypothesis $F^{j-1}$ is a forest such that 
$2(\kappa^{j-1} - \kappa) \geq n^{j-1} - n \geq \sum d^{j-1} - \sum d$ in the beginning 
of the $j$th iteration of the while loop.
Since $\sum d^{j-1} > \sum d$, it follows that $\kappa^{j-1} \geq \kappa + 1 \geq 2$.
Since the first loop is being executed, \Cref{obs:VC-dtau-one} implies
that $d_\tau\geq2$.
A component obtained from $r$
initial stars by $r-1$ edge merges retains at least $2r-2(r-1)=2$ leaves outside $D$.
Consequently, each of the at least two current components contains a suitable leaf, and the
edge merge is well-defined.
By \Cref{obs:edgemerge}, $F^j$ is a forest at the end of the iteration.
In addition, $\sum d^j = \sum d^{j-1} - 2$ and $n^j = n^{j-1} - 2$, while $\kappa^j = \kappa^{j-1} - 1$.
Hence, $2(\kappa^j - \kappa) \geq n^j - n \geq \sum d^j - \sum d$ holds at the end of the iteration.
\end{proof}

Let $\ell$ be the number of iterations that are performed by first while loop.
(Note that it may be the case that $\ell = 0$.)

\begin{observation}
\label{obs:AfterEdgeMerges}
\begin{inparaenum}[(i)]
\item $\ell = \frac{1}{2} \paren{\sum d^0 - \sum d}$, 
\item $\sum d^\ell = \sum d = \sum d^0 - 2\ell$,
\item $n^\ell = n^0 - 2\ell$, 
\item $\kappa^\ell = \kappa^0 - \ell$, and 
\item $d^\ell_i = 1$, for every $i > \tau$.
\end{inparaenum}
\end{observation}
\begin{proof}
This follows from \Cref{obs:edgemerge}.
\end{proof}

It remains to show that the second while loop transforms $\sum d - \sum_{i=1}^\tau d_i$ leaves into 
$n - \tau$ vertices with degrees $d_{\tau+1},\ldots,d_n$.

\begin{observation}
\label{obs:afteredgemerge}
$\kappa^\ell - \kappa = n^\ell - n = \sum_{i=\tau+1}^n (d_i-1)$.
\end{observation}
\begin{proof}
We have that
\(
\kappa^\ell - \kappa
= n^\ell - \frac{1}{2} \sum d^\ell - n + \frac{1}{2} \sum d
= n^\ell - n
\).
Also, 
\[
n^\ell - n 
= n^0 - 2\ell - n
= \tau + \sum_{i=1}^\tau d_i - \paren{\sum d^0 - \sum d} - n 
= \tau + \sum_{i=1}^\tau d_i - \sum_{i=1}^\tau d_i + \sum_{i=\tau+1}^n d_i - n 
= \sum_{i=\tau+1}^n d_i - (n - \tau)
\]
as required.
\end{proof}

\begin{lemma}
\label{lemma:whileleafmerge}
If $\tau \geq n - \omega_1$, then $d^\ell = d$.
Otherwise, if $\tau < n - \omega_1$, then $F^{\ell + q(k)}$ is a forest such that
\begin{compactitem}
\item $d^{\ell + q(k)} = (d_1,\ldots,d_k,1^{n^{\ell + q(k)} - k})$,
\item $n^{\ell + q(k)} = n^0 - 2\ell - q(k)$,
\item $\kappa^{\ell + q(k)} = \tau - \ell - q(k)$.
\end{compactitem}
for every $k \in \set{\tau,\ldots,n-\omega_1}$, where $q(k) = \sum_{i=\tau+1}^k (d_i-1)$.
\end{lemma}
\begin{proof}
If $\tau \geq n - \omega_1$, then $d^\ell = d$ by \Cref{obs:AfterEdgeMerges}.
Otherwise, we prove the claim by induction on the number of iterations of the second while loop.

We first verify that every requested leaf merge is available. Since the second loop is executed,
$d_{\tau+1} \geq2$, and therefore every vertex in $D$ has degree at least two. At the beginning of
the iteration that constructs $v_k$, all previously assigned vertices outside $D$ also have degree
at least two. Hence, every current tree has at least two leaves, and all of them are unassigned
vertices outside $D$. Moreover, by \Cref{obs:afteredgemerge}, the number of current trees is
$\kappa+\sum_{i=k}^{n}(d_i-1)\geq1+(d_k-1)=d_k$. Thus, one can select $u$ and perform the required
$d_k-1$ leaf merges with leaves in distinct other trees.

In the base case, $k = \tau$ and in this case the claim follows from \Cref{obs:AfterEdgeMerges}.

For the induction step, assume that the claim holds for $k-1$.
Due to \Cref{obs:leafMerge} we know that $F$ remains a forest and that $n(F)$
and $\kappa(F)$ decrease by $d_k - 1$ in each iteration of the second while loop.
Hence, 
\[
n^{\ell + q(k)}
= n^{\ell + q(k-1)} - d_k + 1 
= n^0 - 2\ell - q(k) 
~,
\]
and
\[
\kappa^{\ell + q(k)}
= \kappa^{\ell + q(k-1)} - d_k + 1 
= \tau - \ell - q(k) 
~. \qedhere
\]
\end{proof}

We are ready to state the main result of the section.

\begin{theorem}
\label{thm:VertexCover}
Let $d$ be a forestic sequence and let $\tau=\tau^{\calF}(d)$.
\Cref{alg:VertexCover} computes a forest realization $F$ of $d$, 
such that $D = \set{v_1,\ldots,v_\tau}$ is a vertex cover in $F$.
Moreover, $D$ is an independent set, if no edge merges are used.
\end{theorem}
\begin{proof}
\Cref{lemma:whileleafmerge,lemma:VCalgFirstLoop} show that both loops are well-defined, preserve the forest property,
and terminate with degree sequence $d$. Thus, $F$ is a forest realization of $d$.

We prove by induction on $j$ that $D$ is a vertex cover of $F^j$.
Clearly, $D$ is a vertex cover of $F^0$.
For the induction step, assume that $D$ is a vertex cover of $F^{j-1}$.
An edge merge or a leaf merge does not add edges between vertices in $V \setminus D$,
therefore $D$ is a vertex cover of $F^j$.
In its final step, the algorithm relabels the remaining leaves.
\end{proof}

As for the running time, the initial star forest can be 
constructed in $O(\sum_{i=1}^\tau d_i) = O(\sum d) = O(n)$ time.
We describe how the algorithm chooses the vertices for its operations.
The algorithm performs the merges by maintaining one large component $C$ and a set of stars.
Initially $C$ is the star whose center is $v_1$.
We assume that the leaves of each vertex from $D$ are numbered, and that 
each one of these vertices holds an index of the next available leaf.
When an edge merge is performed it is done on the next available leaf of $v_i \in D \cap V(C)$, 
such that $i$ is minimal and on a leaf of $v_j \in D \setminus V(C)$,
where  $j$ is minimal. 
(For example, the first edge merge would involve $v_1$ and $v_2$.)
In the second while loop, the vertex $u$ is the next available leaf of $v_i \in D \cap V(C)$, 
such that $i$ is minimal. In each iteration of the inner loop $v$ is chosen as a leaf adjacent to $v_j \in D \setminus V(C)$,
where  $j$ is minimal.
Hence, each forest operation can be implemented to run in $O(1)$ time.
There are $(\sum d^0 - \sum d)/2$ edge merges, and $\sum_{i=\tau+1}^n (d_i - 1)$ leaf merges.
Hence, the total running time is $O(\sum d) = O(n)$.


\section{Optimized Total Vertex Cover in Forests}
\label{sec:tvc}

In this section we consider the \totvc problem.
We provide a characterization of the minimum size total vertex cover of a given sequence.
We also present a realization algorithm that is based on a reduction to optimized \vc. 
We note that the reduction simplifies the analysis, but our algorithm can also be viewed as 
using our domination framework.
We also show that a minor adaptation is needed to obtain a characterization and a realization algorithm for
\degr with respect to \pairvc.


\subsection{Properties of Total Vertex Cover}

We start by obtaining necessary conditions for a total vertex cover.

\begin{lemma}
\label{lemma:TotalVertexCoverGeneralNecessity}
Let $d$ be a graphic sequence with a realization $G$ such that $\deg(v_i) = d_i$.
A total vertex cover $D$ in $G$ satisfies 
\(
\sum_{v_i \in D} d_i \geq \sum_{v_i \notin D} d_i + 2\ceil{|D|/2}
\).
\end{lemma}
\begin{proof}
Since $V \setminus D$ is an independent set, it follows that
$|E(D,V \setminus D)| = \sum_{v_i \notin D} d_i$.
Also, $G[D]$ contains no isolated vertices, therefore $\sum_{v_i \in D} \deg_{G[D]}(v_i) \geq |D|$.
Since the sum of degrees is an even number we have that $
2|E(D,D)| = \sum_{v_i \in D} \deg_{G[D]}(v_i) \geq 2\ceil{|D|/2}$.
Overall we get
\[
\sum_{v_i \in D} d_i = |E(D,V\setminus D)|  + 2\cdot |E(D,D)| \geq \sum_{v_i \notin D} d_i + 2\ceil{|D|/2}
~.
\qedhere
\]
\end{proof}

Observe that the last bound is stricter than the bound given in \Cref{obs:tvc-odd}.

The following is a necessary condition on a sequence for the existence of a total vertex cover of size $t$.

\begin{lemma}
\label{lemma:TVCnecessary}
Let $d$ be a forestic sequence with a realization $F$ such that $\deg(v_i) = d_i$.
Also, let $D$ be a total vertex cover in $F$, such that $t = |D|$.
Then
\(
\sum^t_{i=1} d_i \geq \sum_{i=t+1}^n d_i + 2\ceil{t / 2}
\).
\end{lemma}
\begin{proof}
Let $D$ be a total vertex cover of size $t$. 
Due to \Cref{lemma:TotalVertexCoverGeneralNecessity} we know that 
\[
\sum_{i=1}^t d_i 
\geq \sum_{v_i \in D} d_i
\geq \sum_{v_i \notin D} d_i + 2\ceil{t/2}
\geq \sum_{i=t+1}^n d_i + 2\ceil{t/2}
~. \qedhere
\]
\end{proof}

Given a forestic sequence $d$, \Cref{lemma:TVCnecessary} implies that
\begin{equation}
\label{eqn:tvc}
\taut^{\calF}(d)  \geq \min \set{ t : \sum_{i=1}^t d_i \geq \sum_{i=t+1}^n d_i + 2\ceil{t/2}}
~.
\end{equation}
where $\taut^{\calF}(d)$ is the minimum total vertex cover size of any forest realization of sequence $d$.
In what follows we present an algorithm which computes a realization with a total vertex cover whose size 
is equal to the above lower bound. 

We now show a prefix lemma for \totvc.

\begin{lemma}[TotalVC Prefix Lemma]
\label{lemma:swappingTotalVertexCover}
Let $d$ be a forestic sequence with a realization $F$ such that $\deg(v_i) = d_i$.
Also, let $D$ be a total vertex cover in $F$, such that $t = |D|$.
Then, there exists a realization $F'$ of $d$ such that $D' = \set{v_1,\ldots,v_t}$ is a total vertex cover.
\end{lemma}
\begin{proof}
We use the same construction as in the proof of \Cref{swappingVertexCoverLemma}.
Recall that in the proof of \Cref{swappingVertexCoverLemma} we assume that
there exist two vertices $v_j$ and $v_\ell$, such that $v_j \not\in D$, $v_\ell \in D$, and $d_j > d_\ell$. 
We obtain a forest $F'$ from $F$ by transferring $d_j - d_\ell$ edges from $v_j$ to $v_\ell$ without creating cycles.  
Observe that the removed edges connect $v_j$, which is not in the cover, with other vertices.
Hence, $F[D]$ is a subgraph of $F'[D']$. It follows that $F'[D']$ does not contain isolated vertices.
\end{proof}

Before presenting our realization algorithm, we treat one special case separately.

\begin{lemma}
\label{lemma:tvc-special}
Let $d$ be a forestic sequence such that $d_2 =1$.
Then, there exists only one possible realization $F$ (up to isomorphism) whose optimal total vertex cover 
is of size $\taut^{\calF}(d) = n - d_1 + 1$.
\end{lemma}
\begin{proof}
First, the only possible realization of $d$ (up to isomorphism) is a star with $d_1$ leaves (i.e., $K_{1,d_1}$)
and a matching. (If $d_1 = 1$, then the star has only one leaf.) 
The total vertex cover $D$ must contain the center of the star, one of its leaves, 
and the endpoints of all matching edges. Hence, $|D| = 1 + n - d_1$.

Notice that $n-1$ and $d_1$ are either both odd or both even, hence
\[
\sum_{i=1}^{n - d_1 + 1} d_i 
= d_1 + (n - d_1)
= n 
= (d_1 - 1) + (n - d_1 + 1)
= \sum_{i=n-d_1+2}^n d_i + 2\ceil{(n - d_1 + 1)/2}
~.
\]
It follows that $\taut^{\calF}(d) = n - d_1 + 1$.
\end{proof}

Assuming that $d_2 \geq 2$ allows us to rely on the following bounds.

\begin{lemma}
\label{lemma:tau-upper-kappa}
Let $d$ be a forestic sequence such that $d_2 \geq 2$.
Then
\(
\taut^{\calF}(d) \leq \nBTwo + 2(\kappa -1) 
~.
\)
\end{lemma}
\begin{proof}
It is known that a forestic sequence can be realized by the union of a caterpillar graph and a matching
(see, e.g., \cite{BBPRR25-2pbe}).
One possible total vertex cover for $F$ is the union of all the vertices in the central path of the caterpillar
and all the matching vertices.
Overall this total vertex cover contains $\nBTwo + 2(\kappa -1)$ vertices.
\end{proof}

\begin{observation}
\label{obs:d2geq2}
Let $d$ be a graphic sequence, such that $d_2 \geq 2$.
Then, $d_1 + d_2 \geq 4$. 
Also, if $\taut = \min \set{ t : \sum_{i=1}^t d_i \geq \sum_{i=t+1}^n d_i + 2\ceil{t/2}}$ is odd, 
then $d_1 + d_2 + d_3 \geq 6$.
\end{observation}
\begin{proof}
First, $d_1 \geq d_2 \geq 2$ implies $d_1 + d_2 \geq 4$.
Next, assume that $\taut$ is odd, and $d_1 + d_2 + d_3 < 6$.
Then it must be that $d_1 = d_2 = 2$ and $d_3 = 1$.
In this case, $n$ must be even. 
The two possible realizations in this case are either a $P_4$ and a matching or two copies of $P_3$ and a matching.
In both cases, $\taut = n-2$. A contradiction.
\end{proof}


\subsection{Computing a Realization}
\label{sec:tvc-computing}

In this section we assume that $d$ is a forestic sequence such that $d_2 \geq 2$.
We provide an algorithm for computing a total vertex cover of size 
$\taut = \min \set{ t : \sum_{i=1}^t d_i \geq \sum_{i=t+1}^n d_i + 2\ceil{t/2}}$.
This will prove that $\taut^{\calF}(d) = \min \set{ t : \sum_{i=1}^t d_i \geq \sum_{i=t+1}^n d_i + 2\ceil{t/2}}$.

As a first step, we give a reduction from optimized \totvc to optimized \vc.
Given a forestic sequence $d$, let $\taut = \min \set{ t : \sum_{i=1}^t d_i \geq \sum_{i=t+1}^n d_i + 2\ceil{t/2}}$.
We define a sequence $d'$ and an integer $\tau'$ as follows.
Let $\tau' = \floor{\taut/2} - \delta$, where 
\[
\delta = \max \paren{\set{ 0} \cup \set{ i : 1 \leq i \leq \floor{\taut/2} \text{ and } d_{\taut- 2i + 1} = 1 } }
~.
\]
Next, define
\begin{equation}
\label{eqn:TotalReduction}
d' = d'' \circ (d_{\taut + 1}, \ldots, d_n)
~,
\end{equation}
where 
\[
d''_i =
\begin{cases}
d_{2i-1} + d_{2i} - 2    &  1 \leq i \leq \tau' \text{ and } \taut \text{ is even}, \\
d_{2i} + d_{2i+1} - 2   &  1 < i \leq \tau' \text{ and } \taut \text{ is odd}, \\
d_1 + d_2 + d_3 - 4   &  i = 1 \text{ and } \taut \text{ is odd}. \\
\end{cases}
\]
\Cref{obs:d2geq2} implies that $d_1 + d_2 \geq 4$, if $\taut$ is even, 
and that $d_1 + d_2 + d_3 \geq 6$, if $\taut$ is odd.
Moreover, observe that $n' = n - \taut + \tau' = n - \ceil{\taut / 2} - \delta$ and that 
$\sum_{i=1}^{\tau'} d'_i = \sum_{i=1}^{\taut} d_i - 2\ceil{\taut / 2}$.

For example, given $d = (4,3^5,2,1^9)$ with $\taut = 6$ then we get $d' = (5,4^2,2,1^9)$ and $\tau' = 3$. 

\begin{observation}
\label{obs:reduction-nonincreasing}
Let $d_2 \geq 2$.
Then, $d'$ is non-increasing.
\end{observation}
\begin{proof}
We start with the prefix of $d'$.
First, if $\tau' = 1$, then we are done. Otherwise,
since $d_1 \geq 2$, if $\taut$ is odd, then by \Cref{obs:d2geq2} 
\begin{align*}
d'_1 & = d_1 + d_2 + d_3 - 4 \geq 2 + d_4 + d_5 - 4 = d'_2 ~, \\
d'_i & = d_{2i} + d_{2i+1} - 2 \geq d_{2i+2} + d_{2i + 3} - 2 = d'_{i+1} ~, &  \text{for every } 1 < i < \tau'
~.
\end{align*}
Similarly, if $\taut$ is even, then 
\begin{align*}
d'_i & = d_{2i-1} + d_{2i} - 2 \geq d_{2i+1} + d_{2i + 2} - 2 = d'_{i+1} ~, & \text{for every } i < \tau'
~.
\end{align*}
In addition, recall that $d_{\taut-2\delta-1} \geq 2$.
Hence, if $\tau' \geq 2$ or $\taut$ is even, we have that 
\[
d'_{\tau'} = d_{\taut - 2\delta - 1} + d_{\taut - 2\delta} - 2 \geq d_{\taut - 2\delta} \geq d_{\taut + 1}
~. 
\]
Otherwise, $d'_1 = d_1 + d_2 + d_3 - 4 \geq d_3$.
\end{proof}

We show that $d'$ is forestic.

\begin{observation}
\label{obs:reduction-forestic}
Let $d_2 \geq 2$.
If $d$ is a forestic sequence, then $d'$ is a forestic sequence.
\end{observation}
\begin{proof}
We show that $d'$ is forestic by calculating the degree sum. 

If $\delta \geq 1$, we have that $\taut \geq \nBTwo + 2\delta$ by definition.
Moreover, $\taut \leq \taut^{\calF}(d) \leq 2(\kappa -1) + \nBTwo$ due to \Cref{lemma:tau-upper-kappa}.
It follows that $\delta \leq \kappa - 1$. The latter inequality also holds, if $\delta = 0$.
Hence, $\sum d = 2n - 2\kappa \leq 2n -2\delta - 2$, for any value of $\delta$.
It follows that 
\[
\sum d' 
= \sum_{i=1}^{\tau'} d'_i + \sum_{i=\tau'+1}^{n'} d'_i 
= \sum_{i=1}^{\taut} d_i - 2\ceil{\taut / 2} + \sum_{i=\taut+1}^n d_i 
= \sum_{i=1}^n d_i - 2\ceil{\taut / 2}
\leq 2n - 2\delta - 2 - 2\ceil{\taut / 2}
= 2n' - 2
~. \qedhere
\]
\end{proof}

Next, we show that the reduction is valid.

\begin{lemma}
\label{lemma:VCiffTVC}
$\sum_{i=1}^{\tau'} d'_i  \geq \sum_{i=\tau'+1}^{n'} d'_i$ 
if and only if 
$\sum^{\taut}_{i=1} d_i \geq  \sum_{i=\taut+1}^n d_i + 2\ceil{\taut / 2}$.
\end{lemma}
\begin{proof}
If $\sum^{\taut}_{i=1} d_i \geq  \sum_{i=\taut+1}^n d_i + 2\ceil{\taut / 2}$, then
\[
\sum_{i=1}^{\tau'} d'_i 
=      \sum_{i=1}^{\taut} d_i - 2\ceil{\taut / 2} 
\geq \paren{ \sum_{i=\taut+1}^n d_i + 2\ceil{\taut/2} } - 2\ceil{\taut / 2} 
=      \sum_{i=\taut+1}^n d_i 
=      \sum_{i=\tau'+1}^{n'} d'_i
~.
\]
If $\sum_{i=1}^{\tau'} d'_i  \geq \sum_{i=\tau'+1}^{n'} d'_i$, then
\[
\sum^{\taut}_{i=1} d_i 
=       \sum_{i=1}^{\tau'} d'_i  + 2\ceil{\taut / 2} 
\geq  \sum_{i=\tau'+1}^{n'} d'_i + 2\ceil{\taut / 2} 
=       \sum_{i=\taut+1}^n d_i + 2\ceil{\taut / 2}
~.\qedhere
\]
\end{proof}

\begin{lemma}
\label{lemma:TVCoptimalVC}
If $\taut = \min \set{ t : \sum_{i=1}^t d_i \geq \sum_{i=t+1}^n d_i + 2\ceil{t/2}}$,
then $\tau' = \tau^{\calF}(d')$.
\end{lemma}
\begin{proof}
Assume that  $\tau' > \tau^{\calF}(d') \geq 1$.
Hence, $\tau^{\calF}(d') \leq \tau' - 1$, and thus $\sum_{i=1}^{\tau'-1} d'_i  \geq \sum_{i=\tau'}^{n'} d'_i$.

First assume that $\delta > 0$.  We have that
\begin{align*}
\sum^{\taut-2}_{i=1} d_i
    =      \sum_{i=1}^{\taut} d_i  - (d_{\taut-1} + d_{\taut})
& =      \sum_{i=1}^{\tau'} d'_i  +  2\ceil{\taut / 2} - (d_{\taut-1} + d_{\taut}) \\
& =      \sum_{i=1}^{\tau'-1} d'_i  + d'_{\tau'} + 2\ceil{\taut / 2} -  (d_{\taut-1} + d_{\taut}) \\
& \geq \sum_{i=\tau'}^{n'} d'_i  + d'_{\tau'} + 2\ceil{\taut / 2} -  (d_{\taut-1} + d_{\taut}) \\
& =      \sum_{i=\tau'+1}^{n'} d'_i  + 2d'_{\tau'} + 2\ceil{\taut / 2} -  (d_{\taut-1} + d_{\taut}) \\
& =      \sum_{i=\taut+1}^n d_i  + 2d'_{\tau'} + 2\ceil{\taut / 2} -  (d_{\taut-1} + d_{\taut}) \\
& =      \sum_{i=\taut-1}^n d_i  + 2(d_{\taut-2\delta-1} + d_{\taut-2\delta} - 2)  + 2\ceil{\taut / 2} - 2 (d_{\taut-1} + d_{\taut}) 
~.
\end{align*}
Since $d_{\taut-2\delta-1} + d_{\taut-2\delta}  > d_{\taut-1} + d_{\taut} = 2$, we get that
$\sum^{\taut-2}_{i=1} d_i \geq \sum_{i=\taut-1}^n d_i  + 2\ceil{(\taut-2) / 2}$, 
in contradiction to the minimality of $\taut$.

Next, assume $\delta=0$. We have that
\begin{align*}
\sum^{\taut-1}_{i=1} d_i
    =      \sum_{i=1}^{\taut} d_i  - d_{\taut}
 =      \sum_{i=1}^{\tau'} d'_i  +  2\ceil{\frac{\taut}{2}} - d_{\taut} 
 =      \sum_{i=1}^{\tau'-1} d'_i  +  d'_{\tau'} + 2\ceil{\frac{\taut}{2}} - d_{\taut} 
& \geq \sum_{i=\tau'}^{n'} d'_i  + d'_{\tau'} +  2\ceil{\frac{\taut}{2}} - d_{\taut} \\
& =      \sum_{i=\tau'+1}^{n'} d'_i + 2d'_{\tau'} +  2\ceil{\frac{\taut}{2}} - d_{\taut} \\
& =      \sum_{i=\taut+1}^n d_i  + 2d'_{\tau'} + 2\ceil{\frac{\taut}{2}} - d_{\taut} \\
& =      \sum_{i=\taut}^n d_i  + 2d'_{\tau'} + 2\ceil{\frac{\taut}{2}} - 2d_{\taut}
~.
\end{align*}
Since $d'_{\tau'} = d_{\taut - 1} + d_{\taut} - 2$ and $d_{\taut-1} \geq 2$, it follows that 
$\sum^{\taut-1}_{i=1} d_i \geq \sum_{i=\taut}^n d_i  + 2\ceil{(\taut-1) / 2}$,
again in contradiction to the minimality of $\taut$.
\end{proof}

\begin{lemma}
\label{lemma:VertexCoverRealizationReduction}
Let $d$ be a forestic sequence, such that $d_2 \geq 2$, and let 
$\taut = \min \set{ t : \sum_{i=1}^t d_i \geq \sum_{i=t+1}^n d_i + 2\ceil{t/2}}$.
Then there exists a forest $F$ that realizes $d$ with a total vertex cover $\set{v_1,\ldots,v_{\taut}}$.
\end{lemma}
\begin{proof}
Let $d'$ and $\tau'$ be the output of the reduction on $d$ and $\taut$.
We know that $d'$ is non-increasing and forestic due to \Cref{obs:reduction-nonincreasing,obs:reduction-forestic}.
\Cref{lemma:TVCoptimalVC} implies that $\tau' = \floor{\taut/2} - \delta = \tau^{\calF}(d')$.
%
\Cref{thm:VertexCover} implies that there exists a forest realization $F'$ of $d'$ with vertex cover
$D' = \set{v_1,\ldots,v_{\tau'}}$.

We construct a forest $F$ from $F'$ by first splitting $\tau'$ vertices and then adding a matching of size $\delta$.
The splits work as follows. If $\taut$ is even, then split $v_i$ into two connected vertices, for $i \in \set{1,\ldots,\tau'}$, 
where one vertex is connected to $d_{2i-1}-1$ neighbors and the other to $d_{2i}-1$ neighbors.
If $\taut$ is odd and $i \geq 2$, then split $v_i$ into two connected vertices, 
where one is connected to $d_{2i}-1$ neighbors and the other to $d_{2i+1}-1$ neighbors.
Otherwise, $v_1$ is split into a path of three vertices that are connected to $d_2 - 1$, $d_1 - 2$, and $d_3 - 1$
neighbors, respectively (Recall that $d_1 \geq 2$, as we assumed $d_2 \geq 2$). 
Next, we add to $F$ a matching of size $\delta$.
Let $D$ be the set that contains the $\taut -2\delta$ vertices that are obtained by the splitting process 
and the $2\delta$ vertices of the matching.
Observe that $F$ is a forest that realizes $d$ and $D$ is a total vertex cover of $F$.
\end{proof}

The following is implied by \Cref{lemma:TVCnecessary,lemma:VertexCoverRealizationReduction,lemma:tvc-special}.
\begin{corollary}
\label{corollary:tvc-characterization}
If $d$ is a forestic sequence then
\(\displaystyle
\taut^{\calF} (d)  = \min \set{ t : \sum_{i=1}^t d_i \geq \sum_{i=t+1}^n d_i + 2\ceil{t/2}}
\).
\end{corollary}

The next result is a realization algorithm.

\begin{theorem}
\label{thm:TotalVertexCover}
Let $d$ be a forestic sequence and let $\taut = \taut^{\calF}(d)$.
Then there is a linear time algorithm that computes a realization $F$ of $d$ with total vertex cover $D$ of size $\taut$.
\end{theorem}
\begin{proof}
If $d_2 = 1$, then applying \Cref{lemma:tvc-special} is enough to prove the theorem.
Otherwise, we may assume that $d_2 \geq 2$.
The proof of \Cref{lemma:VertexCoverRealizationReduction} induces the following algorithm.
Given a forestic sequence $d$, compute $\taut$, $d'$ and $\tau'$.
Next, execute \Cref{alg:VertexCover} on $d'$, $\tau'$, 
and the corresponding star forest $F'_0$ to obtain a realization $F'$ with vertex cover
$D' = \set{v_1,\ldots,v_{\tau'}}$. 
Afterwards, construct a realization $F$ of $d$ and a total vertex cover $D$ as described in the proof of 
\Cref{lemma:VertexCoverRealizationReduction}.

The correctness of the above algorithm is implied by \Cref{lemma:VertexCoverRealizationReduction}.
As for the running time, the computation of $\taut$, $\tau'$ and $d'$ takes $O(n)$ time.
Since $n' \leq n$, \Cref{alg:VertexCover} also requires $O(n') = O(n)$ time.
Finally, obtaining $F$ and $D$ from $F'$ and $D'$ takes $O(n)$ time.
\end{proof}

Running examples of the algorithm are given in \Cref{fig:tvc-algorithm} and \Cref{fig:tvc-algorithm2}.
Note that the reduced sequence $d'$ in \Cref{fig:tvc-example1-left} is realized in \Cref{fig:running-vc}.
Also, the reduced sequence $d'$ in \Cref{fig:tvc-example2-left} is realized in \Cref{fig:running-vc2}.

\input{figures/TotalVertexCoverExample.tex}

\input{figures/TotalVertexCoverExample2.tex}


\subsection{Using the Domination Framework}
In this section we show that the algorithm given in the previous section can be 
described using our domination framework.

Algorithm~\textsf{Realization-TVC} is composed of two phases. 
In the first phase, the algorithm starts with the initial star forest and uses \edgemerge operations 
in order to create a perfect matching on the $\taut$ covering vertices, if $\taut$ is even.
If $\taut$ is odd, then one of the matching edges becomes a $P_3$.
In the second phase, it executes \vertexcover (\Cref{alg:VertexCover}), 
but instead of starting with $d$ and the initial star forest,
the input consists of the sequence $d'$, the prefix size $\tau' =\tau^{\calF}(d')$,
and the altered initial forest that was constructed in phase 1.
Algorithm~\textsf{Realization-TVC} simulates Algorithm~\vertexcover
on the forest $F$ such that each component with $k$ leaves is treated as a star with $k$ leaves.
Observe that when $d_2 = 1$, the execution of \vertexcover in Step~5 is redundant,
and the correctness of \Cref{alg:TotalVertexCover} is implied by \Cref{lemma:tvc-special}.

Given a vertex $v$ with a leaf neighbor, we denote by $l(v)$ an arbitrary leaf of $v$.

\medskip

\begin{algorithm}[H]
\caption{\small\sf Realization-TVC}
\label{alg:TotalVertexCover}
\DontPrintSemicolon
\SetKwInOut{Input}{input}
\SetKwInOut{Output}{output}
\Input{$(d,\taut)$}
\Output{$F$}
Let $F$ be the initial star forest of $d$ and $D = \set{v_1,\dots,v_{\taut}}$ \\
\lFor{$i=0,\ldots, \floor{\taut/2} - 2$} 
{
	$F \gets \edgemerge (F,l(v_{\taut - 2i}),l(v_{\taut - 2i - 1}))$
}
$F \gets \edgemerge(F,l(v_1),l(v_2))$\\
\lIf{$\taut$ is odd}
{
	$F \gets \edgemerge(F,l(v_1),l(v_3))$
}
\Return $\vertexcover(d',\tau',F)$
\end{algorithm}

\medskip

Observe that the preliminary edge merges that are performed in Lines~1-4 simulate the reduction 
from \totvc to \vc that was given above.
%
Moreover, if the execution of \vertexcover in \Cref{alg:TotalVertexCover}
follows the same sequence of edge merges and leaf merges as the execution of 
\vertexcover in the algorithm which is given in \Cref{thm:TotalVertexCover}, 
then the computed forest will be the same.
By \Cref{thm:TotalVertexCover}, the call to \Cref{alg:VertexCover} on $(d',\tau',F)$
is valid, and its vertex cover can be enlarged to the first $\tau'$ vertices.
Each subsequent merge on the reduced forest has a corresponding merge on the altered forest
$F$. Choosing these corresponding operations yields precisely the forest obtained by the
splitting construction in the proof of \Cref{thm:TotalVertexCover}.

We add running examples of the algorithm in \Cref{fig:tvc-algorithm-alternate} and \Cref{fig:tvc-algorithm-alternate2}.

\input{figures/TotalVertexCoverExampleNotTrueAlgorithm.tex}

\input{figures/TotalVertexCoverExampleNotTrueAlgorithm2.tex}


\subsection{Optimized Paired Vertex Cover}

In this section we deal with \degr in forests with respect to \pairvc.

\begin{theorem}
\label{thm:PairedVertexCover}
Let $d$ be a forestic sequence.
Then there is a linear time algorithm that computes a forest realization $F$ of $d$ with a paired vertex cover of size 
$\taup^{\calF}(d) = 2\ceil{\taut^{\calF}(d)/2}$.
\end{theorem}
\begin{proof}
We have that $\taup^{\calF}(d) \geq \taut^{\calF}(d)$ due to \Cref{obs:pairedlstotal}.
Moreover, since every paired vertex cover has even cardinality, 
we have that $\taup^{\calF}(d) \geq 2\ceil{\taut^{\calF}(d)/2}$.

Recall the proof of \Cref{thm:TotalVertexCover}.
If $\taut^{\calF}(d)$ is even, then the total vertex cover is also a paired vertex cover by construction.
Otherwise, $\taut^{\calF}(d)$ is odd. Let $D$ be the total vertex cover computed 
as described in the proof of \Cref{thm:TotalVertexCover}.
Also, recall that the path $v_2 - v_1 - v_3$ is contained in the realization $F$.
In this case we claim that there is a neighbor of $v_2$ that is not in $D$.
Assume towards contradiction that all neighbors of $v_2$ are in $D$.
Then, $D' = D \setminus \set{v_2}$ is also a total vertex cover,
in contradiction to $D$ being a minimum size total vertex cover.
Hence, there is a vertex $u \in N(v_2) \setminus D$.
Observe that $\set{(v_1,v_3),(v_2,u), (v_4,v_5), \ldots, (v_{|D|-1}, v_{|D|})}$ is a matching in $F[D \cup \set{u}]$.
Therefore, $D \cup \set{u}$ is a minimum paired vertex cover.
It follows that $\taup^{\calF}(d) = 2\ceil{\taut^{\calF}(d)/2}$.
\end{proof}

\begin{corollary}
\label{thm:edgedomset}
Let $d$ be a forestic sequence.
Then there is a linear time algorithm that computes a forest realization $F$ of $d$ with an edge dominating set of size 
$\gammae^{\calF}(d) = \taup^{\calF}(d)/2 = \ceil{\taut^{\calF}(d)/2}$.
\end{corollary}


\section{Optimized Dominating Set in Forests}
\label{sec:domset}

In this section we use our realization framework to design an efficient algorithm that given a forestic 
sequence computes a forest realization with a minimum dominating set.
As in \Cref{sec:vc}, the algorithm is composed of two steps. 
The first computes $\gamma = \gamma^{\calF}(d)$ using an algorithm that is implied by the characterization given in~{\cite{GHR18}}.
Then, a realization is computed using our realization framework.


\subsection{Properties of Dominating Set}

Recall that, given a forestic sequence $d$, 
$\gamma = \gamma^{\calF}(d)$ is the minimum dominating set size over all forest realizations of $d$. 
Gentner et al.~\cite{GHR18} showed that 

\begin{theorem}[\cite{GHR18}]
\label{thm:domset-characterization}
Let $d$ be a forestic sequence.
Then, $\gamma^{\calF}(d) = \min \{ \tau^{\calF}(d) , \gammakk(d) \}$,  where
\[
\gammakk(d) = \min \set{ k : \sum_{i=1}^k d_i \geq n-k \text{ and } 0 <  \sum_{i=k+1}^n d_i - \sum_{i=1}^k d_i \leq 2(\nBTwo - k) -2}
\]
\end{theorem}

\begin{observation}
\label{obs:k1Greaterk2}
If $\gammakk(d)$ is well-defined, then 
$\gammakk(d) < \tau^{\calF}(d)$.
\end{observation}

In the next lemma we show that one may assume that an optimal dominating set in a forest is composed of 
vertices with the largest degrees. 
Gentner et al.~\cite{GHR18} proved this lemma to obtain the characterization of $\gamma^{\calF}$.
We provide a (different) proof for completeness.

\begin{lemma}[DS Prefix Lemma \cite{GHR18}]
\label{swappingDominatingSetLemma}
Let $d$ be a forestic sequence, and let $F$ be a forest realization of $d$, where $\deg(v_i) = d_i$.
Also, let $D$ be a dominating set in $F$, such that $t = |D|$.
Then, there exists a realization $F'$ of $d$, such that $D' = \set{v_1,\ldots,v_t}$ is a dominating set.
\end{lemma}
\begin{proof}
Let $F$ and $D$ be a forest realization of $d$  and a dominating set that maximizes $\sum_{v_i \in D} d_i$.
If $\sum_{v_i \in D} d_i = \sum_{i=1}^t d_i$, then we are done.
Otherwise, there exist two vertices $v_j$ and $v_\ell$, such that $v_j \not\in D$, $v_\ell \in D$, and $d_j > d_\ell$.
We show that $D \setminus \set{v_\ell} \cup \set{v_j}$ is a dominating set in some forest realization $F'$ of $d$, 
and this leads to a contradiction.

First, assume that $v_\ell$ is a leaf, and let $v_p$ be its only neighbor.
If $v_p \in D$ or $v_j = v_p \not\in D$, then we are done.
If $v_p \neq v_j$, and $d_p > 1$, we use $D \setminus \set{v_\ell} \cup \set{v_p}$.
Otherwise, we obtain $F'$ by transferring $d_j - d_\ell$ edges from $v_j$ to $v_\ell$ without creating cycles.
Then we obtain our goal by switching labels between $v_j$ and $v_\ell$.

Since $F$ is a forest, we have that $\abs{N(v_j) \cap N(v_\ell)} \leq 1$.
In addition, since $v_j \not\in D$ there exists $v_{j'} \in N(v_j)$ such that $v_{j'} \in D$.
If $v_j$ and $v_\ell$ are not in the same tree in $F$, then $N(v_j) \cap N(v_\ell) = \emptyset$.
Let $U \subseteq N(v_j) \setminus \{v_{j'}\}$ such that $|U| = d_j - d_\ell$.
If $v_j$ and $v_\ell$ belong to the same tree, then there is a path $v_j = v_{i_0}, v_{i_1}, \ldots, v_{i_k} = v_{\ell}$. 
(Note that it may be the case that $v_{i_1} = v_\ell$.)
Notice that $v_\ell$ is not a leaf, since otherwise $v_{i_{k-1}}$, whose degree is at least $2$, can 
replace $v_\ell$ in $D$. Hence, we may assume that $\deg_F(v_\ell) \geq 2$.
Let $U \subseteq N(v_j) \setminus \{v_{j'},v_{i_1}\}$, such that $|U| = d_j - d_\ell$.

In both cases we define
\(
E' = E \setminus (\set{v_j} \times U) \cup (\set{v_\ell} \times U)
\).
Observe that no cycle was created, since the path from $v_j$ to $v_\ell$ remains as it was.
The realization $F'$ is obtained by switching labels between $v_j$ and $v_\ell$.
Clearly, $F'$ is a realization of $d$. Moreover, since $D$ is a dominating set in $F$, 
$D' = D \setminus \set{v_\ell} \cup \set{v_j}$ is a dominating set in $F'$.
\end{proof}

\begin{lemma}
\label{lemma:omega1bound}
Let $d$ be a forestic sequence.
Let $F = (V,E)$ be a forest realization of $d$, and let $D$ be a dominating set of size $t$.
If $\sum_{i=t+1}^n d_i > \sum_{i=1}^t d_i$, then $\omega_1 \leq n - t - 2$.
\end{lemma}
\begin{proof}
By \Cref{swappingDominatingSetLemma} assume that $D = \set{v_1,\ldots,v_t}$.
Suppose toward contradiction that $d_{t+2} = 1$. 
In this case $\sum_{i = t+1}^n d_i = d_{t+1} + n - t - 1$.
Hence, $d_{t+1} + n - t - 1 > \sum_{i=1}^t d_i$.
All degree-1 vertices must be neighbors of vertices in $D$. 
Thus $N(v_{t+1}) \subseteq D$, and we get a contradiction.
\end{proof}

As mentioned above, it was shown in~{\cite{GHR18}} that
$\gamma^{\calF}(d) \geq \min \{ \tau^{\calF}(d), \gammakk(d) \}$.
Next, we provide a simpler alternative proof.
The next lemma describes a necessary condition for the existence of a forest realization 
with a dominating set of certain size.

\begin{lemma}[\cite{GHR18}]
\label{lemma:DominationSetNecessity}
Let $d$ be a forestic sequence.
Let $F = (V,E)$ be a forest realization of $d$, and let $D$ be a dominating set of size $t$.
Either $\sum_{i=1}^t d_i \geq \sum_{i=t+1}^n d_i$, or
\[
\sum_{i=t+1}^n d_i - \sum_{i=1}^t d_i \leq 2(\nBTwo - t) - 2
\]
\end{lemma}
\begin{proof}
By \Cref{swappingDominatingSetLemma} we assume that $D$ is composed of the $t$ highest degree vertices.

If $\sum_{i=1}^t d_i \geq \sum_{i=t+1}^n d_i$, then we are done.
Hence, we assume that $\sum_{i=t+1}^n d_i > \sum_{i=1}^t d_i$.

We partition $V$ into three disjoint subsets: $D$, $\bar{D}_{=1}$, $\bar{D}_{\geq2}$, 
where $D$ is the dominating set, $\bar{D}_{=1} = \set{v_i \not\in D : d_i = 1}$ and $\bar{D}_{\geq2}= V \setminus (D \cup \bar{D}_{=1})$.
$\bar{D}_{\geq 2} \neq \emptyset$, since $\omega_{\geq 2} \geq t+1$ due to \Cref{lemma:omega1bound}.

All degree-1 vertices must be connected to a vertex in $D$. Hence, $|E(D,\bar{D}_{=1})| = n - \nBTwo$.
Each vertex in $\bar{D}_{\geq 2}$ must also be connected to a vertex in $D$.
It follows that 
\[
2|E(\bar{D}_{\geq 2},\bar{D}_{\geq2})| 
\geq \sum_{i=t+1}^{\nBTwo} d_i - \paren{ \sum_{i=1}^t d_i - (n - \nBTwo) }
=    \sum_{i=t+1}^n d_i - \sum_{i=1}^t d_i 
>    0
~.
\]
See depiction in \Cref{fig:DomSetPartition}.
Since $F[\bar{D}_{\geq 2}]$ is a forest, it must be that $2|E(\bar{D}_{\geq 2},\bar{D}_{\geq2})| \leq 2(\nBTwo - t) - 2$.
The lemma follows.
\end{proof}

\input{figures/DomSetFigure.tex}


Looking at \Cref{lemma:DominationSetNecessity} and \Cref{obs:ds-size} we infer that:
\[
\gamma^{\calF}(d) \geq \min \{ \tau^{\calF}(d), \gammakk(d) \}
~.
\]

In the next section we present an algorithm which computes a realization with a dominating set of size
$\min \{ \tau^{\calF}(d), \gammakk(d) \}$
and this can be viewed as a constructive proof of an upper bound on $\gamma^{\calF}$.
We note that the proof given in Gentner et al.~\cite{GHR18} uses an existence claim based on a characterization 
of bi-sequence ranges by Gale and Ryser~\cite{Gale57, Ryser57}. 
One can design an algorithm that computes a realization of bi-sequence ranges
by using a reduction to \textsc{Maximum Flow} with lower bounds.
However, their proof is based on a realizing graph with minimum number of components, 
which is not guaranteed by the above-mentioned realization algorithm.
Hence, as opposed to the case of \textsc{Vertex Cover}, the characterization from~{\cite{GHR18}} does not
imply a realization algorithm.


\subsection{Computing a Realization}

We describe an algorithm that, given a forestic sequence $d$, computes a forest realization 
with a dominating set of size $\gamma^{\calF}(d)$.

The first step of our algorithm is to compute $\gamma = \gamma^{\calF}(d)$.
First, consider the case where $\gamma = \tau^{\calF}(d)$.
In this case we can use our algorithm for \textsc{Vertex Cover}.

\begin{corollary}
\label{VC-ComputesDomination}
Let $d$ be a forestic sequence. 
If $\gamma = \tau^{\calF}(d)$, 
then \Cref{alg:VertexCover} computes a realization $G$ of $d$ with a minimum dominating set $D$.
In addition, $V\setminus D$ is an independent set.
\end{corollary}

Next we consider the case where $\gammakk(d)$ is well-defined, 
and thus $\gamma = \gammakk(d)$.
We then change our framework to handle the situation. The key difference between this case and the previous one would be the sum of degrees. The difference between the sum of degrees of the initial star forest and the required sequence $d$ is:
\[
2\cdot \sum_{i=1}^\gamma d_i - \sum_{i=1}^n d_i = \sum_{i=1}^\gamma d_i - \sum_{i=\gamma +1}^n d_i
~.
\]
In the case of $\gamma = \tau^{\calF}$ this difference is non-negative, 
but in the case of $\gamma = \gammakk(d)$ it is negative.
We deal with it by switching from \edgemerge operations to \edgeaddition operations.

\Cref{alg:DominatingSet} is our algorithm for \textsc{Dominating Set} for the case where 
$\gamma = \gammakk(d)$.
It is based on the framework given in \Cref{sec:framework}. 
As \Cref{alg:VertexCover}, it starts with the initial star forest $F_0$ whose star centers are in the set
$D = \set{v_1,\ldots,v_\gamma}$.
As opposed to \Cref{alg:VertexCover}, the goal of the first while loop is to increase the degree sum to $\sum d$.
This is done using \edgeaddition operations.  After the first while loop, the rest of the construction 
is done using \leafmerge operations.
Starting with $g = \gamma+1$, the algorithm performs \edgeaddition operations in order to obtain 
a vertex with degree $d_g$, where $d_g$ is the largest degree that requires realization.
Notice that leaves that were used in \edgeaddition operations have degree two.
Observe that some iterations of the first while loop do not use edge additions.
This occurs when $d_g = 2$ and there are already unassigned degree-2 vertices in the forest.
Also notice that it may be the case that the first while loop terminates in the middle of realizing $d_g$.
In this case, the second while loop continues to work on $d_g$.

\medskip

\begin{algorithm}[H]
\caption{\small\sf Realization-DS}
\label{alg:DominatingSet}
\DontPrintSemicolon
\SetKwInOut{Input}{input}
\SetKwInOut{Output}{output}
\Input{$(d,\gamma,F_0)$}
\Output{$F$}
$F \gets F_0$ \\
$g \gets \gamma + 1$\\
\While{$\sum \deg^F < \sum d$}{
   Find an unassigned vertex $v \in V \setminus D$ with largest degree such that $\deg^F(v) \leq d_g$ \\
   \While{$\deg^F(v) < d_g$ \text{and} $\sum \deg^F < \sum d$}{
        Find leaf $w\in V \setminus D$ that is not in the same tree as $v$ \\
        $F \gets \edgeaddition(F, w, v)$
    }
    \If{$\deg^F(v) = d_g$}{
        Assign $v$ to index $g$ and relabel $v$ as $v_g$ \\
        $g \gets g+1$ 
    }
}
\While{$g \leq n$}{
    Find an unassigned vertex $v \in V \setminus D$ with largest degree such that $\deg^F(v) \leq d_g$ \\
    \While {$\deg^F(v) < d_g$}{
        Find leaf $w\in V \setminus D$ that is not in the same tree as $v$ \\
        $F \gets \leafmerge(F,w, v)$
    }
    Assign $v$ to index $g$ and relabel $v$ as $v_g$ \\
    $g \gets g+1$ \\
}
return $F$
\end{algorithm}

\medskip

A running example of the algorithm is given in \Cref{fig:ds-algorithm1} and \Cref{fig:ds-algorithm2}.

\input{figures/DominatingSetExample.tex}

\input{figures/DominatingSetExample2.tex}

In what follows we show that the algorithm computes a forest $F$ that realizes $d$
and that $D = \set{v_1,\ldots,v_\gamma}$ is a dominating set in $F$.

Let $F^0$ be the initial star forest that is created by the algorithm.
For the analysis, let $F^j$ be $F$ after the $j$th operation (\edgeaddition or \leafmerge).
Define $n^j$, $d^j$, and $\kappa^j$ to be the number of vertices in $F^j$, 
the degree sequence of $F^j$, and the number of components in $F^j$, respectively.

Notice that $\nBTwo^0 = \gamma$ and $\kappa^0 = \gamma$.
As a first step, we observe that the degree deficit is bounded.

\begin{observation}
\label{obs:EnoughComponentsDS}
If $\gammakk(d)$ is well-defined, then $0 < \sum d - \sum d^0 \leq 2(\gamma - \kappa)$.
\end{observation}
\begin{proof}
First, $\sum d - \sum d^0 > 0$ by the definition of $\gammakk(d)$.
Also, since $\kappa^0 = \gamma$, we have that
\[
2(\gamma - \kappa)
= 2\kappa^0 - 2\kappa
= (2n^0 - \sum d^0) - (2n - \sum d)
= 2(n^0 - n) + (\sum d - \sum d^0).
\]
The observation follows, since $n^0 = \gamma + \sum_{i=1}^\gamma d_i \geq n$.
\end{proof}

Next, we show that $F$ remains a forest during and after the first while loop.
We classify \edgeaddition operations into two types, the first is when two leaves are connected by an edge, 
and the second is when a leaf is connected to a non-leaf.
In the first iteration of the first while loop, $v$ is a leaf and it is connected to a leaf. 
Hence, the first operation of the while loop is always a leaf to leaf \edgeaddition.
Afterwards, as long as $d_g>2$, leaf to non-leaf \edgeaddition operations are performed.
Then, as long as $d_g = 2$, leaf to leaf \edgeaddition operations are performed.
Hence, there are three possible scenarios depending on the input sequence:
\begin{enumerate}
\item If $\omega_{2} = 0$, there is only one leaf to leaf edge addition (the first edge addition).
\item If $d_{\gamma+1} = 2$, there are only leaf to leaf edge additions.
\item If $d_{\gamma+1} > 2$ and $\omega_2 > 0$, then the first operation is a leaf to leaf edge addition, 
      and it is followed by a sequence of leaf to non-leaf edge additions. 
      Afterwards, there is a (possibly empty) sequence of leaf to leaf edge additions.
\end{enumerate}

We define two parameters for counting the two types of edge additions throughout the first loop. 
Let $\addnl$ be the number of non-leaf to leaf edge additions during the first loop,
and let $\ell$ be the number of \edgeaddition operations throughout the first loop.
If the loop terminates due to the condition of the while, then
\[
\ell 
= \frac{1}{2} \paren{ \sum d - \sum d^0 }
= \frac{1}{2} \paren{ \sum_{i=\gamma+1}^n d_i - \sum_{i=1}^\gamma d_i }
~.
\]
 
We show that $F$ remains a forest during the first $\addnl +1$ \edgeaddition operations
(or iterations of the first inner while loop).

\begin{lemma}
\label{lemma:delta0NL-DS}
Assume that $\gammakk(d)$ is well-defined.
During the first $\addnl+1$ \edgeaddition operations of the first while loop, we have that
\begin{inparaenum}[(i)]
\item $F^j$ is a forest,
\item $\kappa^j = \gamma-j \geq 1$,
\item $\nBTwo^j = \gamma+j+1$, and
\item $\nBTwo^j \leq \nBTwo$.
\end{inparaenum}
\end{lemma}
\begin{proof}
We prove the claims by induction on the number of \edgeaddition operations performed 
by the first while loop. Note that the first iteration is executed since $\sum d - \sum d^0 > 0$.

The base case is after the first operation which connects two leaves.
\Cref{obs:edgeaddition,obs:EnoughComponentsDS} imply that $F^1$ is a forest with $\kappa^1 = \gamma - 1 \geq 1$ 
and $\nBTwo^1 = \gamma + 2$. Also, due to \Cref{lemma:omega1bound}, $\nBTwo^1 = \gamma + 2 \leq \nBTwo$.

As for the induction step, assume that $F^j$ is a forest with $\nBTwo^j = \gamma + j + 1$ non-leaves 
and $\kappa^j = \gamma - j \geq 1$ trees.
Now consider the $(j+1)$th operation. 
Observe that $j+1 \leq \addnl+1 \leq \frac{1}{2} \left( \sum d - \sum d^0\right)$.
\Cref{obs:edgeaddition,obs:EnoughComponentsDS} imply that
\[
\kappa^j 
=    \gamma - j 
\geq \gamma - \frac{1}{2} \left( \sum d - \sum d^0\right) + 1
> 1
~. 
\]
\Cref{obs:edgeaddition} implies that 
$F^{j+1}$ is a forest with $\kappa^{j+1} = \kappa^j - 1 = \gamma - (j+1) > 0$ trees.
Since $j+1 \leq \ell_{NL} + 1$, the \edgeaddition operation is performed on a non-leaf and a leaf, which means that 
the number of non-leaves is increased by one, therefore $\nBTwo^{j+1} = \nBTwo^j + 1 = \gamma + (j+1) + 1$.
Also because $j+1 \leq \frac{1}{2} \left( \sum d - \sum d^0\right)$ and \Cref{lemma:DominationSetNecessity}, we get that:
\[
\nBTwo^{j+1} 
=     \gamma + (j+1) + 1 
\leq \gamma + 1 + \frac{1}{2} \paren{ \sum d - \sum d^0 } 
=     \gamma + 1 + \frac{1}{2} \paren{ \sum_{i=\gamma+1}^n d_i - \sum_{i=1}^\gamma d_i }
\leq \nBTwo
\qedhere
\] 
\end{proof}

There are two possible scenarios after the first $\addnl + 1$ operations of the first while loop.
In the first, we have reached that point where the degree deficit has become zero, namely 
$\sum d^{\addnl + 1} = \sum d$. In this case the algorithm exits the first while loop and continues to the second while loop.

This leads to one upper bound on $\addnl$.

\begin{observation}
\label{observation:NLBound-Phase1}
$\addnl \leq \ell - 1$. 
\end{observation}
\begin{proof}
An upper bound of $\addnl$ is the maximum number of edge additions that can occur throughout the first loop, 
which is $\ell = \frac{1}{2} \paren{ \sum_{i=\gamma+1}^n d_i - \sum_{i=1}^\gamma d_i }$.
Also there is a first leaf to leaf edge addition. Overall we get the bound.
\end{proof}

Let $g^*$ be the value of $g$ immediately after the first while loop terminates, 
equivalently, immediately before the second while loop starts. 
Thus, all indices smaller than $g^*$ have been assigned. 
If the vertex processed in the last iteration was not assigned, then it is intended for index $g^*$. 
Let $\mu=\nBTwo^\ell$.

The next lemma describes the sequence $d^\ell$ after the last iteration of the first loop 
assuming that $1+\addnl = \ell$.

\begin{lemma}
\label{lemma:SequenceAfterFirstLoopPhase1-DS}
Assume that $\ell=1+\addnl$.
Then $\mu=\gamma+\ell+1$, and after the first loop the sequence is
\[
d^\ell=
\left(
d_1,\ldots,d_{g^*-1},
\ell+1-\sum_{i=\gamma+1}^{g^*-1}(d_i-2),
2^{\mu-g^*},
1^{n^0-\mu}
\right)
~.
\]
where $2\leq d_{g^*}^\ell\leq d_{g^*}$.
\end{lemma}
\begin{proof}
First, $n^\ell = n^0$ due to \Cref{obs:edgeaddition}.

Since $\ell=1+\addnl$, the $\ell$ operations consist of the initial leaf-to-leaf edge addition 
and exactly $\ell-1 = \addnl$ non-leaf-to-leaf edge additions. 
The first operation creates two new nonleaves, and each of the remaining $\ell-1$ operations 
creates one new non-leaf. Therefore,
\[
\mu
= \gamma+2+(\ell-1)
= \gamma+\ell+1
~.
\]

The vertices assigned to the indices $\gamma+1,\ldots,g^*-1$ realize the corresponding degrees of $d$. 
After such a vertex has degree $2$, realizing degree $d_i$ uses exactly $d_i-2$ non-leaf-to-leaf edge additions. 
Consequently, after realizing these indices, the number of non-leaf-to-leaf edge additions not accounted for by them is
$(\ell-1) - \sum_{i=\gamma+1}^{g^*-1} (d_i - 2)$.

If the vertex processed in the last iteration was not assigned,
these remaining operations were applied to that vertex after it had degree $2$. 
If it was assigned in the last iteration, the displayed difference is zero, 
and the next unassigned non-leaf has degree $2$. In either case, the degree occupying position $g^*$ is
\[
2 + \paren{ (\ell-1) - \sum_{i=\gamma+1}^{g^*-1}(d_i-2) }
= \ell+1-\sum_{i=\gamma+1}^{g^*-1} (d_i - 2)
~.
\]
If the last processed vertex was not assigned, the selection rule and the inner-loop guard give 
$2 \leq d_{g^*}^\ell \leq d_{g^*}$. If it was assigned, then $d_{g^*}^\ell = 2$. 
Moreover, the other endpoint of the final edge addition is an unassigned degree-two vertex. 
Hence $g^*\leq\mu\leq\nBTwo$ by \Cref{lemma:delta0NL-DS}, and therefore $d_{g^*}\geq2=d_{g^*}^\ell$.

Apart from the first $g^*$ positions, the number of remaining nonleaves is $\mu-g^*$, 
and all of them have degree $2$, while the remaining $n^0-\mu$ vertices have degree $1$. 
The lemma follows.
\end{proof}

In the second scenario the first while loop moves to perform leaf to leaf edge additions.
This happens when the degrees $d_1,\ldots,d_p$ are realized, 
where $p$ is defined as follows. 
It is either the last index outside the dominating prefix whose value is at least $3$,
or $\gamma+1$, if the above index does not exist. 
Formally, $p = \max{\set{\gamma + 1, \omega_{\geq 3}}}$.
In this case the first while loop continues to realize $d$ starting from $d_{p+1}= 2$
using leaf to leaf edge additions.

This leads to a second upper bound,

\begin{observation}
\label{obs:NLBound-Phase2}
\(\displaystyle
\ell_{NL} \leq \sum_{i=\gamma+1}^p (d_i-2)
~.
\)
\end{observation}
\begin{proof}
An upper bound on $\addnl$ is
the number of edge additions needed to realize all degrees in the subsequence $\gamma + 1,\ldots, p$ minus the initial one. 
As mentioned before,
the number of edge additions required to realize a degree $d_i$, for $i>\gamma+1$ is $d_i - 2$,
because the initial degree of the vertex which is used to realize $d_i$ is two.
On the other hand, the vertex realizing $d_{\gamma+1}$ starts with a degree one.
Hence,  
\[
\addnl + 1 \leq d_{\gamma+1} - 1 + \sum_{i = \gamma+2}^p (d_i-2) = \sum_{i=\gamma+1}^p (d_i-2) + 1
~.
\qedhere
\]
\end{proof}

Define 
\[
\addnl^* \triangleq \sum_{i=\gamma+1}^p (d_i-2) 
~.
\]
Observe that $\addnl < \addnl^*$ implies that $\addnl + 1 = \ell$.
We note that in \Cref{fig:ds-algorithm1}
$\ell = 1$, $\addnl = 0$, and thus $\addnl+1=\ell$ holds, 
whereas in \Cref{fig:ds-algorithm2} $\ell = 3$, $\addnl = \addnl^* = 1$, and thus $\addnl^*+1<\ell$ holds.
Indeed, in \Cref{fig:ds-algorithm1} we have
$g^*=5$, $\mu=6$, and $d_{g^*}^\ell=2$. 
\Cref{lemma:SequenceAfterFirstLoopPhase1-DS} gives
$d^\ell=(4^4,2^2,1^{14})$, exactly as shown in the figure.
Due to the termination condition of the first while loop, the last two observations lead to the following:

\begin{observation}
\label{observation:NLCount}
\(\displaystyle
\addnl = \min{ \set{\ell - 1,  \addnl^* } }
~.
\)
\end{observation}

The next lemma deals with the second scenario where the first while loop performs leaf to leaf edge additions.
In this case, $\addnl = \addnl^*$.
Notice that there may be iterations with no edge addition operations.
This happens when $d_g = d^j_g = 2$.
Leaf to leaf edge additions are done when $d_g = 2$ and $d^j_g = 1$.

\begin{lemma}
\label{lemma:delta0LL-DS}
Assume that $\addnl^* + 1 < \ell$.
After the first $j$ operations of the first while loop, for $j \in \set{\addnl^*+1,\ldots,\ell}$, we have that
\begin{inparaenum}[(i)]
\item $F^j$ is a forest,
\item $\kappa^j = \gamma - j \geq 1$,
\item $\nBTwo^j = \gamma - \addnl^* + 2j$, and
\item $\nBTwo^j \leq \nBTwo$.
\end{inparaenum}
\end{lemma}
\begin{proof}
We prove the claims by induction on the number of operations performed by the first while loop.
The base case, $j = \addnl^*+1$, is implied by \Cref{lemma:delta0NL-DS}. 

As for the induction step, assume that $F^j$ is a forest with $\nBTwo^j = \gamma - \addnl^* + 2j$ non-leaves
and $\kappa^j = \gamma - j \geq 1$ trees.
Now consider the $(j+1)$th operation. 
Observe that $j+1 \leq \frac{1}{2} \left( \sum d - \sum d^0\right)$.
\Cref{obs:edgeaddition} and \Cref{obs:EnoughComponentsDS} imply that
\[
\kappa^j 
=    \gamma - j 
\geq \gamma - \frac{1}{2} \left( \sum d - \sum d^0\right) + 1
> 1
~. 
\]
\Cref{obs:edgeaddition}, implies that 
$F^{j+1}$ is a forest with $\kappa^{j+1} = \kappa^j - 1 = \gamma - (j+1) > 0$ trees.
Since the edge addition operation is performed on two leaves, the number of non-leaves is increased by two,
Hence $\nBTwo^{j+1} = \nBTwo^j + 2 = \gamma -  \addnl^* + 2(j+1)$.
Because $j+1 \leq \ell$, it follows that $\nBTwo^{j+1} \leq \gamma - \addnl^*  + 2\ell$.
By the definition of $p$, we have that $d_{p+1} \leq 2$, and thus $\sum_{i=p+1}^n d_i = 2(n-p) - \omega_1$.
Hence, 
\begin{align}
\notag
\gamma - \addnl^*  + 2\ell 
 =     \gamma - \sum_{i=\gamma+1}^p (d_i -2) +  \sum_{i=\gamma+1}^n d_i - \sum_{i=1}^\gamma d_i 
& =    2p - \gamma + 2(n-p) - \omega_1 -  \sum_{i=1}^\gamma d_i \\
& =    \nBTwo + n - \gamma -  \sum_{i=1}^\gamma d_i 
~.
\label{eqn:geqtwo}
\end{align}
Therefore, 
\(
\nBTwo^{j+1} 
\leq  \nBTwo + n - \gamma -  \sum_{i=1}^\gamma d_i
\leq \nBTwo
\).
\end{proof}

The next lemma describes $d^\ell$ upon exiting the first while loop, for the case where 
$\addnl^* + 1 < \ell$.
Recall that $\mu=\nBTwo^\ell$.

\begin{lemma}
\label{lemma:SequenceAfterFirstLoopPhase2-DS}
Assume that $\addnl^* + 1 < \ell$. 
Then
\(
d^\ell=(d_1,\ldots,d_\mu,1^{n^0-\mu})
\).
\end{lemma}
\begin{proof}
We recall that $n^\ell = n^0$.
\Cref{lemma:delta0LL-DS} and \eqref{eqn:geqtwo} imply that 
\[
\mu
=\nBTwo^\ell 
= \gamma - \addnl^* + 2\ell 
= \nBTwo +  n - \gamma -  \sum_{i=1}^\gamma d_i
~.
\]
Before this phase, all target degrees greater than two have been realized. 
Each subsequent leaf-to-leaf edge addition creates two degree-two entries: the processed endpoint is assigned, 
while the other endpoint remains as an unassigned degree-two entry. 
Consequently, the first $\mu=\nBTwo^\ell$ entries are
$d_1,\ldots,d_\mu$, and all remaining entries are one. 
\end{proof}

We note that when $\ell = \addnl^*+1$, the final edge addition completes and assigns $d_p$. 
This is the borderline between the two cases.

The next lemma is a first step towards proving the correctness after the second while loop.

\begin{lemma}
\label{lemma:BudgetEqualsExcess-DS}
\(\displaystyle
\kappa^\ell - \kappa = n^\ell -n = \sum_{i=\gamma + 1}^{\nBTwo} (d_i - d_i^\ell)
\).
\end{lemma}
\begin{proof}
Since $\sum d^\ell = \sum d$ upon termination of the first while loop, 
we have that
\[
\kappa^\ell - \kappa
= \paren{ n^\ell - \frac{1}{2} \sum d^\ell } - \paren{ n - \frac{1}{2} \sum d }
= n^\ell - n
~.
\]

First, assume that $\ell = \addnl + 1$.
In this case, \Cref{lemma:SequenceAfterFirstLoopPhase1-DS}, together with
$\mu=\gamma+\ell+1$ and the definition of $\ell$, gives
\begin{align*}
\sum_{i=\gamma+1}^{\nBTwo} (d_i-d_i^\ell)
& = d_{g^*} - d_{g^*}^\ell + \sum_{i=g^*+1}^{\mu}(d_i-2) + \sum_{i=\mu+1}^{\nBTwo}(d_i-1) \\
& = d_{g^*} - (\ell +1) + \sum_{i=\gamma+1}^{g^*-1} (d_i - 2) + \sum_{i=g^*+1}^{\mu}(d_i-2) + \sum_{i=\mu+1}^{\nBTwo}(d_i-1) \\
& = 1-\ell + \sum_{i=\gamma+1}^{\mu}(d_i-2) + \sum_{i=\mu+1}^{\nBTwo}(d_i-1) \\
& = \sum_{i=\gamma+1}^{\nBTwo}d_i +1-\ell -2(\mu-\gamma) -(\nBTwo-\mu) \\
& = \sum_{i=\gamma+1}^{\nBTwo}d_i +\gamma-\nBTwo-2\ell \\
& = \sum_{i=\gamma+1}^{n}(d_i-1)-2\ell
~,
\end{align*}
and by the definition of $\ell$, we have that
\[
\sum_{i=\gamma+1}^{n}(d_i-1)-2\ell
= \sum_{i=1}^{\gamma}d_i-(n-\gamma)
= n^0-n
= n^\ell-n
~.
\]

Next, assume that $\addnl + 1 < \ell$.
Due to \Cref{lemma:SequenceAfterFirstLoopPhase2-DS},
\[
\sum_{i=\gamma + 1}^{\nBTwo} (d_i - d_i^\ell)
= \sum_{i=\mu + 1}^{\nBTwo} (2 - 1)
= \nBTwo - \mu
= - n + \gamma +  \sum_{i=1}^\gamma d_i 
= n^\ell - n 
~.
\qedhere
\]
\end{proof}

Recall that $g^*$ is the value of $g$ upon starting the second while loop.
Consider the last iteration of the first while loop. 
If $d_g$ was assigned to $v$ (i.e., realized) during the last iteration, then $g^* = g + 1$, otherwise $g^* = g$.
If $\ell > \addnl + 1$, then a leaf to leaf edge addition was performed in the last iteration, and thus 
$g^* = \mu$.

The next lemma shows the correctness of the realization after termination of the algorithm.

\begin{lemma}
\label{lemma:whileleafmerge-DS}
After $j$ iterations of the second while loop, $F^{\ell + q(j)}$ is a forest such that
\begin{compactitem}
\item $n^{\ell + q(j)} = n^0 - q(j)$,
\item $\kappa^{\ell + q(j)} = \gamma - \ell - q(j)$, and
\item $d^{\ell+q(j)}_i = d_i$, for every $i \in \set{1,\ldots,g^*+j-1}$.
\end{compactitem}
where $q(j) = \sum_{i=g^*}^{g^*+j - 1}(d_i - d_i^\ell)$. 
At termination, for $j = n - g^* + 1$, we have that $d^{\ell+q(j)} = d$.
\end{lemma}
\begin{proof}
We prove the claim by induction on the number of iterations of the second while loop.

For the base case $j = 0$, observe that $q(j) = 0$, hence the claim follows from either
\Cref{lemma:delta0NL-DS,lemma:SequenceAfterFirstLoopPhase1-DS} (if $\ell = \addnl + 1$) or
\Cref{lemma:delta0LL-DS,lemma:SequenceAfterFirstLoopPhase2-DS} (if $\ell > \addnl + 1$).

For the induction step, assume that the claim holds for $j-1$.
Consider iteration $j$ of the second while loop, and let $h = g^* + j - 1$.
Iteration $j$ starts with selecting an unassigned vertex $v$ with the largest degree, such that $\deg^F(v) \leq d_h$. 
The iteration increases the selected vertex from degree $d_h^\ell$ to $d_h$ 
and deletes $d_h-d_h^\ell$ unassigned degree-$1$ vertices.
At the end of the iteration $v$ is relabeled as $v_h$.
Note that iterations with zero operations are possible, 
in which the only action is the assignment of $d_h$ to $v_h$.
In such an iteration $j$, we have $q(j) = q(j-1)$.

The defining inequality in \Cref{thm:domset-characterization} implies that $\nBTwo \geq \gamma+2$, 
which means that $d_\gamma \geq 2$. 
Thus the leaf-availability argument in the proof of \Cref{lemma:whileleafmerge} applies.
That is, every tree in the current forest, throughout iteration $j$, has at least two leaves.

\Cref{obs:leafMerge} and \Cref{lemma:BudgetEqualsExcess-DS} imply that both
$\kappa^{\ell+q(j)} - \kappa$ and $n^{\ell+q(j)} - n$ remain non-negative.
Moreover, we know that $F$ remains a forest and that $n(F)$ and $\kappa(F)$ 
decrease by $d_h - d_h^\ell$ in iteration $j$ of the second while loop.
Hence, by the induction hypothesis
\[
n^{\ell + q(j)}
= n^{\ell + q(j-1)} - (d_h - d_h^\ell )
= n^0 - q(j) 
~,
\]
and
\[
\kappa^{\ell + q(j)}
= \kappa^{\ell + q(j-1)} - (d_h - d_h^\ell )
= \gamma - \ell - q(j) 
~.
\]
As for the third claim, $d^{\ell+q(j-1)}_i = d_i$, for every $i \in \set{1,\ldots,h-1}$ by the induction hypothesis.
Since iteration $j$ does not change entries $i$, for $i<h$, this also holds for $d^{\ell+q(j)}$.
In addition, iteration $j$ increases entry $h$ from $d_h^{\ell + q(j-1)}$ to $d_h^{\ell + q(j)} = d_h$.
The lemma follows.
\end{proof}

\begin{theorem}
\label{thm:DominatingSet}
Let $d$ be a forestic sequence and let $\gamma=\gamma^{\calF}(d)$.
There exists an algorithm that computes a forest realization $F$ of $d$, 
such that $D = \set{v_1,\ldots,v_\gamma}$ is a dominating set in $F$.
Moreover, either $D$ is an independent set or $V \setminus D$ is an independent set.
\end{theorem}
\begin{proof}
If $\gamma^{\calF}(d) = \tau^{\calF}(d)$, then the theorem follows from \Cref{VC-ComputesDomination}.
In this case, $D$ is a vertex cover, and thus $V \setminus D$ is an independent set.

Otherwise, assume that $\gamma^{\calF}(d) = \gammakk(d)$ and let $F$ be the computed solution by \Cref{alg:DominatingSet}.
$F$ is a forest due to \Cref{lemma:whileleafmerge-DS}.
We prove that $D$ is an independent dominating set in $F^j$ by induction on $j$.
Clearly, $D$ is an independent dominating set of $F^0$.
For the induction step, assume that $D$ is an independent dominating set of $F^{j-1}$.
An edge addition or a leaf merge does not create newly uncovered vertices in $V \setminus D$, 
and also neither operation connects dominating vertices by edges,
therefore $D$ is an independent dominating set of $F^j$.
\end{proof}

As for the running time of \Cref{alg:DominatingSet}, the initial star forest can be 
constructed in $O(\sum_{i=1}^\gamma d_i) = O(\sum d)$ time.
The algorithm performs the merges by maintaining one large component $C$ and a set of stars,
similarly to the implementation of \Cref{alg:VertexCover}.
Recall that in leaf merge phase of \Cref{alg:VertexCover}, 
the next available leaf of $v_i \in D \cap V(C)$, such that $i$ is minimal, was chosen.
Then, each iteration of the inner loop, a leaf of $v_j \in D \setminus V(C)$, where  $j$ is minimal, merged into
the chosen vertex.
The same approach is taken here both for edge additions and leaf merges. 
However, in this algorithm the chosen vertex $v$ may be a vertex of degree $2$.
Hence, each forest operation can be implemented to run in $O(1)$ time.
In addition, there are $\frac{1}{2} (\sum d - \sum d^0)$ edge additions, and $\sum_{i=1}^\gamma d_i - (n-\gamma)$ leaf merges. 
%
Hence, the total running time is $O(\sum d) = O(n)$.


\section{Optimized Total Dominating Set in Forests}

In this section we consider \degr with respect to \totdomset.
We provide a characterization of the minimum size total dominating set of 
a forest realization of a given sequence, and we also present a realization algorithm.
Our realization algorithm is based on a reduction to optimized \domset. 
As in the case of \totvc, the reduction simplifies the analysis, but our algorithm can also be viewed as 
using our domination framework.
As in \Cref{sec:tvc}, we show that a minor adaptation is needed to obtain a characterization and a realization algorithm for
\degr with respect to \pairdomset.


\subsection{Properties of Total Dominating Set}

This section contains several properties and necessary conditions for a total dominating set.
We start with a prefix lemma for \totdomset.

\begin{lemma}[Total DS Prefix Lemma]
\label{lemma:swappingTotalDominatingSet}
Let $d$ be a forestic sequence with a realization $F$ such that $\deg(v_i) = d_i$.
Also, let $D$ be a total dominating set in $F$, such that $t = |D|$.
Then, there exists a realization $F'$ of $d$ such that $D' = \set{v_1,\ldots,v_t}$ is a total dominating set.
\end{lemma}
\begin{proof}
We use an argument similar to the one used
in the proof of \Cref{swappingDominatingSetLemma}.
Recall that in the proof of \Cref{swappingDominatingSetLemma} we assume that
there exist two vertices $v_j$ and $v_\ell$, such that $v_j \not\in D$, $v_\ell \in D$, and $d_j > d_\ell$. 

First, consider the case where $v_j$ and $v_\ell$ lie in the same tree.
If $d_\ell = 1$, then $v_\ell$ has a unique neighbor $v_p$, such that $d_p > 1$ and $v_p \in D$.
Since $v_j \not\in D$, it follows that $v_p \neq v_j$. 
Let $v_q \in N(v_p) \setminus \set{v_\ell}$ be the neighbor of $v_p$ on the unique path from $v_p$ to $v_j$.
If $v_j = v_q$, then we are done (without relabeling).
Otherwise $d_q > 1$.
If $v_q \notin D$, then the set $D \setminus \set{v_\ell} \cup \set{v_q}$ is also a total dominating set of $F$ of size $t$, 
and we are done (without relabeling).
Else, $D \setminus \set{v_\ell} \cup \set{v_j}$ (no relabeling) is a total dominating set of size $t$.

Thus for the rest of the proof we assume that $d_\ell \geq 2$.
The rest of the proof follows the proof of \Cref{swappingDominatingSetLemma}.
We obtain a forest $F'$ from $F$ by transferring $d_j - d_\ell$ edges from $v_j$ to $v_\ell$ without creating cycles.  
Remember that we keep $v_{j'}\in N(v_j) \cap  D$ as a neighbor of $v_j$.
Observe that the removed edges connect $v_j$, which is not in the dominating set, with other vertices.
Hence, $F[D]$ is a subgraph of $F'[D]$. 
It follows that $F'[D']$ does not contain isolated vertices.
\end{proof}

We give a necessary condition
similar to~\Cref{lemma:DominationSetNecessity} in \Cref{sec:domset}.

\begin{lemma}
\label{lemma:TotalDominationSetNecessity}
Let $d$ be a forestic sequence.
Let $F = (V,E)$ be a forest realization of $d$, and let $D$ be a total dominating set of size $t$.
Either $\sum_{i=1}^t d_i \geq \sum_{i=t+1}^n d_i + 2\ceil{t/2}$, or
\[
\sum_{i=t+1}^n d_i - \sum_{i=1}^t d_i \leq 2(\nBTwo - t) - 2 - 2\ceil{t/2}
~.
\]
\end{lemma}

\begin{proof}
By \Cref{lemma:swappingTotalDominatingSet} we assume that $D$ is composed of the $t$ highest degree vertices.

If $\sum_{i=1}^t d_i \geq \sum_{i=t+1}^n d_i +2\ceil{\frac{t}{2}}$, then we are done. 
Hence, we assume that $\sum_{i=t+1}^n d_i + 2\ceil{\frac{t}{2}} > \sum_{i=1}^t d_i$.

We first claim that $d_{t+1} > 1$.
Suppose that $d_{t+1} = 1$, then $\sum_{i=t+1}^n d_i + 2\ceil{\frac{t}{2}}= n - t +2\ceil{\frac{t}{2}} > \sum_{i=1}^t d_i$,
which means that $D$ cannot be a total dominating set due to \Cref{obs:tds-odd}. A contradiction.
It follows that $\omega_{\geq 2} \geq t+1$.
 
We partition $V$ into three disjoint subsets as was done in the proof of \Cref{lemma:DominationSetNecessity}:
$D$, $\bar{D}_{=1}$, $\bar{D}_{\geq2}$,
where $\bar{D}_{=1} = \set{v_i \not\in D : d_i = 1}$ and $\bar{D}_{\geq2} = V \setminus (D \cup \bar{D}_{=1})$.
Notice that $\bar{D}_{\geq 2} \neq \emptyset$, since $\omega_{\geq 2} \geq t+1$.

All degree-1 vertices must be connected to a vertex in $D$. Hence, $|E(D,\bar{D}_{=1})| = n - \nBTwo$.
Each vertex in $\bar{D}_{\geq 2}$ must also be connected to a vertex in $D$. In addition, 
there are at least $\ceil{\frac{t}{2}}$ edges connecting vertices of $D$ with each other.
It follows that 
\[
2|E(\bar{D}_{\geq 2},\bar{D}_{\geq2})| 
\geq \sum_{i=t+1}^{\nBTwo} d_i - \paren{ \sum_{i=1}^t d_i - (n - \nBTwo) - 2\ceil{\frac{t}{2}}}
=    \sum_{i=t+1}^n d_i + 2\ceil{\frac{t}{2}} - \sum_{i=1}^t d_i 
>    0
~.
\]
See depiction in \Cref{fig:DomSetPartition}.
Since $F[\bar{D}_{\geq 2}]$ is a forest, it must be that 
\(
2|E(\bar{D}_{\geq 2},\bar{D}_{\geq2})| 
\leq 2(\nBTwo - t) - 2
\).
The lemma follows.
\end{proof}

Putting together \Cref{obs:tvcIstds} and \Cref{lemma:TotalDominationSetNecessity} we infer that:
\begin{equation}
\label{eqn:tds}
\gammat^{\calF}(d)  \geq \min \set{ \taut^{\calF}(d), \gammakk_t(d) }
~,
\end{equation}
where 
\begin{align*}
\taut^{\calF}(d) & = \min \set{ k : \sum_{i=1}^k d_i - \sum_{i=k+1}^n d_i - 2\ceil{k/2} \geq 0} \\
\gammakk_t(d) &
= \min \set{ k : \sum_{i=1}^k d_i \geq n + (2\ceil{k/2} - k) \text{ and } 0 < \sum_{i=k+1}^n d_i - \sum_{i=1}^k d_i + 2\ceil{k/2} \leq 2(\nBTwo - k) - 2}
\end{align*}

\begin{observation}
\label{obs:k1Greaterk2Total}
If $\gammakk_t(d)$ is well-defined, then $\gammakk_t(d) < \taut^{\calF}(d)$.
\end{observation}

\begin{observation}
\label{obs:Totalk2d2}
If $\gammakk_t(d)$ is well-defined, then $\gammakk_t(d) \leq \nBTwo - 2$.
\end{observation}
\begin{proof}
If $\gammakk_t(d)$ is well-defined, then $0 < 2(\nBTwo - \gammakk_t(d)) - 2$.
\end{proof}


\subsection{Computing a Realization}

In this section we describe an algorithm which, given a forestic sequence $d$, computes a forest realization
with a total dominating set of size $\min \set{ \taut^{\calF}(d), \gammakk_t(d)}$.
The correctness proof of this algorithm serves as a matching upper bound 
to \eqref{eqn:tds} on $\gammat^{\calF}(d)$.
The first step of our algorithm is to compute $\gammat = \min \set{ \taut^{\calF}(d), \gammakk_t(d)}$.

First, consider the case where $\gammat = \taut^{\calF}(d)$.
In this case we can use our algorithm for \textsc{Total Vertex Cover}.

\begin{corollary}
\label{TVC-ComputesDomination}
Let $d$ be a forestic sequence. 
If $\gammat = \taut^{\calF}(d)$, then by \Cref{thm:TotalVertexCover} there is an algorithm 
which computes a realization $G$ of $d$ with a minimum total dominating set $D$.
In addition, $V\setminus D$ is an independent set.
\end{corollary}

Next we consider the case where $\gammakk_t(d)$ is well-defined, and thus $\gammat = \gammakk_t(d)$.
As a first step, we recall the reduction from \Cref{sec:tvc-computing} and use it to reduce from 
optimized \totdomset to optimized \domset.
We use the reduction that is defined in \Cref{eqn:TotalReduction},
where $\taut$, $\tau'$, and $\delta$ are replaced by $\gammat$, $\gamma' = \floor{\gammat/2}$, and $0$, respectively.
Here $\gamma'$ counts the contracted blocks. If $\gammat$ is odd, the first
three vertices form one block and the remaining $\gammat-3$ vertices form
pairs, so $\gamma'=1+(\gammat-3)/2=\floor{\gammat/2}$.
That is,
\begin{equation}
d' = d'' \circ (d_{\gammat + 1}, \ldots, d_n)
~,
\end{equation}
where 
\[
d''_i =
\begin{cases}
d_{2i-1} + d_{2i} - 2    &  1 \leq i \leq \gamma' \text{ and } \gammat \text{ is even}, \\
d_{2i} + d_{2i+1} - 2   &  1 < i \leq \gamma' \text{ and } \gammat \text{ is odd}, \\
d_1 + d_2 + d_3 - 4   &  i = 1 \text{ and } \gammat \text{ is odd}. \\
\end{cases}
\]
Notice that $d_{\gammat} \geq 2$ due to  \Cref{obs:Totalk2d2}.
Since $d$ is non-increasing, every entry among $d_1,\ldots,d_{\gammat}$ is at least $2$.
Therefore, every contracted prefix entry is at least $2$, i.e., $d'_i\geq 2$, for $1\leq i\leq\gamma'$.
The suffix $(d_{\gammat+1},\ldots,d_n)$ is unchanged, and hence
$\omega'_{\geq 2}=\gamma'+(\nBTwo-\gammat)$; consequently,
$\omega'_{\geq 2}-\gamma'=\nBTwo-\gammat$.
Moreover, observe that $n' = n - \gammat + \gamma' = n - \ceil{\gammat / 2}$ and that 
$\sum_{i=1}^{\gamma'} d'_i = \sum_{i=1}^{\gammat} d_i - 2\ceil{\gammat / 2}$.

The proofs of the following observations are similar to those of
\Cref{obs:reduction-nonincreasing,obs:reduction-forestic}

\begin{observation}
\label{obs:reduction-nonincreasing-ds}
$d'$ is non-increasing.
\end{observation}

\begin{observation}
\label{obs:reduction-forestic-ds}
If $d$ is a forestic sequence, then $d'$ is a forestic sequence.
\end{observation}

We proceed to prove that the reduction is valid.

\begin{lemma}
\label{lemma:DSiffTDS1}
$\sum_{i=1}^{\gamma'} d'_i  \geq n' - \gamma'$ 
if and only if 
$\sum_{i=1}^{\gammat} d_i \geq n + (2\ceil{\gammat/2} - \gammat)$.
\end{lemma}
\begin{proof}
If $\sum_{i=1}^{\gammat} d_i \geq n + (2\ceil{\gammat/2} - \gammat)$, then
\[
\sum_{i=1}^{\gamma'} d'_i
=      \sum_{i=1}^{\gammat} d_i - 2\ceil{\gammat / 2} 
\geq \paren{ n + (2\ceil{\gammat /2} - \gammat)} - 2\ceil{\gammat / 2} 
=     n - \gammat
=     n' - \gamma'
~.
\]
If $\sum_{i=1}^{\gamma'} d'_i  \geq n' - \gamma'$, then
\[
\sum^{\gammat}_{i=1} d_i 
=       \sum_{i=1} ^ {\gamma'} d'_i  + 2\ceil{\gammat / 2} 
\geq  n' - \gamma' + 2\ceil{\gammat / 2} 
=       n - \gammat + 2\ceil{\gammat / 2} 
=       n  + (2\ceil{\gammat / 2} - \gammat)
~.
\qedhere
\]
\end{proof}

\begin{lemma}
\label{lemma:DSiffTDS2}
$0 < \sum_{i=\gamma'+1}^{n'} d'_i - \sum_{i=1}^{\gamma'} d'_i \leq 2(\omega'_{\geq 2} - \gamma') - 2$
if and only if 
$0 < \sum_{i=\gammat+1}^n d_i - \sum_{i=1}^{\gammat} d_i  + 2\ceil{\gammat/2} \leq 2(\nBTwo - \gammat) - 2$.
\end{lemma}
\begin{proof}
If $0 < \sum_{i=\gammat+1}^n d_i - \sum_{i=1}^{\gammat} d_i  + 2\ceil{\gammat/2} \leq 2(\nBTwo - \gammat) - 2$,
then
\[
0
<
\sum_{i=\gamma'+1}^{n'} d'_i - \sum_{i=1}^{\gamma'} d'_i
=      \sum_{i=\gammat+1}^n d_i - \sum_{i=1}^{\gammat} d_i + 2\ceil{\gammat / 2} 
\leq 
2(\nBTwo - \gammat) - 2
=
2(\omega'_{\geq 2} - \gamma') - 2
~.
\]
If $0 < \sum_{i=\gamma'+1}^{n'} d'_i - \sum_{i=1}^{\gamma'} d'_i \leq 2(\omega'_{\geq 2} - \gamma') - 2$, then
\[
0
<
\sum_{i=\gammat+1}^n d_i - \sum_{i=1}^{\gammat} d_i + 2\ceil{\gammat/2}
=        \sum_{i=\gamma'+1}^{n'} d'_i  - \sum_{i=1}^{\gamma'} d'_i
\leq  2(\omega'_{\geq 2} - \gamma') - 2
=
2(\nBTwo - \gammat) - 2
~.
\qedhere
\]
\end{proof}

\begin{lemma}
\label{lemma:DSeqTDS}
Let $d$ be a forestic sequence for which $\gammakk_t(d)$ is well-defined. 
Then, $\gamma^{\cal F}(d')=\gammakk(d')=\gamma'$.
\end{lemma}
\begin{proof}
Let $\gammat=\gammakk_t(d)$.
By \Cref{lemma:DSiffTDS1,lemma:DSiffTDS2},
$\gamma'$ satisfies both conditions defining $\gammakk(d')$.
Thus, $\gammakk(d')$ is well-defined and $\gammakk(d') = \gamma^{\cal F}(d') \leq \gamma'$. 
Assume toward a contradiction that $\gamma' > \gammakk(d') \geq1$.

We first show that $\gamma'-1$ also satisfies both conditions in the definition of $\bar{\gamma}(d')$.
First, 
\[
\sum_{i=1}^{\gamma' - 1} d'_i  
=      \sum_{i=1}^{\gammakk(d')} d'_i  + \sum_{i=\gammakk(d')+1}^{\gamma'-1} d'_i 
\geq n' - \gammakk(d') + (\gamma' - 1 - \gammakk(d')) 
\geq n' - (\gamma' -1) 
~.
\]
Due to \Cref{lemma:DSiffTDS2}, 
\[
\sum_{i=\gamma'}^{n'} d'_i - \sum_{i=1}^{\gamma'-1} d'_i
= \paren{ \sum_{i=\gamma'+1}^{n'} d'_i - \sum_{i=1}^{\gamma'} d'_i } + 2d'_{\gamma'}
> 0
~.
\]
Also,
\begin{align*}
\sum_{i=\gamma'}^{n'} d'_i - \sum_{i=1}^{\gamma'-1} d'_i 
& = \sum_{i=\gammakk(d')+1}^{n'} d'_i - \sum_{i=1}^{\gammakk(d')} d'_i - 2\sum_{i=\gammakk(d')+1}^{\gamma'-1} d'_i  \\
& \leq 2(\nBTwo' - \gammakk(d')) - 2 - 4(\gamma' - 1 - \gammakk(d'))  
& \leq 2(\nBTwo' - (\gamma'-1)) - 2
~.
\end{align*}

Next, we show that $\gammat-1$ satisfies both conditions of $\gammakk_t(d)$.
First,
\[
\sum_{i=1}^{\gammat-1} d_i
= \sum_{i=1}^{\gammat} d_i - d_{\gammat} 
= \sum_{i=1}^{\gamma'} d'_i + 2\ceil{\frac{\gammat}{2}} - d_{\gammat} 
= \sum_{i=1}^{\gamma'-1} d'_i + d'_{\gamma'} + 2\ceil{\frac{\gammat}{2}} - d_{\gammat} 
= \sum_{i=1}^{\gamma'-1} d'_i + d_{\gammat-1} + 2\ceil{\frac{\gammat}{2}} - 2 
~.
\]
Since $d_{\gammat-1} \geq 2$ and $n' = n - \ceil{\gammat/2}$, 
\[
\sum_{i=1}^{\gammat-1} d_i
\geq n' - (\gamma' -1) + 2\ceil{\frac{\gammat}{2}} 
=      n - \ceil{\frac{\gammat}{2}} - \paren{ \floor{\frac{\gammat}{2}} - 1 } + 2\ceil{\frac{\gammat}{2}} 
\geq n - (\gammat - 1) + 2\ceil{\frac{\gammat-1}{2}} 
~.
\]

We have that
\begin{align*}
\sum_{i=\gammat}^{n} d_i - \sum_{i=1}^{\gammat-1} d_i + 2\ceil{(\gammat-1)/2} 
& = \sum_{i=1}^{n} d_i - 2 \sum_{i=1}^{\gammat-1} d_i + 2\ceil{(\gammat-1)/2} \\
& = \paren{ \sum_{i=1}^{n'}d'_i + 2\ceil{\gammat/2} } - 2 \paren{\sum_{i=1}^{\gamma'-1}d'_i + d_{\gammat-1} + 2\ceil{\gammat/2} - 2}
       + 2\ceil{(\gammat-1)/2} \\
& = \sum_{i=\gamma'}^{n'} d'_i - \sum_{i=1}^{\gamma'-1} d'_i + 4 + 2\ceil{(\gammat-1)/2} - 2\ceil{\gammat/2} -2d_{\gammat-1} \\
& \leq 2(\nBTwo' - (\gamma' - 1)) - 2\\
& = 2(\nBTwo - (\gammat-1)) - 2
~.
\end{align*}

Finally,
\begin{align*}
\sum_{i=\gammat}^{n} d_i - \sum_{i=1}^{\gammat-1} d_i + 2\ceil{\frac{\gammat-1}{2}} 
& =      \sum_{i=\gammat+1}^{n} d_i + d_{\gammat} - \sum_{i=1}^{\gammat} d_i + d_{\gammat} + 2\ceil{\frac{\gammat-1}{2}} \\
& \geq \sum_{i=\gammat+1}^{n} d_i - \sum_{i=1}^{\gammat} d_i + 2 + 2\ceil{\frac{\gammat-1}{2}} \\
& = \sum_{i=\gammat+1}^{n} d_i - \sum_{i=1}^{\gammat} d_i + 2\ceil{\frac{\gammat+1}{2}} 
& \geq      \sum_{i=\gammat+1}^{n} d_i - \sum_{i=1}^{\gammat} d_i + 2\ceil{\frac{\gammat}{2}} 
& >    0 
~.
\end{align*}
Hence $\gammat-1$ satisfies both conditions of $\gammakk_t(d)$ in contradiction 
to the minimality of $\gammat = \gammakk_t(d)$.
\end{proof}

\begin{lemma}
\label{lemma:DominatingSetRealizationReduction}
Let $d$ be a forestic sequence for which $\gammakk_t(d)$ is well-defined,
and let $\gammat=\gammakk_t(d)$.
Then there exists a forest $F$ that realizes $d$ with a total dominating set $\set{v_1,\ldots,v_{\gammat}}$.
\end{lemma}
\begin{proof}
We know that $d'$ is non-increasing and forestic due to
\Cref{obs:reduction-nonincreasing-ds,obs:reduction-forestic-ds}. 
Moreover, \Cref{lemma:DSeqTDS} 
implies that $\gamma^{\cal F}(d') = \gammakk(d') = \gamma'$.
Hence, by \Cref{thm:DominatingSet}, there is a forest realization $F'$ of $d'$ with dominating set
$D' = \set{v_1,\ldots,v_{\gamma'}}$.

Construct $F$ by splitting every vertex of $D'$ exactly as in
\Cref{lemma:VertexCoverRealizationReduction}, with $\delta=0$, and let $D$ be
the $\gammat$ vertices created by these splits. Splitting vertices preserves
the forest property, and the definition of the contracted degrees shows that
$F$ realizes $d$. Each two-vertex block induces an edge, while in the odd case
the first block induces a three-vertex path, so $F[D]$ has no isolated
vertices. Every vertex outside $D$ corresponds to a vertex outside $D'$ and
retains an edge to a split vertex of $D$. Hence, $D$ is a total dominating set
of $F$.
\end{proof}

The following is implied by Equation~\eqref{eqn:tds}, 
\Cref{TVC-ComputesDomination,corollary:tvc-characterization},
and \Cref{lemma:DominatingSetRealizationReduction}.

\begin{corollary}
\label{thm:totaldomset-characterization}
If $d$ is a forestic sequence, then
\(
\gammat^{\calF}(d) = \min \set{ \taut^{\calF}(d), \gammakk_t(d)}
\).
\end{corollary}

The next result is a realization algorithm.

\begin{theorem}
\label{thm:TotalDominatingSet}
Let $d$ be a forestic sequence.
Then there is a linear time algorithm that computes a realization $F$ of $d$ with total dominating set $D$ 
of size $\gammat = \gammat^{\calF}(d)$.
\end{theorem}
\begin{proof}
First, assume that $\gammat^{\calF}(d) = \taut^{\calF}(d)$.
In this case, the proof follows from \Cref{TVC-ComputesDomination}.

Otherwise, we can assume $\gammat^{\calF}(d) = \gammakk_t(d)$.
The proof of \Cref{lemma:DominatingSetRealizationReduction} induces the following algorithm.
Given $d$, compute $\gammat$, the reduced sequence $d'$, and $\gamma' = \floor{\gammat/2}$. 
Let $F'_0$ be the corresponding initial star forest. Recall that by \Cref{lemma:DSeqTDS},
$\gamma^{\cal F}(d') = \gammakk(d') = \gamma'$.
Execute \Cref{alg:DominatingSet} on $(d',\gamma',F'_0)$ to obtain a realization $F'$ of $d'$ with
dominating set $D' = \set{v_1,\ldots,v_{\gamma'}}$.
Then apply the splitting construction from \Cref{lemma:DominatingSetRealizationReduction}
to obtain $F$ and $D$.

The correctness of the above algorithm is implied by \Cref{lemma:DominatingSetRealizationReduction}.
As for the running time, the computation of $\gammat$, $\gamma'$, and $d'$ takes $O(n)$ time.
Since $n' \leq n$, \Cref{alg:DominatingSet} requires $O(n') = O(n)$ time.
Finally, obtaining $F$ and $D$ from $F'$ and $D'$ takes $O(\gammat) = O(n)$ time.
\end{proof}

\input{figures/TotalDominatingSetExample.tex}


\subsection{Using the Domination Framework}

In this section, similarly to \Cref{sec:tvc}, we present an algorithm which uses only framework operations.
The algorithm in this subsection applies only when $\gammat=\gammakk_t(d)$. 
If $\gammat=\taut^{\calF}(d)$, we instead use \Cref{alg:TotalVertexCover}, as in \Cref{TVC-ComputesDomination}.
The algorithm follows the same steps as \Cref{alg:TotalVertexCover}.
It first performs edge merges that imitate the reduction.
It then runs \dominatingset on the resulting forest while treating the components as stars,
that is each block of dominating vertices created by the preliminary edge merges is treated 
as a single center of the corresponding reduced star. 
After contraction, the resulting forest is exactly the initial star forest for $d'$ and $\gamma'$.
%
\Cref{lemma:DSeqTDS} lets us run \dominatingset in \Cref{alg:TotalDominatingSet} in the correct manner.

\medskip

\begin{algorithm}[H]
\caption{\small\sf Realization-TDS}
\label{alg:TotalDominatingSet}
\SetKwInOut{Input}{input}
\SetKwInOut{Output}{output}
\Input{$(d,\gammat)$}
\Output{$F$}
\DontPrintSemicolon
Let $F$ be the initial star forest of $d$ and $D = \set{v_1,\ldots,v_{\gammat}}$ \\
\lFor{$i=0,\ldots, \floor{\gammat/2} - 2$} 
{
	$F \gets \edgemerge (F,l(v_{\gammat - 2i}),l(v_{\gammat - 2i - 1}))$
}
$F \gets \edgemerge(F,l(v_1),l(v_2))$\\
\lIf{$\gammat$ is odd}
{
	$F \gets \edgemerge(F,l(v_1),l(v_3))$
}
\Return $\dominatingset(d', \gamma', F)$
\end{algorithm}


\medskip

Two running examples are shown in \Cref{fig:tds-algorithm,fig:tds-algorithm2}. 
In each figure, the three subfigures show the initial star forest, the forest after the preliminary edge merges, and the output, respectively. 
For the first example, contracting the black matching edges in \Cref{fig:TDS_Ex1_2,fig:TDS_Ex1_3} yields the initial and output forests, respectively, 
of the dominating-set construction in \Cref{fig:ds-algorithm1} for the reduced sequence $d'=(4^4,3,2,1^{13})$. 
For the second example, the total dominating set has odd size. 
The preliminary edge merges make the black vertices induce 
a matching of five edges and a $P_3$, 
and contraction gives the reduced sequence $d'=(8,6^5,3,2^6,1^{29})$.
It then runs \dominatingset and it follows the same three edge additions and two leaf merges as
shown in \Cref{fig:ds-algorithm2}.

\input{figures/TotalDominatingSetFrameworkExample.tex}

\input{figures/TotalDominatingSetFrameworkExample2.tex}

We again note the fact that if the execution of \dominatingset
(\Cref{alg:TotalDominatingSet}) and the execution of \dominatingset in the
algorithm which is given in \Cref{thm:TotalDominatingSet} behave the same,
then the computed forest will be the same.


\subsection{Optimized Paired Dominating Set}

In this section we give a characterization and realization algorithm for \degr with respect to \pairdomset.

\begin{theorem}
\label{thm:PairedDominatingSet}
Let $d$ be a forestic sequence.
Then there is a linear time algorithm that computes a forest realization of $d$ with paired dominating set of size 
$\gammap^{\calF}(d) = 2\ceil{\gammat^{\calF}(d)/2}$.
\end{theorem}
\begin{proof}
We have that $\gammap^{\calF}(d) \geq \gammat^{\calF}(d)$ due to \Cref{obs:pairedlstotal}.
Since the size of a paired dominating set must be even we have that $\gammap^{\calF}(d) \geq 2\ceil{\gammat^{\calF}(d)/2}$.

Let $\gammat = \gammat^{\calF}(d)$.
First suppose that $\gammat = \taut^{\calF}(d)$. 
By \Cref{thm:PairedVertexCover}, there is a forest realization of $d$ with a paired vertex cover of size $2\ceil{\gammat/2}$. 
By \Cref{obs:tvcIstds}, the same set is a paired dominating set.

Next assume that $\gammat = \gammakk_t(d)$.
Recall the proof of \Cref{thm:TotalDominatingSet}.
If $\gammat$ is even, then the total dominating set is also a paired dominating set by construction.
Otherwise, $\gammat$ is odd.
Let $D$ be the total dominating set computed as described in the proof of \Cref{thm:TotalDominatingSet}.
In this case we claim that there is a neighbor of $v_2$ that is not in $D$.
Assume towards contradiction that all neighbors of $v_2$ are in $D$.
Then, $D' = D \setminus \set{v_2}$ is also a total dominating set, in contradiction to $D$ being a minimum size total dominating set.
Hence, there is a vertex $u \in N(v_2) \setminus D$.
Observe that $\set{(v_1,v_3),(v_2,u),(v_4,v_5),\ldots,(v_{|D|-1},v_{|D|})}$ is a matching in $F[D \cup \set{u}]$.
Therefore, $D \cup \set{u}$ is a paired dominating set, and it follows that $\gammap^{\calF}(d) = \gammat + 1$.

Linear time follows from \Cref{thm:PairedVertexCover,thm:TotalDominatingSet}; in the odd case, $u$ is found by scanning $N_F(v_2)$.
\end{proof}


\clearpage
\bibliography{realizations}
\bibliographystyle{abbrv}


\clearpage
\appendix

\section{Previous Result for Vertex Cover}
\label{sec:vc-previous}

Gentner et al.~\cite{GHR18} gave a characterization for optimized \degr with respect to \textsc{Max Independent Set}.
In terms of \vc, they proved \Cref{thm:vc}, which states that 
\[
\tau^{\calF}(d) = \min \set{t : \sum_{i=1}^t d_i \geq \sum_{i=t+1}^n d_i}
~.
\]
To prove an upper bound on $\tau^{\calF}(d)$ they used an induction on the volume $\sum d$.
Their arguments are constructive and in this section we provide an explicit algorithm which 
is implied by this constructive proof. 
We also provide a running time analysis (that did not appear in~{\cite{GHR18}}).

The following algorithm recursively constructs a forest realization of a given forestic sequence $d$, 
such that $\sum_{i=1}^\tau d_i \geq \sum_{i=\tau+1}^n d_i$.
In the recursion base $d = (1^n)$. 
Since $d$ is forestic, $n$ must be even, 
and it must be that $\tau \geq \frac{n}{2}$. Hence, the algorithm returns a perfect matching.

Otherwise, each recursive call of the algorithm has two options.
\begin{itemize}
\item If $d_1 > d_{\tau +1}$, the algorithm makes a Havel-Hakimi step using $d_n = 1$ as a pivot.
More specifically, the algorithm creates a new sequence $d'$ by decreasing $d_q$ by $1$, 
where $q = \max\set{j : d_j = d_1}$, and removing $d_n$. 
Upon returning from the recursion, the edge $(v_q,v_n)$ is added to the current realization.

\item If $d_1=d_{\tau+1}=\ell$, the algorithm removes the vertices $v_1,\ldots,v_\ell,v_{\tau+1}$ 
and replaces them by one vertex $v''$ of degree $\ell(\ell-1)$. Upon returning from the recursion, 
$v''$ is replaced by the star centered at $v_{\tau+1}$ with leaves $v_1,\ldots,v_\ell$, 
and its former incident edges are distributed equally among those leaves.
\end{itemize}

The following notation is used.
Let $E(v_i) = \set{(v_i, v_j) : v_j \in N(v_i)} = E(\set{v_i}, V \setminus \set{v_i})$.

\smallskip

\begin{algorithm}[H]
\caption{\vertexcoverghr}
\SetKwInOut{Input}{input}
\SetKwInOut{Output}{output}
\Input{$(d,\tau)$}
\Output{$F$}
\DontPrintSemicolon
\label{alg:VC-GHR}
\lIf{$d_1 = 1$}{
    \Return $(\set{v_1,\ldots, v_n}, \set{(v_1,v_{n/2+1}), \ldots, (v_{n/2},v_n)})$
}
\If{$\tau = n$ or $d_1 > d_{\tau+1}$}{
    $q \gets \max\set{j : d_j = d_1}$ \\
    $d' \gets (d_1, \ldots, d_q - 1, \ldots, d_{n-2}, d_{n-1})$ \\
    $\tau' \gets \min\set{\tau,n-1}$\\
    $F' \gets \vertexcoverghr(d', \tau')$ \\
    Let $F$ be obtained by adding $v_n$ and the edge $(v_q,v_n)$ to $F'$. \\
    \Return $F$
}
\Else{
    $d'' \gets (\ell(\ell-1), \ell^{\tau-\ell},d_{\tau+2}, \ldots, d_n)$, where $\ell \triangleq d_{\tau+1} = d_1$ \\
    $\tau'' \gets \tau-\ell+1$ \\
    $F'' \gets \vertexcoverghr(d'',\tau'')$ \\
    Label $V(F'')$ by $\set{v'',v_{\ell+1},\ldots,v_\tau,v_{\tau+2},\ldots,v_n}$ \\
    $E(F) \gets E(F'') \setminus E(v'')$ \\
    Partition $E(v'')$ into $\ell$ subsets $E_1,\ldots,E_\ell$, each with $\ell-1$ edges\\
    \lFor{$i=1$ to $\ell$}{
         $E(F) \gets E(F) \cup \set{(v_i,v_j) : (v'',v_j) \in E_i}$
    }
    $E(F) \gets E(F) \cup \set{(v_1,v_{\tau+1}),\ldots,(v_\ell,v_{\tau+1})}$ \\
    \Return $F$
}
\end{algorithm}

\smallskip

The next lemma proves the correctness of \Cref{alg:VC-GHR}.

\begin{lemma}
Running \Cref{alg:VC-GHR} with a forestic sequence $d$ and $\tau$ such that $\sum_{i=1}^\tau d_i \geq \sum_{i=\tau+1}^n d_i$ 
returns a forest realization, $F$, of $d$ in which $\set{v_1,\ldots,v_\tau}$ is a vertex cover of $F$.
\end{lemma}
\begin{proof}
We prove the lemma by induction on the depth of the recursion.
In the base case $d=(1^n)$. Since $d$ is forestic, $n$ must be even. 
Also, $\tau = \sum_{i=1}^{\tau} d_i  \geq \sum_{i=\tau+1}^{n} d_i = n - \tau$
implies that $\tau \geq n/2$. 
The algorithm returns the perfect matching $\set{(v_1,v_{n/2+1}),\ldots,(v_{n/2},v_n)}$, 
which realizes $d$ and is covered by $\set{v_1,\ldots,v_{n/2}} \subseteq\set{v_1,\ldots,v_\tau}$.

As for the inductive step, there are two options.
If $\tau = n$ or $d_1 > d_{\tau+1}$, then the algorithm constructs the sequence $d'$.
$d'$ is forestic, since
\[
\sum_{i=1}^{n'} d'_i = \sum_{i=1}^{n} d_i - 2 \leq(2n-2)-2=2n'-2
~.
\]
Also, notice that, if $\tau' < n'$, then
\[
\sum_{i=1}^{\tau'} d'_i 
=      \sum_{i=1}^{\tau} d_i - 1
\geq \sum_{i=\tau+1}^n d_i - 1
=      \sum_{i=\tau'+1}^{n'} d'_i 
~.
\]
Hence, by the induction hypothesis, $F'$ realizes $d'$ and $\set{v_1,\ldots,v_{\tau'}}$ is a vertex cover in $F'$.
A forest $F$ that realizes $d$ is obtained by adding the edge $(v_q,v_n)$ to $F'$. 
Since $q \leq \tau$, the set $\set{v_1,\ldots,v_\tau}$ remains a vertex cover.

If $d_1 = d_{\tau+1}$, then the algorithm constructs $d''$.
Observe that $d''$ is forestic, since
\[
\sum_{i=1}^{n''} d''_i 
=     \sum_{i=1}^{n} d_i - 2\ell
\leq (2n - 2) - 2\ell
=     2n'' - 2
~.
\]
We show that $\ell \leq \tau$ and that $\tau''\geq 1$. 
Since $d$ is forestic and by the selection of $\tau$, we have that 
\[
n-1 
~\geq~ \frac{1}{2} \sum_{i=1} ^n d_i 
~=~      \frac{1}{2} \left( \sum_{i=1} ^ \tau d_i + \sum_{i=\tau+1}^n d_i \right) 
~\geq~ \sum_{i=\tau+1}^n d_i 
~\geq~      d_{\tau+1} + (n - \tau - 1)
~=~ \ell + n - \tau - 1
~.
\]
It follows that $\ell \leq \tau$ and that $\tau'' = \tau - \ell + 1 \geq 1$.
Also, notice that
\[
\sum_{i=1}^{\tau''} d''_i 
=      \sum_{i=1}^{\tau} d_i - \ell
\geq \sum_{i=\tau+1}^n d_i - \ell
=      \sum_{i=\tau''+1}^{n''} d''_i
~.
\]
By induction, $F''$ realizes $d''$ on $\set{v'',v_{\ell+1},\ldots,v_\tau,v_{\tau+2},\ldots,v_n}$ 
with the cover $\set{v'',v_{\ell+1},\ldots,v_\tau}$. 
Replace $v''$ by the star centered at $v_{\tau+1}$ with leaves $v_1,\ldots,v_\ell$, 
distributing its $\ell(\ell-1)$ incident edges equally among the leaves. 
This preserves acyclicity, since the removal of $v''$ separates all its incident subtrees, 
and reconnecting them cannot create a cycle.
The center and leaves have degree $\ell$, and all other degrees are unchanged. 
The new and redistributed edges are covered by
$v_1,\ldots,v_\ell$, and the remaining edges by $v_{\ell+1},\ldots,v_\tau$. 
Thus, $F$ realizes $d$ with the required vertex cover.
\end{proof}

It remains to bound the running time of the algorithm.

\begin{lemma}
The running time of \Cref{alg:VC-GHR} is $O(\sum d) = O(n)$.
\end{lemma}
\begin{proof}
Let $T(\sum d)$ denote the running time of the algorithm on a sequence of volume $\sum d$.
Hence, we have
\[
T(\sum d) = 
\begin{cases}
O(n)						 & d = (1^n), \\
O(1) + T(\sum d - 2)               & \tau = n \text{ or } d_1 > d_{\tau+1} , \\
O(\ell^2) + T(\sum d - 2\ell)  & d_1 = d_{\tau+1}.
\end{cases}
\]
The base case is reached only once, and its $O(n) = O(\sum d)$ cost 
is bounded by the volume of the original input.

If one uses buckets to hold the sequence $d$, it is possible to implement 
the first recursion step in $O(1)$ time. (The initialization takes $O(n)$ time.)
Observe that if $d_1 = d_{\tau+1}$, we have that $d''_1 = \ell(\ell-1)$ and $d''_2 \leq \ell$.
It follows that at least $\ell(\ell-2)$ recursive calls of type~1 are executed after a call of type~2.
For $\ell=2$, the type-2 cost is already $O(1)$.
The lemma follows.
\end{proof}

\end{document}

%% file: figures/VCPrefixLemmaExample.tex
\begin{figure}[t]
\centering
\begin{tabular}{ccc}
\begin{subfigure}{0.3\textwidth}
\centering
\begin{tikzpicture}[scale=0.92]
  \begin{scope}
    \node[thesisNVertex,label={[font=\scriptsize]above:$v_1$}] (a1) at (0,1.4) {3};
    \node[thesisDVertex,label={[font=\scriptsize]left:$v_2$}]  (a2) at (-1.35,0) {2};
    \node[thesisDVertex,label={[font=\scriptsize]left:$v_3$}]  (a3) at (0,0) {2};
    \node[thesisDVertex,label={[font=\scriptsize]right:$v_4$}] (a4) at (1.35,0) {2};
    \node[thesisNVertex,label={[font=\scriptsize]below:$v_5$}] (a5) at (-1.35,-1.3) {1};
    \node[thesisNVertex,label={[font=\scriptsize]below:$v_6$}] (a6) at (0,-1.3) {1};
    \node[thesisNVertex,label={[font=\scriptsize]below:$v_7$}] (a7) at (1.35,-1.3) {1};
    \draw[ultra thick] (a1)--(a2) (a1)--(a3) (a1)--(a4);
    \draw[ultra thick] (a2)--(a5) (a3)--(a6) (a4)--(a7);
  \end{scope}
\end{tikzpicture}
\caption{$(F,D)$.}
\label{fig:vc-prefix-a}
\end{subfigure}

&

\begin{subfigure}{0.3\textwidth}
\centering
\begin{tikzpicture}[scale=0.92]
  \begin{scope}[shift={(5.3,0)}]
    \node[thesisNVertex,label={[font=\scriptsize]above:$v_1$}] (b1) at (0,1.4) {2};
    \node[thesisDVertex,label={[font=\scriptsize]left:$v_2$}]  (b2) at (-1.35,0) {2};
    \node[thesisDVertex,label={[font=\scriptsize]left:$v_3$}]  (b3) at (0,0) {2};
    \node[thesisDVertex,label={[font=\scriptsize]right:$v_4$}] (b4) at (1.35,0) {3};
    \node[thesisNVertex,label={[font=\scriptsize]below:$v_5$}] (b5) at (-1.35,-1.3) {1};
    \node[thesisNVertex,label={[font=\scriptsize]below:$v_6$}] (b6) at (0,-1.3) {1};
    \node[thesisNVertex,label={[font=\scriptsize]below:$v_7$}] (b7) at (1.35,-1.3) {1};
    \draw[ultra thick] (b1)--(b2) (b1)--(b4) (b4)--(b3);
    \draw[ultra thick] (b2)--(b5) (b3)--(b6) (b4)--(b7);
  \end{scope}
\end{tikzpicture}
\caption{$E  \setminus \set{(v_1,v_3)} \cup \set{(v_3,v_4)}$.}
\end{subfigure}

&

\begin{subfigure}{0.3\textwidth}
\centering
\begin{tikzpicture}[scale=0.92]
  \begin{scope}[shift={(10.6,0)}]
    \node[thesisNVertex,label={[font=\scriptsize]above:$v_4$}] (c4) at (0,1.4) {2};
    \node[thesisDVertex,label={[font=\scriptsize]left:$v_2$}]  (c2) at (-1.35,0) {2};
    \node[thesisDVertex,label={[font=\scriptsize]left:$v_3$}]  (c3) at (0,0) {2};
    \node[thesisDVertex,label={[font=\scriptsize]right:$v_1$}] (c1) at (1.35,0) {3};
    \node[thesisNVertex,label={[font=\scriptsize]below:$v_5$}] (c5) at (-1.35,-1.3) {1};
    \node[thesisNVertex,label={[font=\scriptsize]below:$v_6$}] (c6) at (0,-1.3) {1};
    \node[thesisNVertex,label={[font=\scriptsize]below:$v_7$}] (c7) at (1.35,-1.3) {1};
    \draw[ultra thick] (c4)--(c2) (c4)--(c1) (c1)--(c3);
    \draw[ultra thick] (c2)--(c5) (c3)--(c6) (c1)--(c7);
  \end{scope}
\end{tikzpicture}
\caption{$(F',D')$.}
\end{subfigure}
\end{tabular}
\vspace{-5pt}
\caption{An exchange in the VC Prefix Lemma for $d=(3,2^3,1^3)$. Black vertices form the current cover. 
Here $v_j = v_1$, $v_\ell = v_4$, and $U = \set{v_3}$.
Replacing $(v_1,v_3)$ by $(v_3,v_4)$ preserves the forest; swapping the labels $v_1$ and $v_4$ then produces $F'$ with the prefix cover $D'=\set{v_1,v_2,v_3}$.}
\label{fig:vc-prefix-exchange}
\end{figure}
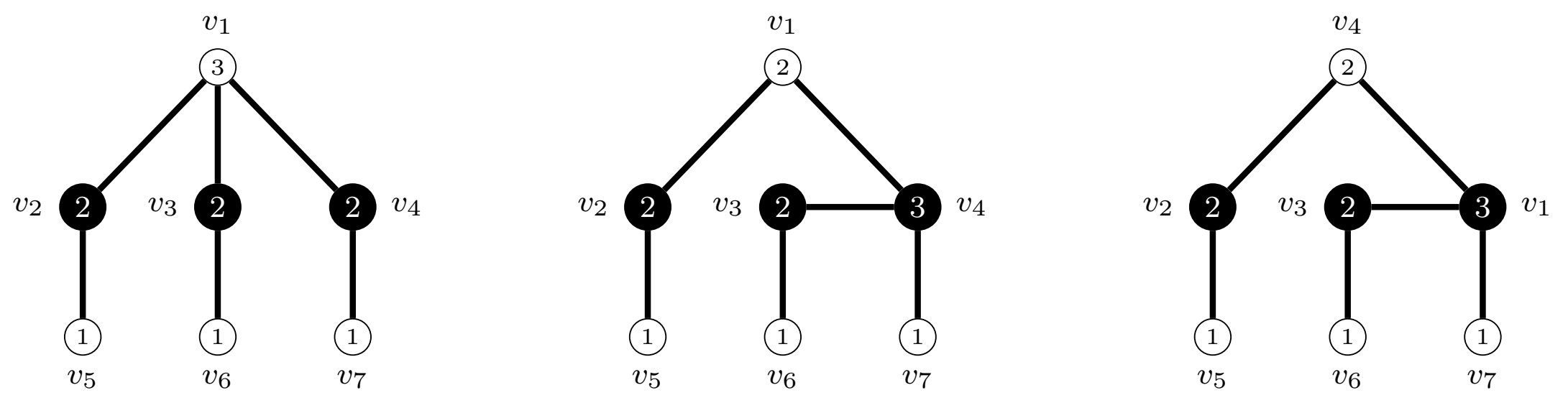

%% file: figures/VertexCoverExample.tex
%
%
\newcommand{\starGadget}[9]{%
    \begin{scope}[shift={({#2},{#3})}, rotate=#4]

        \node (#1-center) [thesisDVertex] at (0,0) {#5};

        \node (#1-up)    [thesisNVertex] at (0,  3) {#6};
        \node (#1-down)  [thesisNVertex] at (0, -3) {#7};
        \node (#1-left)  [thesisNVertex] at (-3, 0) {#8};
        \node (#1-right) [thesisNVertex] at ( 3, 0) {#9};

        \draw[ultra thick] (#1-center) -- (#1-up);
        \draw[ultra thick] (#1-center) -- (#1-down);
        \draw[ultra thick] (#1-center) -- (#1-left);
        \draw[ultra thick] (#1-center) -- (#1-right);

    \end{scope}
}

%
\newcommand{\starGadgetMinusEdge}[8]{%
    \begin{scope}[shift={({#2},{#3})}, rotate=#4]

        \node (#1-center) [thesisDVertex] at (0,0) {#5};

        \node (#1-up)    [thesisNVertex] at (0,  3) {#6};
        \node (#1-down)  [thesisNVertex] at (0, -3) {#7};
        \node (#1-left)  [thesisNVertex] at (-3, 0) {#8};

        \draw[ultra thick] (#1-center) -- (#1-up);
        \draw[ultra thick] (#1-center) -- (#1-down);
        \draw[ultra thick] (#1-center) -- (#1-left);

    \end{scope}
}

\newcommand{\starGadgetMinusTwoEdges}[7]{%
    \begin{scope}[shift={({#2},{#3})}, rotate=#4]

        \node (#1-center) [thesisDVertex] at (0,0) {#5};

        \node (#1-up)    [thesisNVertex] at (0,  3) {#6};
        \node (#1-left)  [thesisNVertex] at (-3, 0) {#7};

        \draw[ultra thick] (#1-center) -- (#1-up);
        \draw[ultra thick] (#1-center) -- (#1-left);

    \end{scope}
}

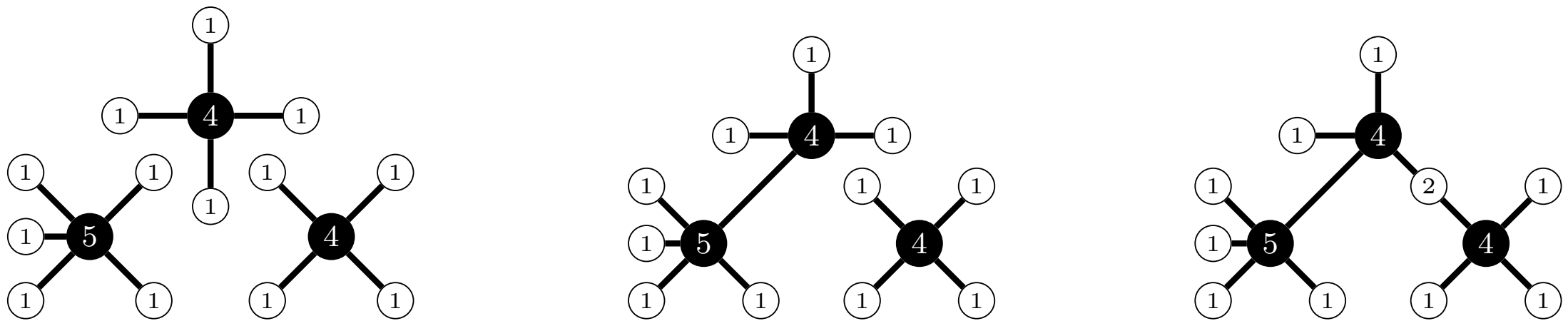
\begin{figure}[t]
\centering

\begin{subfigure}{0.34\textwidth}
\centering
\begin{tikzpicture}[scale=0.28]
	\starGadget{A1}{0}{0}{0}{4}{1}{1}{1}{1}      
	\starGadget{A2}{-4}{-4}{45}{5}{1}{1}{1}{1}     
	\starGadget{A3}{4}{-4}{45}{4}{1}{1}{1}{1}   
	
	\node (A2-Fifth) [thesisNVertex] at (-4-2.121,-4) {1};
	\draw[ultra thick] (A2-center) -- (A2-Fifth);
\end{tikzpicture}
\caption{Initial star forest.}
\end{subfigure}%
\hfill
\begin{subfigure}{0.35\textwidth}
\centering
\begin{tikzpicture}[scale=0.25] 
	\starGadgetMinusEdge{A1}{0}{0}{-90}{4}{1}{1}{1}      
	\starGadgetMinusEdge{A2}{-4}{-4}{45}{5}{1}{1}{1}     
	\starGadget{A3}{4}{-4}{45}{4}{1}{1}{1}{1}   
	
	\draw[ultra thick] (A1-center) -- (A2-center);
	\node (A2-Fifth) [thesisNVertex] at (-4-2.121,-4) {1};
	\draw[ultra thick] (A2-center) -- (A2-Fifth);
\end{tikzpicture}
\caption{After first loop.}
\end{subfigure}%
\hfill
\begin{subfigure}{0.30\textwidth}
\centering
\begin{tikzpicture}[scale=0.25] 
	\starGadgetMinusTwoEdges{A1}{0}{0}{0}{4}{1}{1}      
	\starGadgetMinusEdge{A2}{-4}{-4}{45}{5}{1}{1}{1}     
	\starGadget{A3}{4}{-4}{45}{4}{2}{1}{1}{1}   
	
	\draw[ultra thick] (A1-center) -- (A2-center);
	\draw[ultra thick] (A1-center) -- (A3-up);
	\node (A2-Fifth) [thesisNVertex] at (-4-2.121,-4) {1};
	\draw[ultra thick] (A2-center) -- (A2-Fifth);
\end{tikzpicture}
\caption{Final forest.}
\end{subfigure}

\caption{Construction of $d = (5,4^2, 2, 1^9)$. In this case, $\sum d  = 24 = 2n - 2$ and $\tau = 3$.
Also, $d^0 = (5,4^2,1^{13})$ and $d^1 = (5,4^2,1^{11})$.}
\label{fig:running-vc}
\end{figure}

%% file: figures/VertexCoverExample2.tex
\begin{figure}[t]
\centering

\begin{subfigure}{0.39\textwidth}
\centering
\begin{tikzpicture}[scale=0.27]
    \node (U) [thesisDVertex] at (0,0) {5};
    \node (V) [thesisDVertex] at (10,0) {4};

    \node (U1) [thesisNVertex] at (-3,0) {1};
    \node (U2) [thesisNVertex] at (-2.2,2.4) {1};
    \node (U3) [thesisNVertex] at (0,3) {1};
    \node (U4) [thesisNVertex] at (0,-3) {1};
    \node (U5) [thesisNVertex] at (3,0) {1};

    \node (V1) [thesisNVertex] at (7,0) {1};
    \node (V2) [thesisNVertex] at (8,2.5) {1};
    \node (V3) [thesisNVertex] at (10,3) {1};
    \node (V4) [thesisNVertex] at (13,0) {1};

    \foreach \x in {U1,U2,U3,U4,U5}
        \draw[ultra thick] (U)--(\x);
    \foreach \x in {V1,V2,V3,V4}
        \draw[ultra thick] (V)--(\x);
\end{tikzpicture}
\caption{Initial star forest.}
\end{subfigure}%
\hspace{30pt}
\begin{subfigure}{0.39\textwidth}
\centering
\begin{tikzpicture}[scale=0.27]
    \node (U) [thesisDVertex] at (0,0) {5};
    \node (V) [thesisDVertex] at (10,0) {4};

    \node (U1) [thesisNVertex] at (-3,0) {1};
    \node (U2) [thesisNVertex] at (-2.2,2.4) {1};
    \node (U3) [thesisNVertex] at (0,3) {1};
    \node (U4) [thesisNVertex] at (0,-3) {1};

    \node (V1) [thesisNVertex] at (8,2.5) {1};
    \node (V2) [thesisNVertex] at (10,3) {1};
    \node (V3) [thesisNVertex] at (13,0) {1};

    \draw[ultra thick] (U)--(V);
    \foreach \x in {U1,U2,U3,U4}
        \draw[ultra thick] (U)--(\x);
    \foreach \x in {V1,V2,V3}
        \draw[ultra thick] (V)--(\x);
\end{tikzpicture}
\caption{After first loop and final forest.}
\end{subfigure}%

\caption{Construction of $d = (5,4,1^7)$. In this case, $\sum d  = 16 = 2n - 2$, $\tau = 2$, and $d^0 = (5,4,1^9)$.}
\label{fig:running-vc2}
\end{figure}
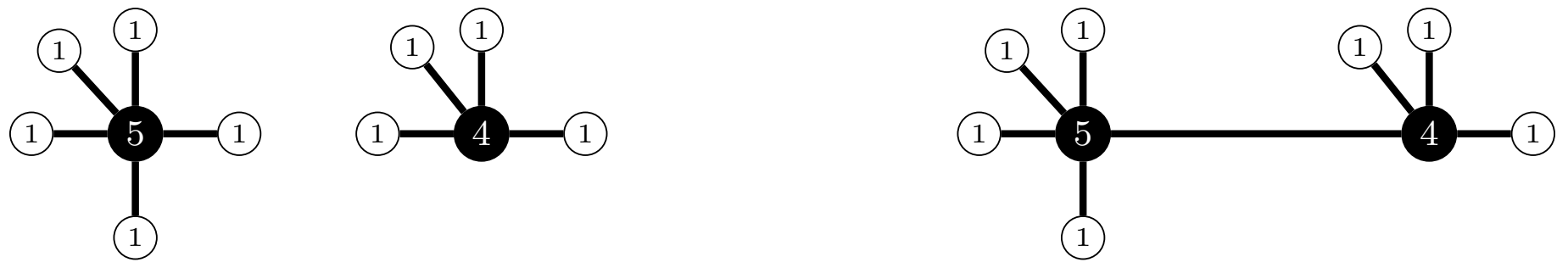

%% file: figures/TotalVertexCoverExample.tex
%
%
\newcommand{\DualStarGadget}[4]{%
    \begin{scope}[shift={({#2},{#3})}, rotate=#4]

        \node (#1-center) [thesisDVertex] at (0,0) {2};

        \node (#1-first)  [thesisNVertex] at (1.5, -3) {1};
        \node (#1-second)  [thesisNVertex] at (-1.5, -3) {1};

        \draw[ultra thick] (#1-center) -- (#1-first);
        \draw[ultra thick] (#1-center) -- (#1-second);
        
    \end{scope}
}
\begin{figure}[t]
\centering

\begin{subfigure}{0.30\textwidth}
\centering
\begin{tikzpicture}[scale=0.25] 
	\starGadgetMinusTwoEdges{A1}{0}{0}{0}{4}{1}{1}      
	\starGadgetMinusEdge{A2}{-4}{-4}{45}{5}{1}{1}{1}     
	\starGadget{A3}{4}{-4}{45}{4}{2}{1}{1}{1}   
	
	\draw[ultra thick] (A1-center) -- (A2-center);
	\draw[ultra thick] (A1-center) -- (A3-up);
	\node (A2-Fifth) [thesisNVertex] at (-4-2.121,-4) {1};
	\draw[ultra thick] (A2-center) -- (A2-Fifth);
\end{tikzpicture}
\caption{Forest for $d' = (5,4^2,2,1^{9})$.}
\label{fig:tvc-example1-left}
\end{subfigure}
\hspace{2cm}
\begin{subfigure}{0.5\textwidth}
\centering
\begin{tikzpicture}[scale=0.25]

\node (A2) [thesisDVertex] at (3,0) {4};
\node (A3) [thesisDVertex] at (6,0) {3};
\node (A4) [thesisDVertex] at (9,3) {3};
\node (A5) [thesisDVertex] at (12,3) {3};
\node (A6) [thesisDVertex] at (15,0) {3};
\node (A7) [thesisDVertex] at (18,0) {3};

\node (B1) [thesisNVertex] at (0,0) {1};
\node (B2) [thesisNVertex] at (0,3) {1};
\node (B3) [thesisNVertex] at (0,-3) {1};
\node (B4) [thesisNVertex] at (9,-3) {1};
\node (B5) [thesisNVertex] at (6, 3) {1};
\node (B6) [thesisNVertex] at (12,6) {1};
\node (B7) [thesisNVertex] at (13.5,1.5) {2};
\node (B8) [thesisNVertex] at (12,-3) {1};
\node (B9) [thesisNVertex] at (21,3) {1};
\node (B10) [thesisNVertex] at (21,-3) {1};

\draw[ultra thick] (A2) -- (B2);
\draw[ultra thick] (A2) -- (B1);
\draw[ultra thick] (A2) -- (B3);
\draw[ultra thick] (A2) -- (A3);
\draw[ultra thick] (A3) -- (A4);
\draw[ultra thick] (A3) -- (B4);
\draw[ultra thick] (A4) -- (B5);
\draw[ultra thick] (A4) -- (A5);
\draw[ultra thick] (A5) -- (B6);
\draw[ultra thick] (A5) -- (B7);
\draw[ultra thick] (B7) -- (A6);
\draw[ultra thick] (A6) -- (B8);
\draw[ultra thick] (A6) -- (A7);
\draw[ultra thick] (A7) -- (B9);
\draw[ultra thick] (A7) -- (B10);
\end{tikzpicture}

\caption{Forest for $d = (4,3^5,2,1^9)$.}
\end{subfigure}

\caption{Construction of $d = (4,3^5,2,1^9)$ with $\taut = 6$, $\delta = 0$, $d' = (5,4^2,2,1^9)$ and $\tau' = 3$.}
\label{fig:tvc-algorithm}
\end{figure}
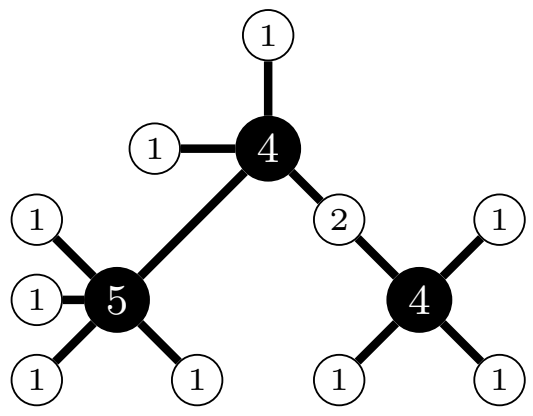
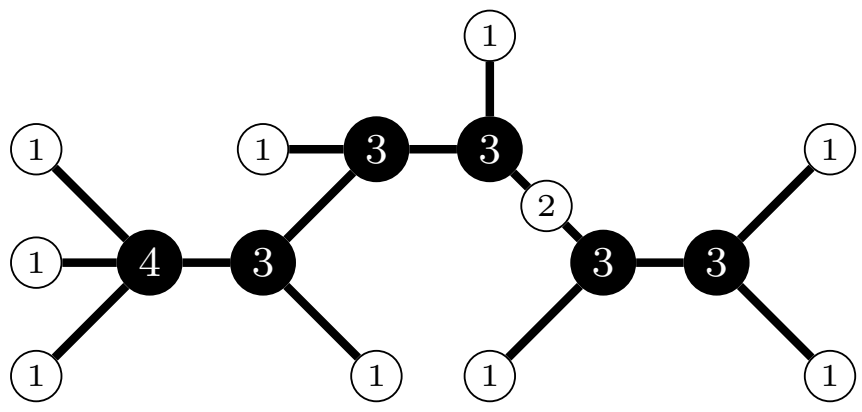

%% file: figures/TotalVertexCoverExample2.tex
\begin{figure}[t]
\centering

\begin{subfigure}{0.39\textwidth}
\centering
\begin{tikzpicture}[scale=0.27]
    \node (U) [thesisDVertex] at (0,0) {5};
    \node (V) [thesisDVertex] at (10,0) {4};

    \node (U1) [thesisNVertex] at (-3,0) {1};
    \node (U2) [thesisNVertex] at (-2.2,2.4) {1};
    \node (U3) [thesisNVertex] at (0,3) {1};
    \node (U4) [thesisNVertex] at (0,-3) {1};
    \node (V1) [thesisNVertex] at (8,2.5) {1};
    \node (V2) [thesisNVertex] at (10,3) {1};
    \node (V3) [thesisNVertex] at (13,0) {1};

    \draw[ultra thick] (U)--(V);
    \foreach \x in {U1,U2,U3,U4}
        \draw[ultra thick] (U)--(\x);
    \foreach \x in {V1,V2,V3}
        \draw[ultra thick] (V)--(\x);
\end{tikzpicture}
\caption{Forest for $d'=(5,4,1^7)$.}
\label{fig:tvc-example2-left}
\end{subfigure}
\hfill
\begin{subfigure}{0.56\textwidth}
\centering
\begin{tikzpicture}[scale=0.235]
    \node (V2) [thesisDVertex] at (0,0) {3};
    \node (V1) [thesisDVertex] at (4,0) {3};
    \node (V3) [thesisDVertex] at (8,0) {3};
    \node (V4) [thesisDVertex] at (12,0) {3};
    \node (V5) [thesisDVertex] at (16,0) {3};

    \node (L21) [thesisNVertex] at (-2.4,2.2) {1};
    \node (L22) [thesisNVertex] at (-2.4,-2.2) {1};
    \node (L1)  [thesisNVertex] at (4,3) {1};
    \node (L3)  [thesisNVertex] at (8,-3) {1};
    \node (L4)  [thesisNVertex] at (12,3) {1};
    \node (L51) [thesisNVertex] at (18.4,2.2) {1};
    \node (L52) [thesisNVertex] at (18.4,-2.2) {1};

    \draw[ultra thick] (V2)--(V1)--(V3)--(V4)--(V5);
    \draw[ultra thick] (V2)--(L21) (V2)--(L22);
    \draw[ultra thick] (V1)--(L1);
    \draw[ultra thick] (V3)--(L3);
    \draw[ultra thick] (V4)--(L4);
    \draw[ultra thick] (V5)--(L51) (V5)--(L52);
\end{tikzpicture}
\caption{Forest for $d=(3^5,1^7)$.}
\end{subfigure}

\caption{Construction of $d=(3^5,1^7)$ with $\taut=5$, $\delta=0$,
$d'=(5,4,1^7)$ and $\tau'=2$.}
\label{fig:tvc-algorithm2}
\end{figure}
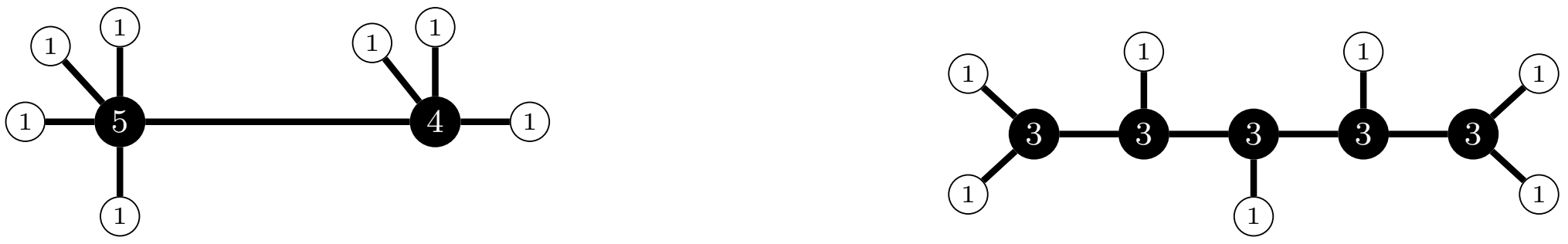

%% file: figures/TotalVertexCoverExampleNotTrueAlgorithm.tex
\begin{figure}[t]
\centering

\begin{subfigure}{0.4\textwidth}
\centering
\begin{tikzpicture}[scale=0.25]
	
	\node (A1) [thesisDVertex] at (0,0) {4};
	\node (A2) [thesisDVertex] at (8,0) {3};
	\node (A3) [thesisDVertex] at (16,0) {3};
	\node (A4) [thesisDVertex] at (0,-8) {3};
	\node (A5) [thesisDVertex] at (8,-8) {3};
	\node (A6) [thesisDVertex] at (16,-8) {3};
	
	\node (B1) [thesisNVertex] at (-3, -3) {1};
	\node (B2) [thesisNVertex] at (-1, -3) {1};
	\node (B3) [thesisNVertex] at (1, -3) {1};
	\node (B4) [thesisNVertex] at (3, -3) {1};
	\node (B5) [thesisNVertex] at (5, -3) {1};
	\node (B6) [thesisNVertex] at (8, -3) {1};
	\node (B7) [thesisNVertex] at (11, -3) {1};
	\node (B8) [thesisNVertex] at (13, -3) {1};
	\node (B9) [thesisNVertex] at (16, -3) {1};
	\node (B10) [thesisNVertex] at (19, -3) {1};
	\node (B11) [thesisNVertex] at (-3, -5) {1};
	\node (B12) [thesisNVertex] at (0, -5) {1};
	\node (B13) [thesisNVertex] at (3, -5) {1};
	\node (B14) [thesisNVertex] at (5, -5) {1};
	\node (B15) [thesisNVertex] at (8, -5) {1};
	\node (B16) [thesisNVertex] at (11, -5) {1};
	\node (B17) [thesisNVertex] at (13, -5) {1};
	\node (B18) [thesisNVertex] at (16, -5) {1};
	\node (B19) [thesisNVertex] at (19, -5) {1};
        
        \draw[ultra thick] (A1) -- (B1);
        \draw[ultra thick] (A1) -- (B2);
        \draw[ultra thick] (A1) -- (B3);
        \draw[ultra thick] (A1) -- (B4);
        \draw[ultra thick] (A2) -- (B5);
        \draw[ultra thick] (A2) -- (B6);
        \draw[ultra thick] (A2) -- (B7);
        \draw[ultra thick] (A3) -- (B8);
        \draw[ultra thick] (A3) -- (B9);
        \draw[ultra thick] (A3) -- (B10);
        \draw[ultra thick] (A4) -- (B11);
        \draw[ultra thick] (A4) -- (B12);
        \draw[ultra thick] (A4) -- (B13);
        \draw[ultra thick] (A5) -- (B14);
        \draw[ultra thick] (A5) -- (B15);
        \draw[ultra thick] (A5) -- (B16);
        \draw[ultra thick] (A6) -- (B17);
        \draw[ultra thick] (A6) -- (B18);
        \draw[ultra thick] (A6) -- (B19);
        
\end{tikzpicture}
\caption{Construction of $d = (4,3^5,1^{19})$.}
\end{subfigure}%
\hspace{30pt}
\begin{subfigure}{0.4\textwidth}
\centering
\begin{tikzpicture}[scale=0.25]
	
	\node (A1) [thesisDVertex] at (0,0) {4};
	\node (A2) [thesisDVertex] at (8,0) {3};
	\node (A3) [thesisDVertex] at (16,0) {3};
	\node (A4) [thesisDVertex] at (0,-8) {3};
	\node (A5) [thesisDVertex] at (8,-8) {3};
	\node (A6) [thesisDVertex] at (16,-8) {3};
	
	\node (B1) [thesisNVertex] at (-3, -3) {1};
	\node (B2) [thesisNVertex] at (-1, -3) {1};

	\node (B4) [thesisNVertex] at (3, -3) {1};
	\node (B5) [thesisNVertex] at (5, -3) {1};

	\node (B7) [thesisNVertex] at (11, -3) {1};
	\node (B8) [thesisNVertex] at (13, -3) {1};

	\node (B10) [thesisNVertex] at (19, -3) {1};
	\node (B11) [thesisNVertex] at (-3, -5) {1};

	\node (B13) [thesisNVertex] at (3, -5) {1};
	\node (B14) [thesisNVertex] at (5, -5) {1};

	\node (B16) [thesisNVertex] at (11, -5) {1};
	\node (B17) [thesisNVertex] at (13, -5) {1};

	\node (B19) [thesisNVertex] at (19, -5) {1};
        
        \draw[ultra thick] (A1) -- (B1);
        \draw[ultra thick] (A1) -- (B2);

        \draw[ultra thick] (A1) -- (B4);
        \draw[ultra thick] (A2) -- (B5);

        \draw[ultra thick] (A2) -- (B7);
        \draw[ultra thick] (A3) -- (B8);

        \draw[ultra thick] (A3) -- (B10);
        \draw[ultra thick] (A4) -- (B11);

        \draw[ultra thick] (A4) -- (B13);
        \draw[ultra thick] (A5) -- (B14);

        \draw[ultra thick] (A5) -- (B16);
        \draw[ultra thick] (A6) -- (B17);

        \draw[ultra thick] (A6) -- (B19);

        \draw[ultra thick] (A1) -- (A4);
        \draw[ultra thick] (A2) -- (A5);
        \draw[ultra thick] (A3) -- (A6);
        
\end{tikzpicture}
\caption{Construction of $d = (4,3^5,1^{13})$.}
\end{subfigure}

\begin{subfigure}{0.6\textwidth}
\centering
\begin{tikzpicture}[scale=0.25]
	
	\node (A1) [thesisDVertex] at (0,0) {4};
	\node (A2) [thesisDVertex] at (8,0) {3};
	\node (A3) [thesisDVertex] at (16,0) {3};
	\node (A4) [thesisDVertex] at (0,-8) {3};
	\node (A5) [thesisDVertex] at (8,-8) {3};
	\node (A6) [thesisDVertex] at (16,-8) {3};
	
	\node (B1) [thesisNVertex] at (-3, -3) {1};
	\node (B2) [thesisNVertex] at (-1, -3) {1};

	\node (B4) [thesisNVertex] at (3, -3) {1};
	\node (B5) [thesisNVertex] at (5, -3) {1};

	\node (B78) [thesisNVertex] at (12, -3) {2};

	\node (B10) [thesisNVertex] at (19, -3) {1};
	\node (B11) [thesisNVertex] at (-3, -5) {1};

	\node (B16) [thesisNVertex] at (11, -5) {1};
	\node (B17) [thesisNVertex] at (13, -5) {1};

	\node (B19) [thesisNVertex] at (19, -5) {1};
        
        \draw[ultra thick] (A1) -- (B1);
        \draw[ultra thick] (A1) -- (B2);

        \draw[ultra thick] (A1) -- (B4);
        \draw[ultra thick] (A2) -- (B5);

        \draw[ultra thick] (A2) -- (B78);
        \draw[ultra thick] (A3) -- (B78);

        \draw[ultra thick] (A3) -- (B10);
        \draw[ultra thick] (A4) -- (B11);

        \draw[ultra thick] (A5) -- (B16);
        \draw[ultra thick] (A6) -- (B17);

        \draw[ultra thick] (A6) -- (B19);

        \draw[ultra thick] (A1) -- (A4);
        \draw[ultra thick] (A2) -- (A5);
        \draw[ultra thick] (A3) -- (A6);
        \draw[ultra thick] (A4) -- (A5);
        
\end{tikzpicture}
\caption{Construction of $d = (4,3^5,2,1^9)$.}
\end{subfigure}%
\caption{Construction of $d =(4,3^5,2,1^9)$ with $\taut = 6$.}
\label{fig:tvc-algorithm-alternate}
\end{figure}
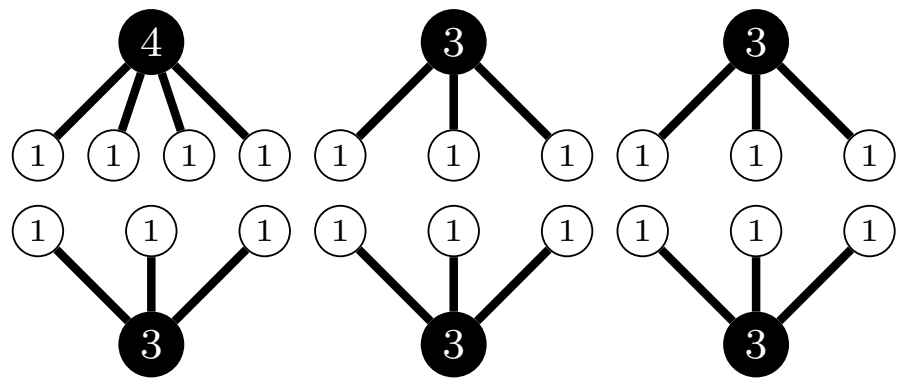
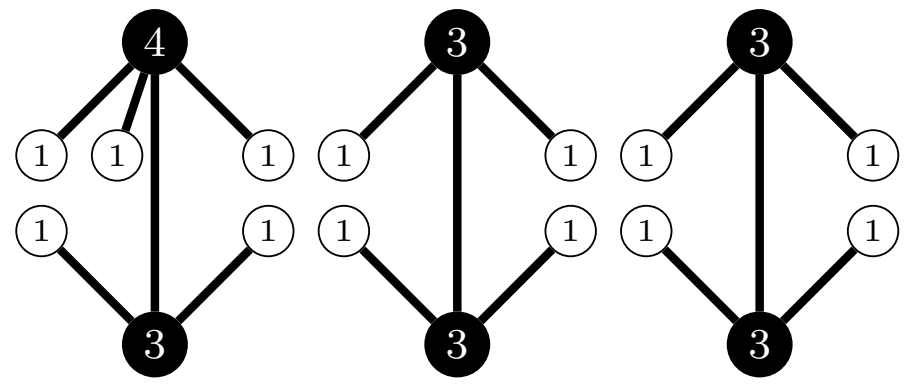
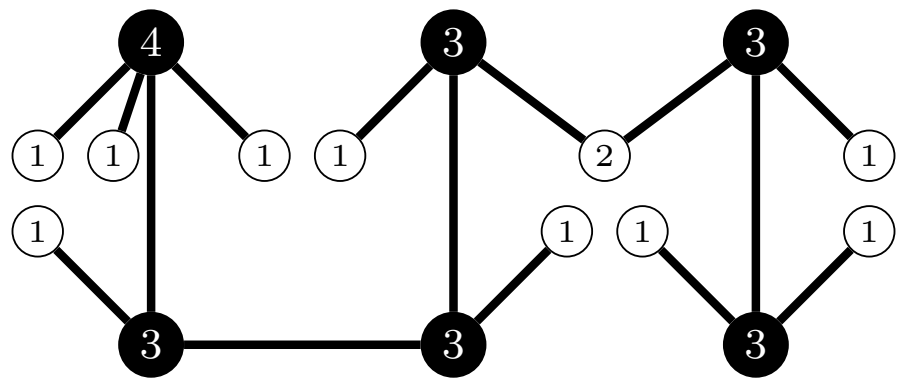

%% file: figures/TotalVertexCoverExampleNotTrueAlgorithm2.tex
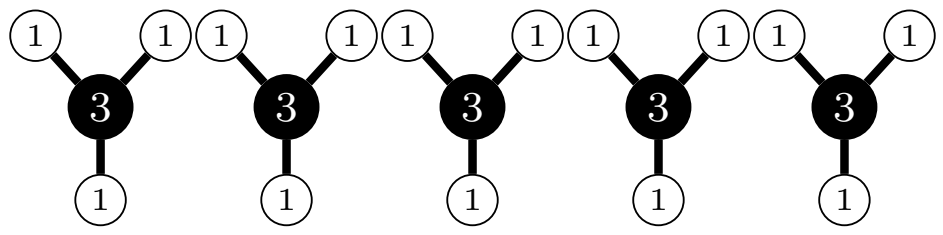
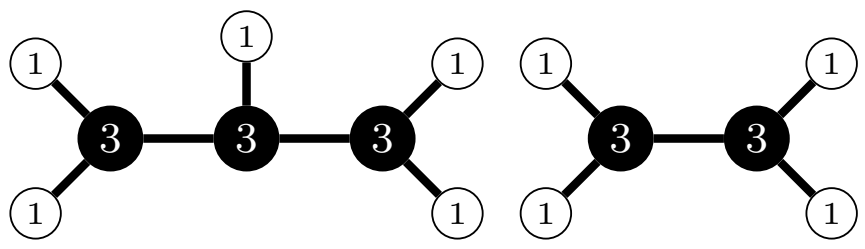
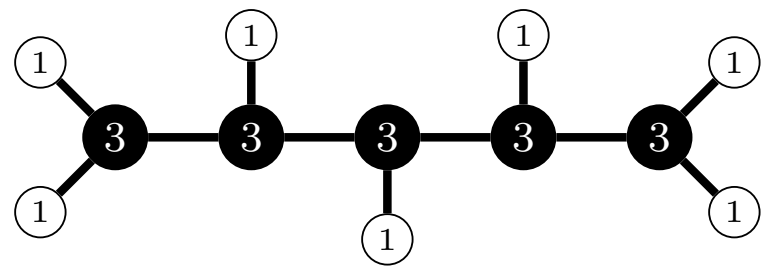
\begin{figure}[t]
\centering

\begin{subfigure}{\textwidth}
\centering
\begin{tikzpicture}[scale=0.205]
    \foreach \i/\x in {1/0,2/6,3/12,4/18,5/24} {
        \node (V\i) [thesisDVertex] at (\x,0) {3};
        \node (L\i a) [thesisNVertex] at (\x-2.1,2.3) {1};
        \node (L\i b) [thesisNVertex] at (\x+2.1,2.3) {1};
        \node (L\i c) [thesisNVertex] at (\x,-3) {1};
        \draw[ultra thick] (V\i)--(L\i a) (V\i)--(L\i b) (V\i)--(L\i c);
    }
\end{tikzpicture}
\caption{Construction of $d=(3^5,1^{15})$.}
\end{subfigure}

\medskip

\begin{subfigure}{0.46\textwidth}
\centering
\begin{tikzpicture}[scale=0.225]
    \node (V2) [thesisDVertex] at (0,0) {3};
    \node (V1) [thesisDVertex] at (4,0) {3};
    \node (V3) [thesisDVertex] at (8,0) {3};
    \node (V4) [thesisDVertex] at (15,0) {3};
    \node (V5) [thesisDVertex] at (19,0) {3};

    \node (L21) [thesisNVertex] at (-2.2,2.2) {1};
    \node (L22) [thesisNVertex] at (-2.2,-2.2) {1};
    \node (L1)  [thesisNVertex] at (4,3) {1};
    \node (L31) [thesisNVertex] at (10.2,2.2) {1};
    \node (L32) [thesisNVertex] at (10.2,-2.2) {1};
    \node (L41) [thesisNVertex] at (12.8,2.2) {1};
    \node (L42) [thesisNVertex] at (12.8,-2.2) {1};
    \node (L51) [thesisNVertex] at (21.2,2.2) {1};
    \node (L52) [thesisNVertex] at (21.2,-2.2) {1};

    \draw[ultra thick] (V2)--(V1)--(V3) (V4)--(V5);
    \draw[ultra thick] (V2)--(L21) (V2)--(L22);
    \draw[ultra thick] (V1)--(L1);
    \draw[ultra thick] (V3)--(L31) (V3)--(L32);
    \draw[ultra thick] (V4)--(L41) (V4)--(L42);
    \draw[ultra thick] (V5)--(L51) (V5)--(L52);
\end{tikzpicture}
\caption{Construction of $d=(3^5,1^9)$.}
\end{subfigure}
\hfill
\begin{subfigure}{0.46\textwidth}
\centering
\begin{tikzpicture}[scale=0.225]
    \node (V2) [thesisDVertex] at (0,0) {3};
    \node (V1) [thesisDVertex] at (4,0) {3};
    \node (V3) [thesisDVertex] at (8,0) {3};
    \node (V4) [thesisDVertex] at (12,0) {3};
    \node (V5) [thesisDVertex] at (16,0) {3};

    \node (L21) [thesisNVertex] at (-2.2,2.2) {1};
    \node (L22) [thesisNVertex] at (-2.2,-2.2) {1};
    \node (L1)  [thesisNVertex] at (4,3) {1};
    \node (L3)  [thesisNVertex] at (8,-3) {1};
    \node (L4)  [thesisNVertex] at (12,3) {1};
    \node (L51) [thesisNVertex] at (18.2,2.2) {1};
    \node (L52) [thesisNVertex] at (18.2,-2.2) {1};

    \draw[ultra thick] (V2)--(V1)--(V3)--(V4)--(V5);
    \draw[ultra thick] (V2)--(L21) (V2)--(L22);
    \draw[ultra thick] (V1)--(L1);
    \draw[ultra thick] (V3)--(L3);
    \draw[ultra thick] (V4)--(L4);
    \draw[ultra thick] (V5)--(L51) (V5)--(L52);
\end{tikzpicture}
\caption{Construction of $d=(3^5,1^7)$.}
\end{subfigure}

\caption{Construction of $d=(3^5,1^7)$ with $\taut=5$.}
\label{fig:tvc-algorithm-alternate2}
\end{figure}

%% file: figures/DomSetFigure.tex
\begin{figure}[t]
\centering
\begin{tikzpicture}[
    every node/.style={font=\small},
    dotD/.style={thesisDVertex,minimum size=6mm,font=\scriptsize},
    dotN/.style={thesisNVertex,minimum size=6mm,font=\scriptsize},
    >=stealth,
    scale=0.95
  ]
\begin{scope}
  \node[dotD] (d1) {$d_1$};
  \node[dotD, below=4mm of d1] (d2) {$d_2$};
  \node at ($(d2)!0.5!(d2)+(0,-6mm)$) {$\vdots$};
  \node[dotD, below=5mm of d2] (dt) {$d_t$};

  \node[draw, ellipse, fit=(d1) (d2) (dt), label=above:$D$,
        inner sep=2mm] (D) {};

\end{scope}

  \begin{scope}[xshift=-50mm,yshift=-5mm]
    \node[dotN] (e1) {1};
    \node[dotN, below=6mm of e1] (e2) {1};
    \node at ($(e1)!0.5!(e2)$) {$\vdots$};
    \node[draw, ellipse, fit=(e1) (e2), label=above:$\bar D_{=1}$,
          inner sep=2mm] (D1) {};
  \end{scope}

  \begin{scope}[xshift=50mm,yshift=-5mm]
    \node[dotN] (f1) {$\geq 2$};
    \node[dotN, below=6mm of f1] (f2) {$\geq 2$};
    \node at ($(f1)!0.5!(f2)$) {$\vdots$};
    \node[draw, ellipse, fit=(f1) (f2), label=above:$\bar D_{\ge2}$,
          inner sep=2mm] (D2) {};
  \end{scope}

    
    \draw[decorate,decoration={brace,amplitude=5pt},thick]
    (e1) -- ([yshift=7mm]D.west) 
    node[midway, above=6pt] {\(=n - \nBTwo \)};
    \draw[-] ([yshift=7mm]D.west) --(e1);
    \draw[-] ([yshift=-6mm]D.west) --(e2);
    
    \node at ($(e1)!0.5!(D.west)$) [yshift=-2mm] {\vdots};

    \draw[decorate,decoration={brace,amplitude=5pt},thick]
    ([yshift=7mm]D.east) -- ([yshift=7mm]D2.west)
    node[midway, above=6pt] {\(\leq \sum_{i=1}^t d_i - (n - \nBTwo)\)};
    \draw[-] ([yshift=7mm]D.east) --([yshift=7mm]D2.west);
    \draw[-] ([yshift=-6mm]D.east) -- ([yshift=-7mm]D2.west);
    
    \node at ($(f1)!0.5!(D.west)$) [yshift=-2mm] {\vdots};

    \path(f1)edge[bend left](f2);
\end{tikzpicture}
\caption{The partition of $V$ into $D, \bar{D}_{=1} \text{ and } \bar{D}_{\geq2}$}
\label{fig:DomSetPartition}
\end{figure}
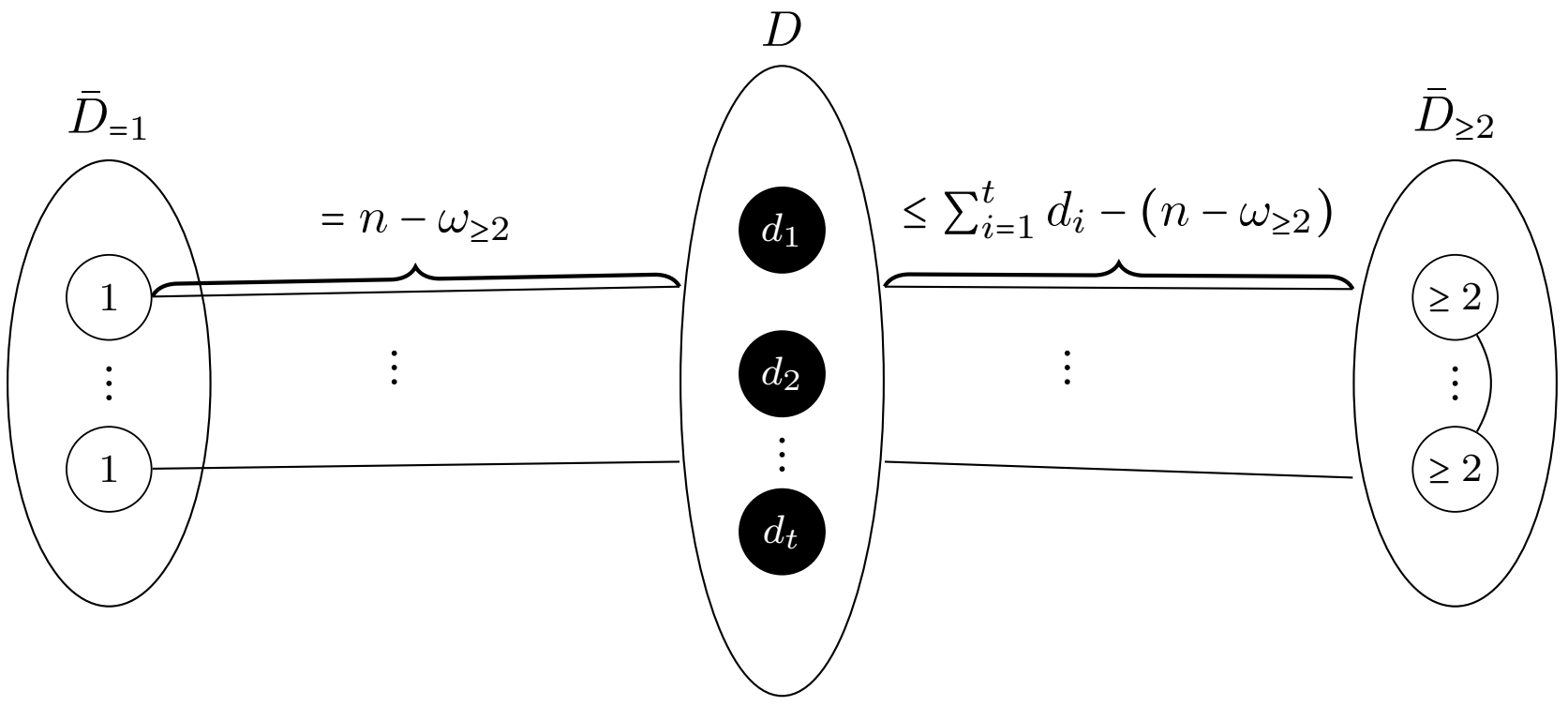

%% file: figures/DominatingSetExample.tex
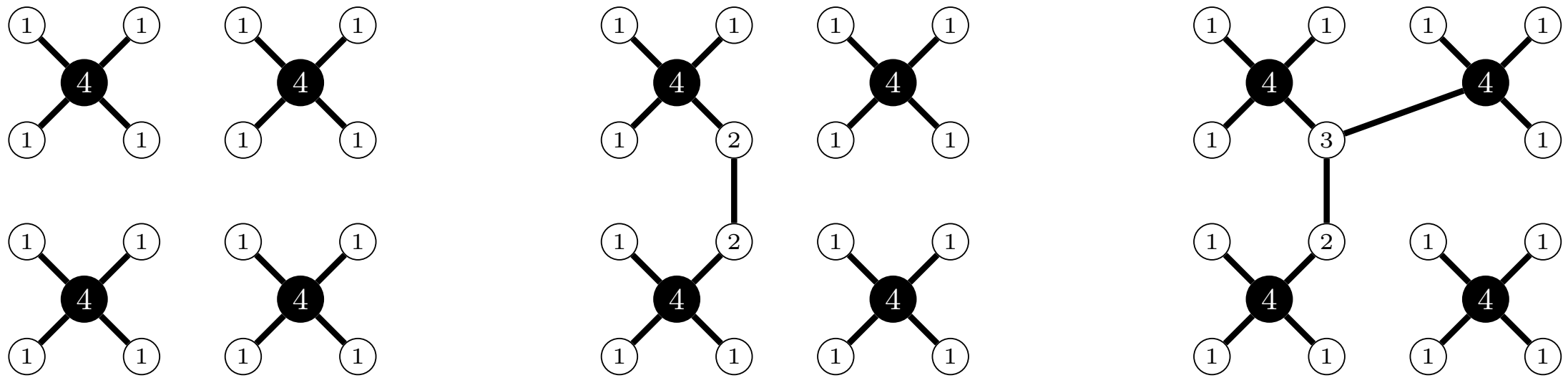
\begin{figure}[t]
\centering
\begin{tabular}{c@{\hspace{20pt}}c@{\hspace{20pt}}c}

\begin{subfigure}{0.3\textwidth}

\centering
\begin{tikzpicture}[scale=0.25]
	\starGadget{A1}{-4}{4}{45}{4}{1}{1}{1}{1}     
	\starGadget{A2}{4}{4}{45}{4}{1}{1}{1}{1}     
	\starGadget{A3}{-4}{-4}{45}{4}{1}{1}{1}{1}     
	\starGadget{A4}{4}{-4}{45}{4}{1}{1}{1}{1}   
\end{tikzpicture}
\caption{Initial star forest.}
\end{subfigure}%

&

\begin{subfigure}{0.3\textwidth}

\centering
\begin{tikzpicture}[scale=0.25]
	\starGadget{A1}{-4}{4}{45}{4}{1}{2}{1}{1}     
	\starGadget{A2}{4}{4}{45}{4}{1}{1}{1}{1}     
	\starGadget{A3}{-4}{-4}{45}{4}{1}{1}{1}{2}     
	\starGadget{A4}{4}{-4}{45}{4}{1}{1}{1}{1}   
		
	\draw[ultra thick] (A1-down) -- (A3-right);
\end{tikzpicture}
\caption{Forest after the first loop.}
\end{subfigure}%

&

\begin{subfigure}{0.3\textwidth}

\centering
\begin{tikzpicture}[scale=0.25]
	\starGadget{A1}{-4}{4}{45}{4}{1}{3}{1}{1}     
	\starGadgetMinusEdge{A2}{4}{4}{-135}{4}{1}{1}{1}     
	\starGadget{A3}{-4}{-4}{45}{4}{1}{1}{1}{2}     
	\starGadget{A4}{4}{-4}{45}{4}{1}{1}{1}{1}   
		
	\draw[ultra thick] (A1-down) -- (A3-right);
	\draw[ultra thick] (A1-down) -- (A2-center);
\end{tikzpicture}
\caption{Output forest.}
\end{subfigure}

\end{tabular}
\caption{Construction of $d = (4^4,3^1,2^1,1^{13})$, $\sum d = 34$, $\gamma = \gammakk(d) = 4$, and $\sum d^0 = 32$.
Also, $d^0 = (4^4,1^{16})$ and  $d^1 = (4^4,2^2,1^{14})$.}
\label{fig:ds-algorithm1}
\end{figure}

%% file: figures/DominatingSetExample2.tex
\usetikzlibrary{calc}

\newcommand{\MergeNodesMidpoint}[2]{%
  \coordinate (midpoint) at ($ (#1)!0.5!(#2) $);%
  \coordinate (#1) at (midpoint);%
  \coordinate (#2) at (midpoint);%
}

\begin{figure}[h]
\centering

\begin{subfigure}{\textwidth}
\centering
\begin{tikzpicture}[scale=0.2]
  \StarTree{-25}{0}{1}{3}{1}{1}{1}
  \StarTree{-15}{0}{2}{3}{1}{1}{1}
  \StarTree{-5}{0}{3}{3}{1}{1}{1}
  \StarTree{5}{0}{4}{3}{1}{1}{1}
  \StarTree{15}{0}{5}{3}{1}{1}{1}
  \StarTree{25}{0}{6}{3}{1}{1}{1}
\end{tikzpicture}
\caption{The initial star forest. $d^0 = (3^6, 1^{18})$.}
\end{subfigure}

\bigskip

\begin{subfigure}{\textwidth}
\centering
\begin{tikzpicture}[scale=0.2]
  \StarTree{0}{0}{1}{3}{1}{2}{1}
  \StarTree{10}{0}{2}{3}{1}{3}{1}
  \StarTree{20}{0}{3}{3}{1}{2}{1}
  \StarTree{30}{0}{4}{3}{1}{1}{1}
  \StarTree{40}{0}{5}{3}{1}{1}{1}
  \StarTree{50}{0}{6}{3}{1}{1}{1}

  \begin{scope}[every edge/.style={draw=black,ultra thick}]
    \path (C2) edge (C1);
    \path (C2) edge (C3);
  \end{scope}
\end{tikzpicture}
\caption{The forest after realizing $d_7 = 3$: $d^2 = (3^7, 2^2, 1^{15})$.}
\end{subfigure}

\bigskip

\begin{subfigure}{\textwidth}
\centering
\begin{tikzpicture}[scale=0.2]
  \StarTree{0}{0}{1}{3}{1}{2}{1}
  \StarTree{10}{0}{2}{3}{1}{3}{1}
  \StarTree{20}{0}{3}{3}{1}{2}{2}
  \StarTree{30}{0}{4}{3}{2}{1}{1}
  \StarTree{40}{0}{5}{3}{1}{1}{1}
  \StarTree{50}{0}{6}{3}{1}{1}{1}

  \begin{scope}[every edge/.style={draw=black,ultra thick}]
    \path (C2) edge (C1);
    \path (C2) edge (C3);
    \path (D3) edge (B4);
  \end{scope}
\end{tikzpicture}
\caption{The forest after finishing the first loop and realizing $d_{10} = 2$: $d^3 = (3^7, 2^4, 1^{13})$.}
\end{subfigure}

\bigskip

\begin{subfigure}{\textwidth}
\centering
\begin{tikzpicture}[scale=0.2]
  \StarTree{0}{0}{1}{3}{1}{2}{1}
  \StarTree{10}{0}{2}{3}{1}{3}{1}
  \StarTree{20}{0}{3}{3}{1}{2}{2}
  \StarTree{30}{0}{4}{3}{2}{1}{2}
  \StarTreeSharedLeft{36}{0}{5}{D4}{3}{1}{2}
  \StarTreeSharedLeft{42}{0}{6}{D5}{3}{1}{1}
  
  \begin{scope}[every edge/.style={draw=black,ultra thick}]
    \path (C2) edge (C1);
    \path (C2) edge (C3);
    \path (D3) edge (B4);
  \end{scope}
\end{tikzpicture}
\caption{The output forest of the algorithm: $d = (3^7, 2^6, 1^{9})$.}
\end{subfigure}
\caption{Construction of $d = (3^7, 2^6, 1^9)$, $\sum d = 42$, $\gamma = \gammakk(d) = 6$, and $\sum d^0 = 36$.}
\label{fig:ds-algorithm2}
\end{figure}
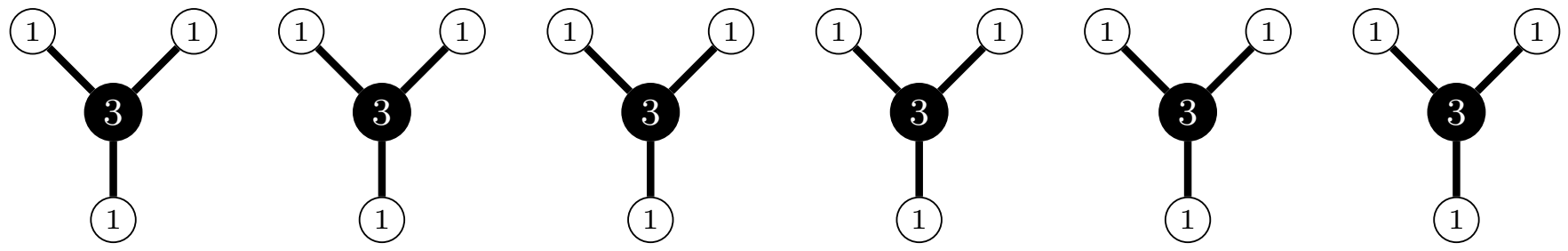
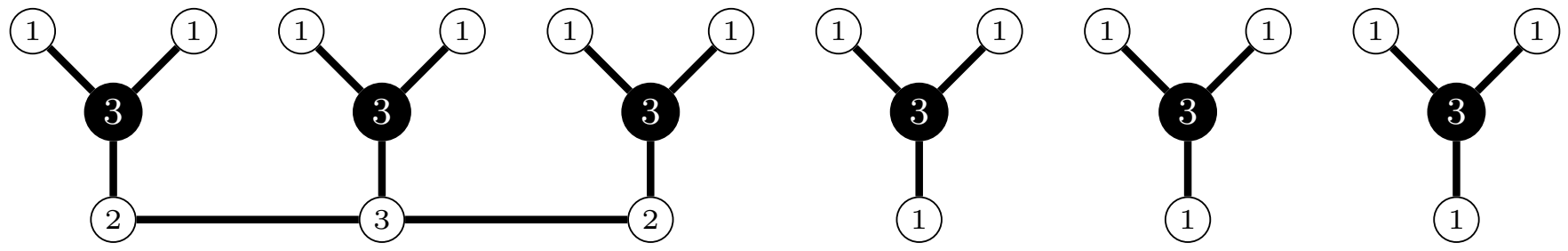
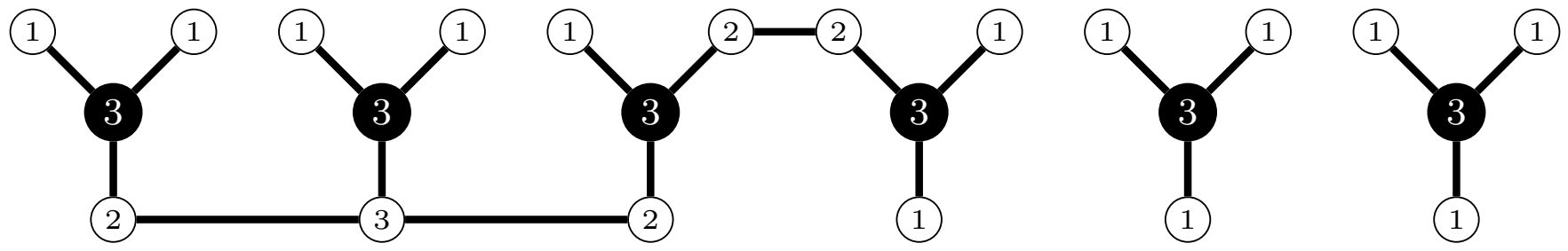
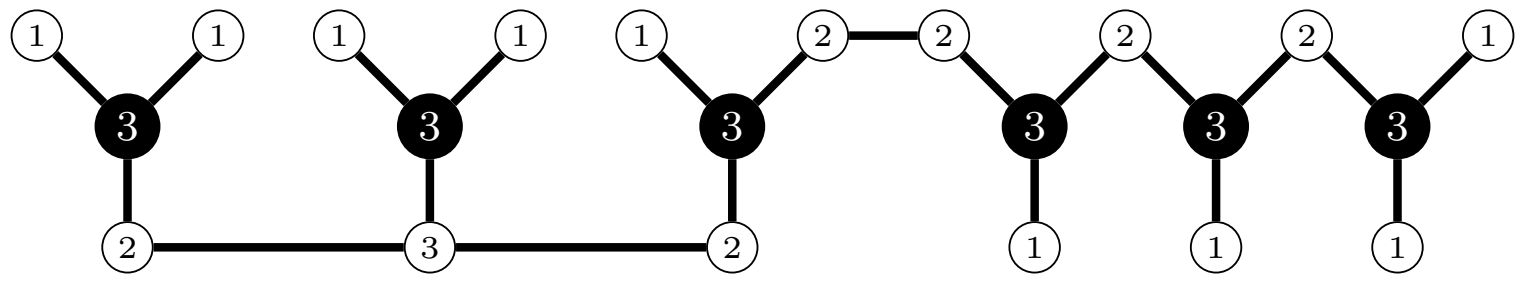

%% file: figures/TotalDominatingSetExample.tex


%% file: figures/TotalDominatingSetFrameworkExample.tex
\tikzset{
    tdsDVertex/.style={thesisDVertex},
    tdsNVertex/.style={thesisNVertex},
    tdsPendantEdge/.style={draw=black,line width=0.7pt},
    tdsCoreEdge/.style={draw=black,line width=1.15pt}
}

\newcommand{\tdsEvenInitialPair}[3]{%
    \begin{scope}[shift={({#2},{#3})}]
        \node (#1-left)  [tdsDVertex] at (-1.1,0) {3};
        \node (#1-right) [tdsDVertex] at ( 1.1,0) {3};

        \node (#1-lu) [tdsNVertex] at (-2.6, 1.15) {1};
        \node (#1-ld) [tdsNVertex] at (-2.6,-1.15) {1};
        \node (#1-li) [tdsNVertex] at (-0.2, 1.25) {1};
        \node (#1-ru) [tdsNVertex] at ( 2.6, 1.15) {1};
        \node (#1-rd) [tdsNVertex] at ( 2.6,-1.15) {1};
        \node (#1-ri) [tdsNVertex] at ( 0.2,-1.25) {1};

        \draw[tdsPendantEdge] (#1-left)  -- (#1-lu);
        \draw[tdsPendantEdge] (#1-left)  -- (#1-ld);
        \draw[tdsPendantEdge] (#1-left)  -- (#1-li);
        \draw[tdsPendantEdge] (#1-right) -- (#1-ru);
        \draw[tdsPendantEdge] (#1-right) -- (#1-rd);
        \draw[tdsPendantEdge] (#1-right) -- (#1-ri);
    \end{scope}
}

\newcommand{\tdsEvenMatchedPair}[4]{%
    \begin{scope}[shift={({#2},{#3})}]
        \node (#1-left)  [tdsDVertex] at (-1.1,0) {3};
        \node (#1-right) [tdsDVertex] at ( 1.1,0) {3};

        \node (#1-lu) [tdsNVertex] at (-2.6, 1.15) {1};
        \ifnum#4=1
            \node (#1-ld) [tdsNVertex] at (-2.6,-1.15) {1};
        \fi
        \node (#1-ru) [tdsNVertex] at ( 2.6, 1.15) {1};
        \node (#1-rd) [tdsNVertex] at ( 2.6,-1.15) {1};

        \draw[tdsCoreEdge] (#1-left) -- (#1-right);
        \draw[tdsPendantEdge] (#1-left) -- (#1-lu);
        \ifnum#4=1
            \draw[tdsPendantEdge] (#1-left) -- (#1-ld);
        \fi
        \draw[tdsPendantEdge] (#1-right) -- (#1-ru);
        \draw[tdsPendantEdge] (#1-right) -- (#1-rd);
    \end{scope}
}

\begin{figure}[t]
\centering
\captionsetup[subfigure]{justification=centering}

\begin{tabular}{c@{\hspace{10pt}}c@{\hspace{10pt}}c}
\begin{subfigure}{0.3\textwidth}
\centering
\begin{tikzpicture}[scale=0.32]
    \path[use as bounding box] (-6.3,-3.7) rectangle (6.3,3.7);
    \tdsEvenInitialPair{A}{-3.4}{ 2.25}
    \tdsEvenInitialPair{B}{ 3.4}{ 2.25}
    \tdsEvenInitialPair{C}{-3.4}{-2.25}
    \tdsEvenInitialPair{D}{ 3.4}{-2.25}
\end{tikzpicture}
\caption{Construction of $d^S=(3^8,1^{24})$.}
\end{subfigure}%
&
\begin{subfigure}{0.3\textwidth}
\centering
\begin{tikzpicture}[scale=0.32]
    \path[use as bounding box] (-6.3,-3.7) rectangle (6.3,3.7);
    \tdsEvenMatchedPair{A}{-3.4}{ 2.25}{1}
    \tdsEvenMatchedPair{B}{ 3.4}{ 2.25}{1}
    \tdsEvenMatchedPair{C}{-3.4}{-2.25}{1}
    \tdsEvenMatchedPair{D}{ 3.4}{-2.25}{1}
\end{tikzpicture}
\caption{Construction of $d^0=(3^8,1^{16})$.}
\label{fig:TDS_Ex1_2}
\end{subfigure}
&
\begin{subfigure}{0.3\textwidth}
\centering
\begin{tikzpicture}[scale=0.32]
    \path[use as bounding box] (-6.3,-3.7) rectangle (6.3,3.7);
    \tdsEvenMatchedPair{A}{-3.4}{ 2.25}{1}
    \tdsEvenMatchedPair{B}{ 3.4}{ 2.25}{0}
    \tdsEvenMatchedPair{C}{-3.4}{-2.25}{1}
    \tdsEvenMatchedPair{D}{ 3.4}{-2.25}{1}

    \draw[tdsCoreEdge] (A-rd) -- (C-ru);
    \draw[tdsCoreEdge] (B-left) -- (A-rd);
    \node[tdsNVertex] at (A-rd) {3};
    \node[tdsNVertex] at (C-ru) {2};
\end{tikzpicture}
\caption{Construction of $d=(3^9,2,1^{13})$.}
\label{fig:TDS_Ex1_3}
\end{subfigure}

\end{tabular}
\caption{Construction of $d=(3^9,2,1^{13})$ with $\gammat=\gammakk_t(d)=8$.}
\label{fig:tds-algorithm}
\end{figure}
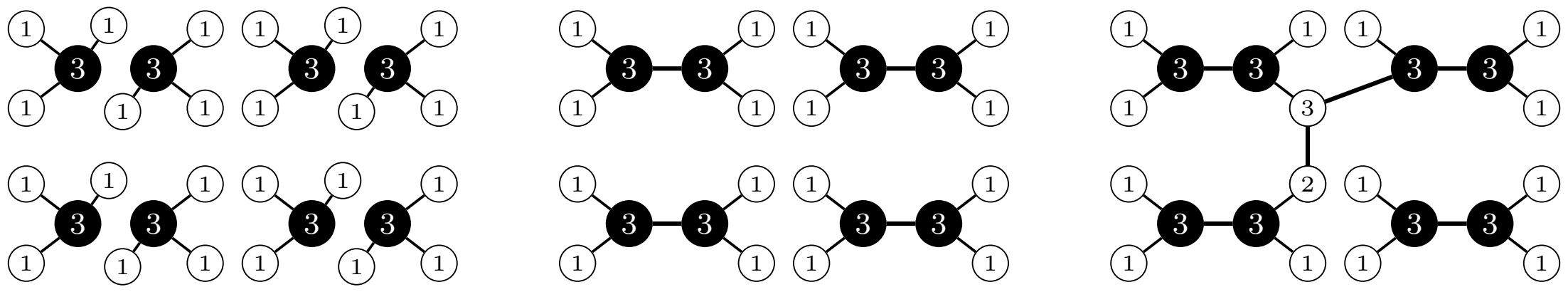

%% file: figures/TotalDominatingSetFrameworkExample2.tex
\newcommand{\tdsOddInitialPair}[3]{%
    \begin{scope}[shift={({#2},{#3})}]
        \node (#1-left)  [tdsDVertex] at (-1.4,0) {4};
        \node (#1-right) [tdsDVertex] at ( 1.4,0) {4};

        \node (#1-llu) [tdsNVertex] at (-3.0, 1.5) {1};
        \node (#1-llm) [tdsNVertex] at (-3.2, 0)   {1};
        \node (#1-lld) [tdsNVertex] at (-3.0,-1.5) {1};
        \node (#1-li)  [tdsNVertex] at (-0.2, 1.25) {1};
        \node (#1-rru) [tdsNVertex] at ( 3.0, 1.5) {1};
        \node (#1-rrm) [tdsNVertex] at ( 3.2, 0)   {1};
        \node (#1-rrd) [tdsNVertex] at ( 3.0,-1.5) {1};
        \node (#1-ri)  [tdsNVertex] at ( 0.2,-1.25) {1};

        \draw[tdsPendantEdge] (#1-left)  -- (#1-llu);
        \draw[tdsPendantEdge] (#1-left)  -- (#1-llm);
        \draw[tdsPendantEdge] (#1-left)  -- (#1-lld);
        \draw[tdsPendantEdge] (#1-left)  -- (#1-li);
        \draw[tdsPendantEdge] (#1-right) -- (#1-rru);
        \draw[tdsPendantEdge] (#1-right) -- (#1-rrm);
        \draw[tdsPendantEdge] (#1-right) -- (#1-rrd);
        \draw[tdsPendantEdge] (#1-right) -- (#1-ri);
    \end{scope}
}

\newcommand{\tdsOddInitialTriple}[3]{%
    \begin{scope}[shift={({#2},{#3})}]
        \node (#1-left)   [tdsDVertex] at (-3,0) {4};
        \node (#1-center) [tdsDVertex] at ( 0,0) {4};
        \node (#1-right)  [tdsDVertex] at ( 3,0) {4};

        \node (#1-llu) [tdsNVertex] at (-4.8, 1.5) {1};
        \node (#1-llm) [tdsNVertex] at (-5.0, 0)   {1};
        \node (#1-lld) [tdsNVertex] at (-4.8,-1.5) {1};
        \node (#1-li)  [tdsNVertex] at (-1.6, 1.2) {1};
        \node (#1-cu)  [tdsNVertex] at ( 0, 2.1)   {1};
        \node (#1-cd)  [tdsNVertex] at ( 0,-2.1)   {1};
        \node (#1-ciL) [tdsNVertex] at (-1.4,-1.2) {1};
        \node (#1-ciR) [tdsNVertex] at ( 1.4, 1.2) {1};
        \node (#1-ri)  [tdsNVertex] at ( 1.6,-1.2) {1};
        \node (#1-rru) [tdsNVertex] at ( 4.8, 1.5) {1};
        \node (#1-rrm) [tdsNVertex] at ( 5.0, 0)   {1};
        \node (#1-rrd) [tdsNVertex] at ( 4.8,-1.5) {1};

        \draw[tdsPendantEdge] (#1-left)   -- (#1-llu);
        \draw[tdsPendantEdge] (#1-left)   -- (#1-llm);
        \draw[tdsPendantEdge] (#1-left)   -- (#1-lld);
        \draw[tdsPendantEdge] (#1-left)   -- (#1-li);
        \draw[tdsPendantEdge] (#1-center) -- (#1-cu);
        \draw[tdsPendantEdge] (#1-center) -- (#1-cd);
        \draw[tdsPendantEdge] (#1-center) -- (#1-ciL);
        \draw[tdsPendantEdge] (#1-center) -- (#1-ciR);
        \draw[tdsPendantEdge] (#1-right)  -- (#1-ri);
        \draw[tdsPendantEdge] (#1-right)  -- (#1-rru);
        \draw[tdsPendantEdge] (#1-right)  -- (#1-rrm);
        \draw[tdsPendantEdge] (#1-right)  -- (#1-rrd);
    \end{scope}
}

\newcommand{\tdsOddPair}[5]{%
    \begin{scope}[shift={({#2},{#3})}]
        \node (#1-left)  [tdsDVertex] at (-1.4,0) {4};
        \node (#1-right) [tdsDVertex] at ( 1.4,0) {4};

        \node (#1-llu) [tdsNVertex] at (-3.0, 1.5) {1};
        \ifnum#4=1
            \node (#1-llm) [tdsNVertex] at (-3.2,0) {1};
        \fi
        \node (#1-lld) [tdsNVertex] at (-3.0,-1.5) {1};
        \node (#1-rru) [tdsNVertex] at ( 3.0, 1.5) {1};
        \ifnum#5=1
            \node (#1-rrm) [tdsNVertex] at (3.2,0) {1};
        \fi
        \node (#1-rrd) [tdsNVertex] at ( 3.0,-1.5) {1};

        \draw[tdsCoreEdge] (#1-left) -- (#1-right);
        \draw[tdsPendantEdge] (#1-left) -- (#1-llu);
        \ifnum#4=1
            \draw[tdsPendantEdge] (#1-left) -- (#1-llm);
        \fi
        \draw[tdsPendantEdge] (#1-left) -- (#1-lld);
        \draw[tdsPendantEdge] (#1-right) -- (#1-rru);
        \ifnum#5=1
            \draw[tdsPendantEdge] (#1-right) -- (#1-rrm);
        \fi
        \draw[tdsPendantEdge] (#1-right) -- (#1-rrd);
    \end{scope}
}

\newcommand{\tdsOddTriple}[3]{%
    \begin{scope}[shift={({#2},{#3})}]
        \node (#1-left)   [tdsDVertex] at (-3,0) {4};
        \node (#1-center) [tdsDVertex] at ( 0,0) {4};
        \node (#1-right)  [tdsDVertex] at ( 3,0) {4};

        \node (#1-cu)  [tdsNVertex] at ( 0, 2.1)   {1};
        \node (#1-cd)  [tdsNVertex] at ( 0,-2.1)   {1};
        \node (#1-llu) [tdsNVertex] at (-4.8, 1.5) {1};
        \node (#1-llm) [tdsNVertex] at (-5.0, 0)   {1};
        \node (#1-lld) [tdsNVertex] at (-4.8,-1.5) {1};
        \node (#1-rru) [tdsNVertex] at ( 4.8, 1.5) {1};
        \node (#1-rrm) [tdsNVertex] at ( 5.0, 0)   {1};
        \node (#1-rrd) [tdsNVertex] at ( 4.8,-1.5) {1};

        \draw[tdsCoreEdge] (#1-left) -- (#1-center);
        \draw[tdsCoreEdge] (#1-center) -- (#1-right);
        \draw[tdsPendantEdge] (#1-center) -- (#1-cu);
        \draw[tdsPendantEdge] (#1-center) -- (#1-cd);
        \draw[tdsPendantEdge] (#1-left) -- (#1-llu);
        \draw[tdsPendantEdge] (#1-left) -- (#1-llm);
        \draw[tdsPendantEdge] (#1-left) -- (#1-lld);
        \draw[tdsPendantEdge] (#1-right) -- (#1-rru);
        \draw[tdsPendantEdge] (#1-right) -- (#1-rrm);
        \draw[tdsPendantEdge] (#1-right) -- (#1-rrd);
    \end{scope}
}

\newcommand{\tdsOddJointPair}[3]{%
    \begin{scope}[shift={({#2},{#3})}]
        \node (#1-left)  [tdsDVertex] at (-1.4,0) {4};
        \node (#1-right) [tdsDVertex] at ( 1.4,0) {4};

        \node (#1-llu)   [tdsNVertex] at (-3.0, 1.5) {1};
        \node (#1-llm)   [tdsNVertex] at (-3.2, 0)   {1};
        \node (#1-joint) [tdsNVertex] at ( 0,-3.7)   {3};
        \node (#1-rru)   [tdsNVertex] at ( 3.0, 1.5) {1};
        \node (#1-rrm)   [tdsNVertex] at ( 3.2, 0)   {1};
        \node (#1-rrd)   [tdsNVertex] at ( 3.0,-1.5) {1};

        \draw[tdsCoreEdge] (#1-left) -- (#1-right);
        \draw[tdsPendantEdge] (#1-left) -- (#1-llu);
        \draw[tdsPendantEdge] (#1-left) -- (#1-llm);
        \draw[tdsPendantEdge] (#1-left) -- (#1-joint);
        \draw[tdsPendantEdge] (#1-right) -- (#1-rru);
        \draw[tdsPendantEdge] (#1-right) -- (#1-rrm);
        \draw[tdsPendantEdge] (#1-right) -- (#1-rrd);
    \end{scope}
}

\begin{figure}[t]
\centering
\captionsetup[subfigure]{justification=centering}

\begin{subfigure}{0.48\textwidth}
\centering
\begin{tikzpicture}[scale=0.32]
    \path[use as bounding box] (-9.8,-8.4) rectangle (9.3,8.4);
    \tdsOddInitialTriple{T}{-4.5}{ 5.8}
    \tdsOddInitialPair{P2}{ 5.6}{ 5.8}
    \tdsOddInitialPair{P3}{-4.5}{ 0.0}
    \tdsOddInitialPair{P4}{ 4.5}{ 0.0}
    \tdsOddInitialPair{P5}{-4.5}{-5.8}
    \tdsOddInitialPair{P6}{ 4.5}{-5.8}
\end{tikzpicture}
\caption{Construction of $d^S=(4^{13},1^{52})$.}
\end{subfigure}%
\hfill%
\begin{subfigure}{0.48\textwidth}
\centering
\begin{tikzpicture}[scale=0.32]
    \path[use as bounding box] (-9.8,-8.4) rectangle (9.3,8.4);
    \tdsOddTriple{T}{-4.5}{ 5.8}
    \tdsOddPair{P2}{ 5.6}{ 5.8}{1}{1}
    \tdsOddPair{P3}{-4.5}{ 0.0}{1}{1}
    \tdsOddPair{P4}{ 4.5}{ 0.0}{1}{1}
    \tdsOddPair{P5}{-4.5}{-5.8}{1}{1}
    \tdsOddPair{P6}{ 4.5}{-5.8}{1}{1}
\end{tikzpicture}
\caption{Construction of $d^0=(4^{13},1^{38})$.}
\end{subfigure}

\par\medskip

\begin{subfigure}{0.76\textwidth}
\centering
\begin{tikzpicture}[scale=0.30]
    \path[use as bounding box] (-9.8,-8.4) rectangle (9.3,8.4);
    \tdsOddTriple{T}{-4.5}{ 5.8}
    \tdsOddJointPair{P2}{ 5.6}{ 5.8}
    \tdsOddPair{P4}{-4.5}{ 0.0}{0}{1}
    \begin{scope}[rotate around={180:(5.6,0.0)}]
        \tdsOddPair{P3}{ 5.6}{ 0.0}{1}{1}
    \end{scope}
    \begin{scope}[rotate around={180:(-4.5,-5.8)}]
        \tdsOddPair{P5}{-4.5}{-5.8}{0}{0}
    \end{scope}
    \begin{scope}[rotate around={180:(4.5,-5.8)}]
        \tdsOddPair{P6}{ 4.5}{-5.8}{1}{0}
    \end{scope}

    \draw[tdsCoreEdge] (T-rrd) -- (P2-joint);
    \draw[tdsCoreEdge] (P2-joint) -- (P3-lld);
    \draw[tdsCoreEdge] (P3-rrd) -- (P4-rru);

    \node (M45) [tdsNVertex] at (-5.9,-2.9) {2};
    \node (M56) [tdsNVertex] at ( 0.0,-5.8) {2};
    \draw[tdsCoreEdge] (P4-left) -- (M45) -- (P5-right);
    \draw[tdsCoreEdge] (P5-left) -- (M56) -- (P6-right);

    \node[tdsNVertex] at (T-rrd)  {2};
    \node[tdsNVertex] at (P3-lld) {2};
    \node[tdsNVertex] at (P3-rrd) {2};
    \node[tdsNVertex] at (P4-rru) {2};
\end{tikzpicture}
\caption{Construction of $d=(4^{13},3,2^6,1^{29})$.}
\end{subfigure}

\caption{Construction of $d=(4^{13},3,2^6,1^{29})$ with $\gammat=\gammakk_t(d)=13$.}
\label{fig:tds-algorithm2}
\end{figure}
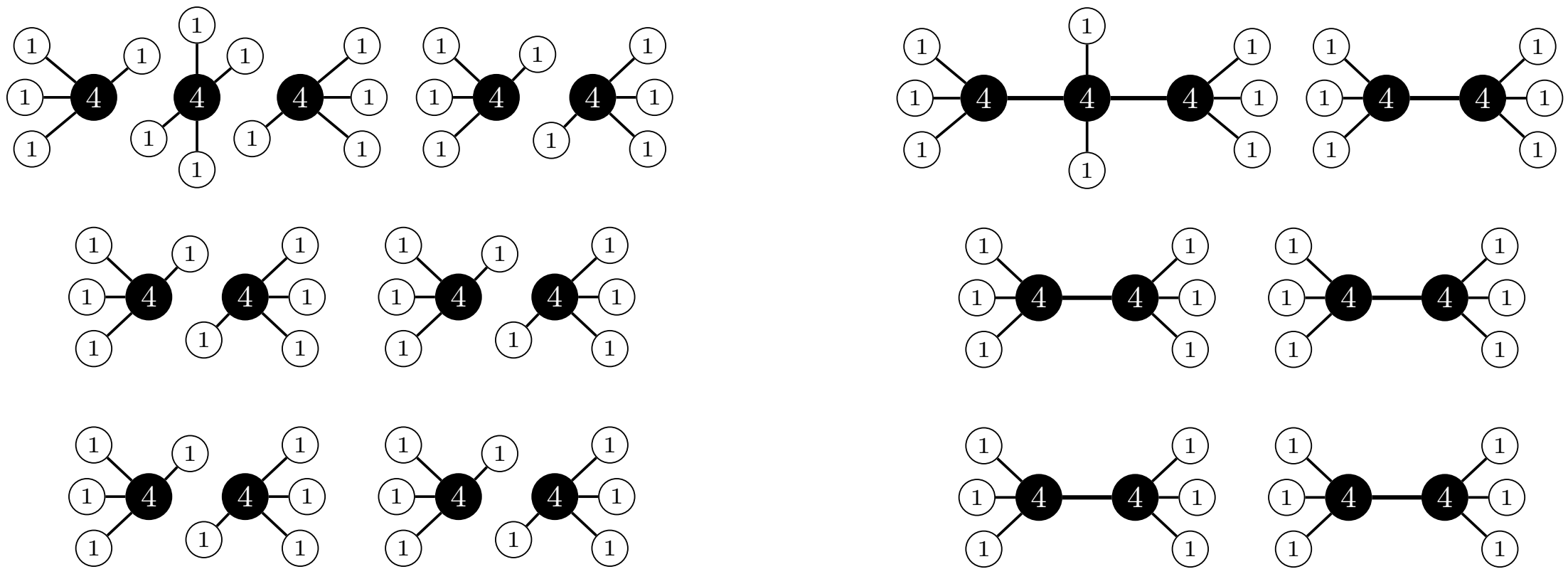
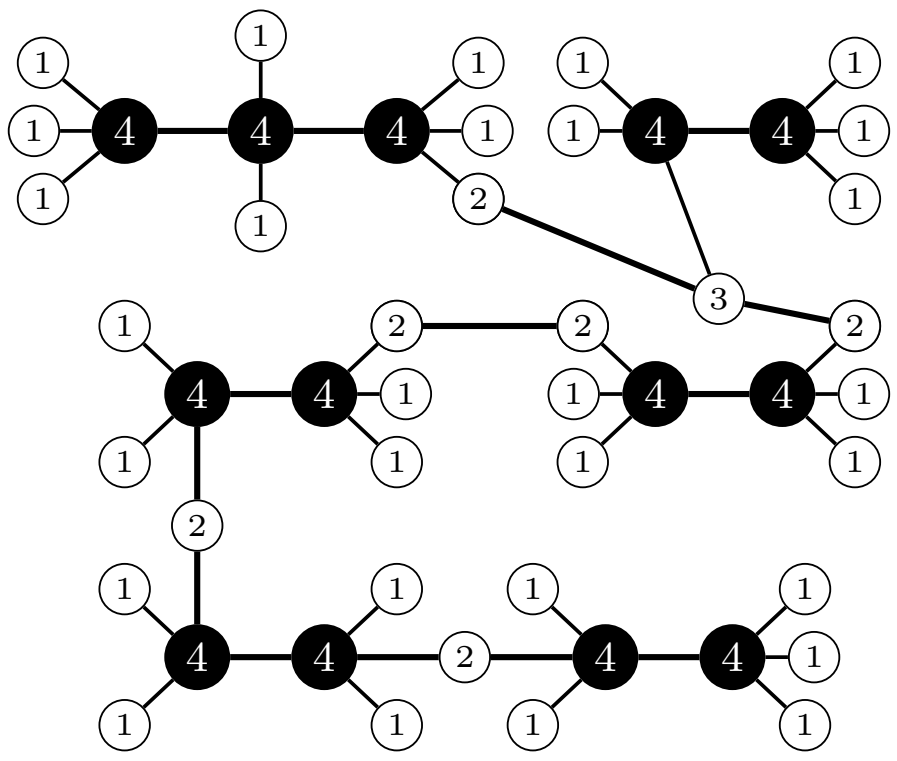